%% file: arXiv_v2.tex
\documentclass[superscriptaddress,twocolumn,nopreprintnumbers,floatfix,nofootinbib,trackchanges,table]{aastex631}
\pdfoutput=1
\usepackage[utf8]{inputenc}
\usepackage{xspace}
\usepackage{acronym}
\usepackage{booktabs}
\usepackage{amsmath,amssymb,graphicx}
\usepackage{color}
\usepackage{xcolor}
\usepackage{tabularx}
\usepackage{multirow}
\usepackage{makecell}
\usepackage{array}
\let\tablenum\relax
\usepackage{longtable}
\usepackage{siunitx}
\usepackage{xparse}
\usepackage{hyperref}
\usepackage{bm}
\definecolor{medium-blue}{rgb}{0,0,1}
\hypersetup{colorlinks, urlcolor={medium-blue}}
\usepackage[hang,flushmargin]{footmisc}

\newcolumntype{Y}{>{\centering\arraybackslash}X}
\newcolumntype{Z}{>{\raggedright\arraybackslash}X}
\newcolumntype{K}{>{\raggedleft\arraybackslash}X}
\newcolumntype{U}{>{\hsize=1.01\hsize}Y}
\newcolumntype{V}{>{\hsize=1.2\hsize}Y}
\newcolumntype{W}{>{\hsize=0.71\hsize}Y}

\usepackage{scrextend}
\usepackage{textgreek}
\usepackage[caption=false]{subfig}

\usepackage{makerobust}

\renewcommand{\today}{\number\day\space\ifcase\month\or
  January\or February\or March\or April\or May\or June\or
  July\or August\or September\or October\or November\or December\fi
  \space\number\year}

\input{macros}

\input{macros_from_json}
\input{acronyms}

\input{gwtc-common-files__acronyms}
\input{gwtc-common-files__commands}

\input{gwtc-common-files__macros__symbol_macros}

\input{gwtc-common-files__macros__misc_macros}

\input{gwtc-common-files__macros__datetime_macros}
\input{gwtc-common-files__macros__quantities_macros}

\definecolor{NOTECOLOR}{rgb}{0.4, 0.2, 0.1}
\definecolor{notecolor}{rgb}{0.4, 0.2, 0.1}

\begin{document}

\title{GWTC-4.0: Population Properties of Merging Compact Binaries}

\input{LSC-Virgo-KAGRA-Authors-Aug-2024-aas.tex}

\date[\relax]{Compiled: \today}

\begin{abstract}

\input{abstract.tex}
\end{abstract}

\pacs{%
04.80.Nn,
04.25.dg,
95.85.Sz,
97.80.-d  
04.30.Db,
04.30.Tv 
}

\input{introduction.tex}
\input{methods.tex}
\input{data.tex}
\input{full_cbc_population.tex}
\input{ns_results.tex}

\input{bbh_results.tex}

\input{conclusion.tex}

\section*{acknowledgments}
\input{gwtc-common-files__LVKack}

\appendix
\input{appendix.tex}

\clearpage

\bibliography{}

\end{document}

%% file: macros.tex
\newcommand{\CIPlusMinus}[1]{{#1[median]^{+#1[error plus]}_{-#1[error minus]}}}

\newcommand{\CIBoundsDash}[1]{{#1[5th percentile]\textendash#1[95th percentile]}}
\NewDocumentCommand{\numRound}{O{2} m}{\num[round-mode=figures, round-precision=#1]{#2}}
\NewDocumentCommand{\numRoundSci}{O{2} m}{\num[scientific-notation=true, round-mode=figures, round-precision=#1]{#2}}

\newcommand{\Ndet}[0]{N_{\rm det}}
\newcommand{\Nfound}[0]{N_{\rm found}}
\newcommand{\Nexp}[0]{N_{\rm exp}}
\newcommand{\Ndraw}[0]{N_{\rm draw}}
\newcommand{\Npe}[0]{N_{\rm PE}}

\newcommand{\parametric}[0]{strongly modeled approach\xspace}
\newcommand{\nonparametric}[0]{weakly modeled approach\xspace}

\newcommand{\parametricAdj}[0]{strongly modeled\xspace}
\newcommand{\nonparametricAdj}[0]{weakly modeled\xspace}

\newcommand{\parametrics}[0]{strongly modeled approaches\xspace}
\newcommand{\nonparametrics}[0]{weakly modeled approaches\xspace}

\newcommand{\Nonparametrics}[0]{Weakly modeled approaches\xspace}

\newcommand{\vamana}[0]{\ac{FM}\xspace}

\newcommand{\PLPeak}{\textsc{Power Law + Peak}\xspace}
\newcommand{\PP}{\textsc{PP}\xspace}
\newcommand{\PDB}[0]{\textsc{FullPop}-4.0\xspace}
\newcommand{\BrokenPLTwoPeaks}[0]{\textsc{Broken Power Law + 2 Peaks}\xspace}
\newcommand{\BPTP}[0]{\textsc{BP2P}\xspace}
\newcommand{\BPOP}[0]{\textsc{BP1P}\xspace}
\newcommand{\BPHP}[0]{\textsc{BP3P}\xspace}
\newcommand{\alphaOT}[0]{$3.5^{+0.6}_{-0.6} \,\Msun$\xspace}
\newcommand{\muOOT}[0]{$10.2^{+0.3}_{-0.6}\,\Msun$\xspace}
\newcommand{\muTOT}[0]{$35^{+1.7}_{-2.9}\,\Msun$\xspace}
\newcommand{\ExtendedBrokenPLTwoPeaks}[0]{\textsc{Extended Broken Power Law + 2 Peaks}\xspace}
\newcommand{\BSpline}[0]{\textsc{B-Spline}\xspace}

\newcommand{\default}[0]{\textsc{Gaussian Component Spins}\xspace}
\newcommand{\effectiveSpinModel}[0]{\textsc{Gaussian Effective Spins}\xspace}
\newcommand{\skewnormalChiEff}[0]{\textsc{Skew-normal Effective Spin}\xspace}

\newcommand{\powerlawRedshift}[0]{\textsc{Power Law Redshift}\xspace}
\newcommand{\qchiefflinearcorrelation}[0]{\textsc{$(q, \chieff)$ Linear}\xspace}
\newcommand{\qchieffsplinecorrelation}[0]{\textsc{$(q, \chieff)$ Spline}\xspace}

\newcommand{\linearcorrelation}[0]{\textsc{Linear}\xspace}
\newcommand{\splinecorrelation}[0]{\textsc{Spline}\xspace}
\newcommand{\copulacorrelation}[0]{\textsc{Copula}\xspace}

\newcommand{\confidenceLevel}{90\%}

\newcommand{\chiEffMinGaussianGTWCThree}{-0.18^{+0.09}_{-0.12}}
\newcommand{\fNegChieffGaussianGTWCThree}{0.28^{+0.12}_{-0.13}}

\newcommand{\PEhyperparam}{\ensuremath{\boldsymbol{\Lambda}}}%

\newcommand{\gwtcthree}{\gwtc[3.0]\xspace}
\newcommand{\gwtctwo}{\gwtc[2.0]\xspace}
\newcommand{\gwtctwopone}{\gwtc[2.1]\xspace}
\newcommand{\gwtcfour}{\thisgwtc\xspace}

%% file: acronyms.tex
\acrodef{EM}[EM]{electromagnetic}
\acrodef{PPISNe}[PPISNe]{pulsational pair-instability supernovae}
\acrodef{PPISN}[PPISN]{pulsational pair-instability supernova}
\acrodef{PISN}[PISN]{pair-instability supernova}
\acrodef{TOV}[TOV]{Tolman--Oppenheimer--Volkoff}
\acrodef{AGN}[AGN]{active galactic nuclei}
\acrodef{PPD}[PPD]{population predictive distribution}
\acrodef{GWTC}[GWTC]{Gravitational Wave Transient Catalog}
\acrodef{AR}[AR]{\textsc{Autoregressive Process}}
\acrodef{BGP}[BGP]{\textsc{Binned Gaussian Process}}
\acrodef{BS}[BS]{\textsc{Basis Spline}}
\acrodef{FM}[FM]{\textsc{Flexible Mixtures}}
\acrodef{HMC}[HMC]{Hamiltonian Monte Carlo}
\acrodef{NUTS}[NUTS]{No-U-Turn Sampler}
\acrodef{IID}[IID]{independently and identically distributed}
\acrodef{NID}[NID]{nonindependently but identically distributed}
\acrodef{SFR}[SFR]{star formation rate}
\acrodef{IAS}[IAS]{Institute for Advanced Study}
\acrodef{OGC}[OGC]{Open Gravitational-wave Catalog}
\acrodef{ER}[ER]{engineering run}
\acrodef{ML}[ML]{Machine Learning}
\acrodef{KDE}[KDE]{kernel density estimation}
\acrodef{GraceDB}[GraceDB]{Gravitational-Wave Candidate Event Database}

\newcommand{\GWPOPULATION}{\textsc{GWPopulation}\xspace}

\newcommand{\JAX}{\textsc{Jax}\xspace}
\newcommand{\NUMPYRO}{\textsc{NumPyro}\xspace}

%% file: gwtc-common-files__acronyms.tex
\usepackage{amstext}

\makeatletter
\@ifpackageloaded{cleveref}{
  \AtBeginDocument{
    \def\ltx@label#1{\cref@label{#1}}%
    \def\label@in@display@noarg#1{\cref@old@label@in@display{#1}}%
    \def\label@in@mmeasure@noarg#1{%
      \begingroup%
        \measuring@false%
        \cref@old@label@in@display{#1}%
      \endgroup}%
  }
}{}
\makeatother

\protected\def\protectedacused{\acused}

\acrodef{LIGO}[LIGO]{Laser Interferometer Gravitational-Wave Observatory}
\acrodef{LHO}[LHO]{\ac{LIGO} Hanford Observatory}
\acrodef{LLO}[LLO]{\ac{LIGO} Livingston Observatory}
\acrodef{KAGRA}[KAGRA]{KAGRA}\acused{KAGRA}
\acrodef{iKAGRA}[iKAGRA]{initial-phase \ac{KAGRA}}
\acrodef{bKAGRA}[bKAGRA]{baseline-design \ac{KAGRA}}
\acrodef{GEO}[GEO]{GEO\,600 \ac{GW} detector}
\acrodef{aLIGO}{Advanced \ac{LIGO}}
\acrodef{A+}{Advanced+ \ac{LIGO}}
\acrodef{Asharp}[\ensuremath{\text{A}^\sharp}]{\ac{LIGO} \acs{Asharp}}
\acrodef{AdV}{Advanced \acl{Virgo}}
\acrodef{AdV+}{Advanced \acl{Virgo}+}
\acrodef{Virgo}{Virgo}\acused{Virgo}
\acrodef{VirgoNEXT}[Virgo\_nEXT]{Virgo\_nEXT}\acused{VirgoNEXT}

\acrodef{LSC}[LSC]{\acs{LIGO} Scientific Collaboration}
\acrodef{LV}[LV]{\acs{LIGO}--\acs{Virgo} Collaboration\protect\protectedacused{LVC}}
\acrodef{LVC}[LV]{\acs{LIGO}--\acs{Virgo} Collaboration\protect\protectedacused{LV}}
\acrodef{LVK}[LVK]{\acs{LIGO}--\ac{Virgo}--\ac{KAGRA} Collaboration}
\acrodef{IGWN}[IGWN]{International \ac{GWH} Observatory Network}

\acrodef{O1}[O1]{first observing run}
\acrodef{O2}[O2]{second observing run}
\acrodef{O3}[O3]{third observing run}
\acrodef{O3a}[O3a]{first half of the third observing run}
\acrodef{O3b}[O3b]{second half of the third observing run}
\acrodef{O3GK}[O3GK]{observing run}
\acrodef{O4}[O4]{fourth observing run}
\acrodef{O4a}[O4a]{first part of the fourth observing run}
\acrodef{O4b}[O4b]{second part of the fourth observing run}
\acrodef{O4c}[O4c]{third part of the fourth observing run}
\acrodef{O5}[O5]{fifth observing run}

\acrodef{BH}[BH]{black hole}
\acrodef{BBH}[BBH]{binary black hole}
\acrodef{BNS}[BNS]{binary neutron star}
\acrodef{IMBH}[IMBH]{intermediate-mass black hole}
\acrodef{NS}[NS]{neutron star}
\acrodef{BHNS}[BHNS]{black hole--neutron star binary}
\acrodefplural{BHNS}[BHNSs]{black hole--neutron star binaries}
\acrodef{NSBH}[NSBH]{neutron star--black hole binary}
\acrodefplural{NSBH}[NSBHs]{neutron star--black hole binaries}
\acrodef{PBH}[PBH]{primordial \ac{BH}}
\acrodef{CBC}[CBC]{compact binary coalescence}

\acrodef{GW}[GW]{gravitational wave\protect\protectedacused{GWH}}
\acrodef{GWH}[GW]{gravitational-wave\protect\protectedacused{GW}}
\acrodef{IFO}[IFO]{interferometer}
\acrodef{SNR}[SNR]{signal-to-noise ratio}
\acrodef{FAR}[FAR]{false alarm rate}
\acrodef{IFAR}[IFAR]{inverse false alarm rate}
\acrodef{FAP}[FAP]{false alarm probability}
\acrodef{PSD}[PSD]{power spectral density}
\acrodef{GR}[GR]{general relativity}
\acrodef{NR}[NR]{numerical relativity}
\acrodef{PN}[PN]{post-Newtonian}
\acrodef{EOB}[EOB]{effective-one-body}
\acrodef{ROM}[ROM]{reduced-order model}
\acrodef{IMR}[IMR]{inspiral--merger--ringdown}
\acrodef{PDF}[PDF]{probability density function}
\acrodef{PE}[PE]{parameter estimation}
\acrodef{CI}[CI]{credible interval}
\acrodef{CL}[CL]{credible level}
\acrodef{EOS}[EoS]{equation of state}
\acrodef{KLD}[KLD]{Kullback--Leibler divergence}
\acrodef{JSD}[JSD]{Jensen--Shannon divergence}
\acrodef{GCN}[GCN]{General Coordinates Network}
\acrodef{GWTC}[GWTC]{Gravitational-Wave Transient Catalog}
\acrodef{GWOSC}[GWOSC]{Gravitational Wave Open Science Center}
\acrodef{WDM}[WDM]{Wilson--Debauchies--Meyer}

\acrodef{CWB}[cWB]{coherent WaveBurst}
\acrodef{LAL}[LAL]{\ac{LIGO} algorithm library}

\acrodef{CHRoCC}{central heating radius of curvature correction}
\acrodef{NonSENS}{non-stationary estimation and noise subtraction}

\acrodef{PTA}{Pulsar Timing Array}

%% file: gwtc-common-files__commands.tex
\newcommand\gwtc[1][?]{\mbox{GWTC\if#1?\else-#1\fi}}
\newcommand\thisgwtcversionmajor{4}
\newcommand\thisgwtcversionminor{0}
\newcommand\thisgwtcversionfull{\thisgwtcversionmajor.\thisgwtcversionminor}
\newcommand\thisgwtcversion\thisgwtcversionfull
\newcommand\thisgwtc{\gwtc[\thisgwtcversion]}

%% file: gwtc-common-files__macros__symbol_macros.tex
\newcommand{\Msun}{\ensuremath{\mathit{M_\odot}}}

\newcommand\perGpcyr{\ensuremath{\mathrm{Gpc}^{-3}\,\mathrm{yr}^{-1}}}

\newcommand{\massone}{\ensuremath{m_1}}

\newcommand{\chieff}{\ensuremath{\chi_\mathrm{eff}}}
\newcommand{\chip}{\ensuremath{\chi_\mathrm{p}}}

\newcommand\PEpdfp{\ensuremath{p}}

\newcommand{\PEparameter}{\ensuremath{\boldsymbol{\theta}}}%

\newcommand\PEpdf[2][?]{\ensuremath{\PEpdfp({#2}\ifx#1?\else | {#1}\fi)}}

\newcommand\PEpriorpdfpi{\ensuremath{\pi}}
\newcommand\PEpdfprior[1]{\ensuremath{\PEpriorpdfpi({#1})}}
\newcommand\PEprior[1][\PEparameter]{\PEpdfprior{#1}}
\newcommand\PEpriorpe[1][\PEparameter]{{\let\keepPEpriorpdfpi\PEpriorpdfpi\def\PEpriorpdfpi{\keepPEpriorpdfpi_{\text{PE}}}\PEprior[#1]\let\PEpriorpdfpi\keepPEpriorpdfpi}}

\newcommand{\VT}{\ensuremath{\langle VT \rangle}\xspace}
\newcommand{\pastro}{\ensuremath{p_{\mathrm{astro}}}\xspace}

%% file: gwtc-common-files__macros__misc_macros.tex
\newcommand{\soft}[1]{\textsc{#1}}

\newcommand{\GSTLAL}{\soft{GstLAL}\xspace}

\newcommand{\CWB}{\soft{cWB}\xspace}

\newcommand{\PYCBC}{\soft{PyCBC}\xspace}
\newcommand{\MBTA}{\soft{MBTA}\xspace}

\newcommand{\BILBY}{\soft{Bilby}\xspace}

\newcommand{\DYNESTY}{\soft{Dynesty}\xspace}

\newcommand{\IMRPhenomXPHMST}{\soft{IMRPhenomXPHM\_SpinTaylor}\xspace}

\newcommand{\SEOBNRFIVEPHM}{\soft{SEOBNRv5PHM}\xspace}

\newcommand{\IMRPhenomPTWONRTidal}{\soft{IMRPhenomPv2\_NRTidal}\xspace}

\newcommand{\SURSEVENDQFOUR}{\soft{NRSur7dq4}\xspace}

%% file: gwtc-common-files__macros__quantities_macros.tex
\DeclareSIUnit\parsec{pc}
\DeclareSIUnit\Mpc{\mega\parsec}
\DeclareSIUnit\yr{yr}
\DeclareSIUnit\GpcCubedYear{\giga\parsec\cubed\yr}

%% file: LSC-Virgo-KAGRA-Authors-Aug-2024-aas.tex
\author[0000-0003-4786-2698]{A.~G.~Abac}
\affiliation{Max Planck Institute for Gravitational Physics (Albert Einstein Institute), D-14476 Potsdam, Germany}
\author{I.~Abouelfettouh}
\affiliation{LIGO Hanford Observatory, Richland, WA 99352, USA}
\author{F.~Acernese}
\affiliation{Dipartimento di Farmacia, Universit\`a di Salerno, I-84084 Fisciano, Salerno, Italy}
\affiliation{INFN, Sezione di Napoli, I-80126 Napoli, Italy}
\author[0000-0002-8648-0767]{K.~Ackley}
\affiliation{University of Warwick, Coventry CV4 7AL, United Kingdom}
\author[0000-0001-5525-6255]{C.~Adamcewicz}
\affiliation{OzGrav, School of Physics \& Astronomy, Monash University, Clayton 3800, Victoria, Australia}
\author[0009-0004-2101-5428]{S.~Adhicary}
\affiliation{The Pennsylvania State University, University Park, PA 16802, USA}
\author{D.~Adhikari}
\affiliation{Max Planck Institute for Gravitational Physics (Albert Einstein Institute), D-30167 Hannover, Germany}
\affiliation{Leibniz Universit\"{a}t Hannover, D-30167 Hannover, Germany}
\author[0000-0002-4559-8427]{N.~Adhikari}
\affiliation{University of Wisconsin-Milwaukee, Milwaukee, WI 53201, USA}
\author[0000-0002-5731-5076]{R.~X.~Adhikari}
\affiliation{LIGO Laboratory, California Institute of Technology, Pasadena, CA 91125, USA}
\author{V.~K.~Adkins}
\affiliation{Louisiana State University, Baton Rouge, LA 70803, USA}
% ...

\collaboration{3000}{The LIGO Scientific Collaboration, the Virgo Collaboration, and the KAGRA Collaboration}

%% file: abstract.tex
\noindent We detail the population properties of merging compact objects using \counts[combined][total] mergers from the cumulative \acl{GWTC} \thisgwtcversionfull{}, which includes three types of binary mergers: \acl{BNS}, \acl{NSBH}, and \acl{BBH} mergers.
We resolve multiple over- and under-densities in the \acl{BH} mass distribution: features persist at primary masses of $10\,\Msun$ and $35\,\Msun$ with a possible third feature at ${\sim}20\,\Msun$.  These are departures from an otherwise power-law-like continuum that steepens above $35\,\Msun$.
\Aclp{BBH} with primary masses near $10\,\Msun$ are more likely to have less massive secondaries, with a mass ratio distribution peaking at $q = \CIPlusMinus{\IsoPeakIID[mass_ratio][peak][peak_location]}$, potentially a signature of stable mass transfer during binary evolution.
\Acl{BH} spins are inferred to be non-extremal, with $90\%$ of \aclp{BH} having $\chi<\MagTruncnormIidTiltIsotropicTruncnormNid[ppd][a][90th percentile]$, and preferentially aligned with binary orbits, implying many merging binaries form in isolation. However, we find a significant fraction, $\gaussianChiEffChiP[param][f_neg_eff][5th percentile]$--$\EpsSkewNormalChiEff[param][f_neg_chi_eff][95th percentile]$, of binaries have negative effective inspiral spins, suggesting many could be formed dynamically in gas-free environments.
We find evidence for correlation between effective inspiral spin and mass ratio, though it is unclear if this is driven by variation in the mode of the distribution or the width.
The \acl{BBH} merger rate increases with redshift as $(1+z)^\kappa$ with $\kappa = \defaultbbh[lamb][median]^{+\defaultbbh[lamb][error plus]}_{-\defaultbbh[lamb][error minus]}$, consistent with the cosmic star formation density.
While there is no evidence of the mass spectrum evolving with redshift, the distribution of effective inspiral spin is found to broaden as redshift increases out to $z\approx 1$.
We infer the local merger rates (i.e., at redshift $z=0$) to be $\FullMassSpectrumMerged[R_BNS][5th_percentile]\textendash\FullMassSpectrumMerged[R_BNS][95th_percentile]\,\perGpcyr$ for \aclp{BNS}, $\FullMassSpectrumMerged[R_NSBH][5th_percentile]\textendash\FullMassSpectrumMerged[R_NSBH][95th_percentile]\,\perGpcyr$ for \aclp{NSBH}, and $\FullMassSpectrumMerged[R_BBH][5th_percentile]\textendash\FullMassSpectrumMerged[R_BBH][95th_percentile]\,\perGpcyr$ for \aclp{BBH}; all values reflect central 90\% credible intervals.

%% file: introduction.tex
\section{Introduction}\label{sec:intro}
\acresetall
\acused{Virgo}
\acused{KAGRA}

\Acp{GW} have revolutionized the study of compact objects and their populations~\citep{LIGOScientific:2016dsl,LIGOScientific:2018jsj,LIGOScientific:2020kqk,KAGRA:2021duu}, with observations now extending beyond redshift $z=1$. Searches for \acp{GW} from compact binary mergers have robustly quantifiable selection effects, which enable detailed inference of population properties with few sources of potential bias~\citep{Mandel:2018mve,Essick:2023upv}.  By studying the growing catalog of compact binary mergers, we aim to uncover both where these systems form, and how they evolve toward merger.  For example, are they formed from stars that are gravitationally bound at birth, possibly as the products of chemically homogeneous evolution~\citep{Mandel:2015qlu,Marchant:2016wow,deMink:2016vkw}, or brought close enough together for \ac{GW}-driven inspiral through phases of common envelope evolution~\citep{Bethe:1998bn,PortegiesZwart:1997ugk,Belczynski:2001uc,Dominik:2014yma} or stable mass transfer~\citep{Hurley:2002rf,Neijssel:2019irh,vanSon:2021zpk}?  Could binaries be assisted in assembly and hardening in the gaseous disks of \ac{AGN}~\citep{McKernan:2012rf,Bartos:2016dgn,Stone:2016wzz,Fragione:2018yrb} or through dynamical interactions in dense stellar environments~\citep{Kulkarni:1993fr,Sigurdsson:1993zrm,PortegiesZwart:1999nm,Ziosi:2014sra}?

From the first three observing runs of the \ac{LVK}, we established that stellar-mass \acp{BH} in merging binaries have a broad distribution of masses, with peaks at primary masses of ${\sim}10\,\Msun$ and ${\sim}35\,\Msun$, that falls off steeply above ${\sim}45\,\Msun$~\citep{KAGRA:2021duu}.  Despite the steep decline, the rate of mergers in the expected \ac{PISN} gap, predicted to begin at ${\sim}45$ -- $50\,\Msun$~\citep{Woosley:2016hmi,Farmer:2019jed}, was found to be non-zero~\citep{LIGOScientific:2020iuh,LIGOScientific:2020ufj}.  Likewise, the rate of mergers in the purported lower mass gap between ${\sim}3$ -- $5\,\Msun$~\citep{Bailyn:1997xt,Ozel:2010su,Farr:2010tu} was found to be small but non-zero~\citep{LIGOScientific:2020zkf,LIGOScientific:2024elc}.  The \ac{NS} mass distribution does not require a peak in the distribution at ${\sim}1.33\,\Msun$, as is seen in the galactic \ac{BNS} population~\citep{Ozel:2016oaf,Farrow:2019xnc}. The \ac{BBH} merger rate was found to definitively increase with redshift.  Spins were found to be small in magnitude, with a non-zero fraction of systems with spin components anti-aligned with the binary orbit.

Advanced LIGO~\citep{LIGOScientific:2014pky}, Advanced Virgo~\citep{VIRGO:2014yos}, and KAGRA~\citep{KAGRA:2020tym} began their \ac{O4} on 2023 May 24 at 15:00 UTC.  The \acf{O4a} ended on 2024 January 16 at 16:00 UTC.  The accompanying \ac{GWTC} version \thisgwtcversionfull{} (hereafter, \gwtcfour)~\citep{GWTC:Introduction,GWTC:Methods,GWTC:Results} contains those events observed in previous observing runs, \acsu{O1}~\citep{LIGOScientific:2016dsl}, \acsu{O2}~\citep{LIGOScientific:2018mvr}, and \acsu{O3}~\citep{LIGOScientific:2020ibl,KAGRA:2021vkt,LIGOScientific:2021usb}, together with the newest observations from \ac{O4a}. We use the updated catalog to infer the population properties of \ac{BNS}, \ac{NSBH}, and \ac{BBH} systems in the local Universe.

\ac{GWTC}-3.0 included 76 candidates with \ac{FAR} $<1\,\text{yr}^{-1}$: 69 \acp{BBH}, 4 \acp{NSBH}, 2 \acp{BNS}, and one event GW190814\_211039 (henceforth, GW190814) that is either a \ac{NSBH} or \ac{BBH}.
\citet{GWTC:Results} identifies 128 candidate signals in \ac{O4a} with a probability of astrophysical compact binary origin of $\pastro\ge0.5$, of which \counts[FAROnePerYear][O4aTotal] (\counts[FAROnePerYear][O4aBBH] \acp{BBH} and \counts[FAROnePerYear][O4aNSBH] \ac{NSBH}) have \ac{FAR} $<1\,\text{yr}^{-1}$. Selecting \acp{BBH} from the catalog using this threshold yields a cumulative \ac{BBH} count of \counts[FAROnePerYear][totalBBH].
With fewer signal candidates from binaries containing at least one \ac{NS}, maintaining a comparable contamination fraction to that of \ac{BBH} mergers (${\sim}5\%$) requires a more conservative threshold (see Section~\ref{sec:data} for more details).  We adopt the same threshold used for similar analyses of \gwtcthree, $\text{FAR}<0.25\,\text{yr}^{-1}$, for analyses that include \ac{NS}-containing populations~\citep{KAGRA:2021duu}. This yields two \ac{NSBH} candidates from O3, and one new candidate, GW230529\_181500 (henceforth, GW230529), detected above this threshold in designated observing time during \ac{O4a}~\citep{LIGOScientific:2024elc}.
Section~\ref{sec:data} provides further details and discussion of threshold choices and candidate inclusion.  Adopting the more conservative $\text{FAR}<0.25\,\text{yr}^{-1}$ threshold reduces the \ac{O4a} \ac{BBH} count to \counts[FAROnePerFourYears][O4aBBH], resulting in a catalog of \counts[FAROnePerFourYears][totalBNS] \acp{BNS}, \counts[FAROnePerFourYears][totalNSBH] \acp{NSBH}, and \counts[FAROnePerFourYears][totalBBH] \acp{BBH} for analyses that include \ac{BNS} and \ac{NSBH} populations in this work.

The remainder of the paper is structured as follows. In Section~\ref{sec:methods} we provide a brief description of our inference techniques (with remaining details in  Appendix~\ref{appendix:technical_details}) and the classes of models used (with full descriptions of the models in Appendices~\ref{appendix:parametric_models_summary} and \ref{appendix:non_parametric_models_summary}, and how they were chosen in Appendix~\ref{appendix:model_comparison}).  Section~\ref{sec:data} describes our dataset and sample selection, including brief descriptions of search techniques, threshold choices, and waveform models used.  In Section~\ref{sec:all_masses} we present measurements of the complete compact-binary mass spectrum (\acp{NS} and \acp{BH}).  In Section~\ref{sec:ns}, we study the properties of the \ac{NS}-containing population in detail, and in Section~\ref{sec:bbh_results} focus on the ensemble properties of \acp{BBH}, including their masses, spins, redshifts, and associated correlations. Section~\ref{sec:conclusion} concludes by summarizing the key results of this work.  Associated data releases provide analysis results and figure generation scripts~\citep{lvk_pop_data_release} and data products for estimating sensitivity~\citep{essick_2025_16740117,essick_2025_16740128}.

%% file: methods.tex
\section{Methods}\label{sec:methods}

We determine the population properties of merging compact binaries using hierarchical Bayesian inference as has been done in previous studies~\citep{LIGOScientific:2018jsj,LIGOScientific:2020kqk,KAGRA:2021duu}. Our aim here is to estimate the posterior distribution $p(\PEhyperparam|d)$ on population-level model parameters $\PEhyperparam$ (also referred to as \textit{hyperparameters}) in light of new data $d$ representing the observed \ac{GW} data from individual merger events. These model-dependent hyperparameters describe the population-level properties of \ac{GW} source parameters $\PEparameter$ (masses, spins, redshifts, etc.).
According to Bayes' theorem,
\begin{equation}
    p(\PEhyperparam | d) = \frac{\mathcal{L}(d|\PEhyperparam)\pi(\PEhyperparam)}{p(d)} \ ,
    \label{eq:bayes_theorem}
\end{equation}
where $\pi(\PEhyperparam)$ is the prior probability distribution and $\mathcal{L}(d|\PEhyperparam)$ is the likelihood---the probability of obtaining the data $d$ given some $\PEhyperparam$. 
The Bayesian \textit{evidence} $p(d)$ ensures that $p(\PEhyperparam | d)$ is properly normalized.

In order to use Bayes' theorem to infer the posterior probability distribution of hyperparameters, we need the likelihood of obtaining the observed catalog of events given a set of hyperparameters and a population model. 
Because \ac{GW} detectors are not equally sensitive to different astrophysical sources, the likelihood must account for selection biases. 
Assuming (i) a Poisson process generates realizations from the source population of which a subset is detected, and (ii) the observed data associated are statistically independent (e.g., no overlapping signals), the likelihood is given by~\citep{Loredo:2004nn, Farr:2013yna, Mandel:2018mve, Thrane:2018qnx,Vitale:2020aaz}
\begin{multline}
    \mathcal{L}(\{d_i\},\Ndet|\PEhyperparam) \propto \\ N(\PEhyperparam)^{\Ndet} e^{-\Nexp(\PEhyperparam)} \prod_{i=1}^{\Ndet}\int \mathrm{d}\PEparameter \, \mathcal{L}(d_i|\PEparameter)\pi(\PEparameter|\PEhyperparam) \ ,
    \label{eq:hierarchical_likelihood}
\end{multline}
where $N(\PEhyperparam)$ is the total number of mergers (detected and undetected) incident on the detectors within the observing period, $\Ndet$ is the number of detections, $d_i$ are the strain data corresponding to the $i$th detection, and $\mathcal{L}(d_i|\PEparameter)$ is the likelihood of the data $d_i$ given the \ac{GW} source parameters $\PEparameter$~\citep{GWTC:Methods}. 
The expected number of detections is $\Nexp(\PEhyperparam) = N\,\xi(\PEhyperparam)$, with $\xi(\PEhyperparam)$ being the expected fraction of the population parameterized by $\PEhyperparam$ that is detectable. Formally, we impose a detection criterion on the data, e.g., a \ac{FAR} threshold, by the selection function $x(d)$ acting on a data segment $d$.
A \ac{GW} event is detectable in $d$ if $x(d) > x_{\rm thr}$, where $x_{\rm thr}$ is the chosen detectability threshold. Then,
\begin{equation}
    \xi(\PEhyperparam) = \int_{x(d) > x_{\rm thr}} \mathrm{d}d\, \mathrm{d}\PEparameter \, p(d|\PEparameter)\pi(\PEparameter|\PEhyperparam) .
    \label{eq:selection_efficiency}
\end{equation}
In Equation~\eqref{eq:selection_efficiency}, the domain of integration is over all data $d$ which surpasses the detection threshold $x_{\rm thr}$ \citep{Essick:2025zed}.
When a prior $\pi(N)\propto 1/N$ is assumed, Equation~\eqref{eq:hierarchical_likelihood} can be analytically marginalized over $N$ leaving the rate-marginalized hierarchical likelihood
\begin{equation}
    \mathcal{L}(\{d\},\Ndet|\PEhyperparam) \propto \prod_{i=1}^{\Ndet}\frac{\int \mathrm{d}\PEparameter \, \mathcal{L}(d_i|\PEparameter)\pi(\PEparameter|\PEhyperparam)}{\xi(\PEhyperparam)}\ .
    \label{eq:rate_marginalized_hierarchical_likelihood}
\end{equation}
This likelihood may also be obtained by marginalizing over a prior on the expected number of detections, $N_{\rm exp} = N\xi(\PEhyperparam)$, rather than on $N$~\citep{Essick:2023upv}.

Having described the hierarchical likelihood in Equation~\eqref{eq:hierarchical_likelihood} and Equation~\eqref{eq:rate_marginalized_hierarchical_likelihood} we have one of the two ingredients necessary for sampling the posterior in Equation~\eqref{eq:bayes_theorem}.
We also must choose a prior over the space of hyperparameters. This depends on the population model, which we describe in more detail below. We specify the priors for each model in Appendix~\ref{appendix:parametric_models_summary}.

Equation~\eqref{eq:hierarchical_likelihood} is the exact form of the likelihood. However, calculating $\xi({\PEhyperparam})$ and integrating over $\PEparameter$ is analytically intractable.
Therefore, we approximate the likelihood via Monte Carlo reweighting using importance sampling.
Our Monte Carlo approximation for the likelihood carries some uncertainty and may not be converged appropriately for some hyperparameter values. Therefore, we discard hyperparameters whose likelihood uncertainty exceeds a chosen threshold \textit{a posteriori} \citep{Talbot:2023pex}.
For more description of this problem and our approach for mitigating it, see Appendix~\ref{appendix:technical_details}.

The Bayesian inference problem is stated here in terms of the population probability density $\pi(\PEparameter|\PEhyperparam)$ and the overall number of merging binaries $N$. 
This can be converted to another astrophysical quantity of interest, the comoving source-frame merger rate density
\begin{equation}
    \mathcal{R}(z) = \frac{\mathrm{d}N}{\mathrm{d}V_\mathrm{c} \mathrm{d}t_\mathrm{s}}(z) = \frac{\mathrm{d}N}{\mathrm{d}t_\mathrm{d}\mathrm{d}z}\left(\frac{\mathrm{d}V_\mathrm{c}}{\mathrm{d}z}\frac{1}{1+z}\right)^{-1},
    \label{eq:comoving_merger_rate}
\end{equation}
where $t_\mathrm{s}$ is the time measured in the comoving source frame, $t_\mathrm{d}$ is the time at the detector (redshift $z=0$) and ${\mathrm{d}V_\mathrm{c}}/{\mathrm{d}z}$ is the differential comoving volume with respect to redshift $z$~\citep[see e.g.,][]{Essick:2025zed}.
The comoving source frame merger rate represents the number of mergers in a unit of comoving volume and source-frame time, conventionally measured in units $\perGpcyr$. 

\input{summary_table.tex}

To construct models for the population distribution of astrophysical \ac{GW} sources, we take one of two approaches. 
The first approach---which we call the \textit{\parametric} (sometimes called the parametric approach)---assumes a specific functional form $\pi(\PEparameter|\PEhyperparam)$ for the astrophysical distribution \textit{a priori}, e.g.,~a Gaussian distribution or a power law. 
A second approach---which we call the \textit{\nonparametric} (elsewhere data-driven, flexible, or nonparametric)---attempts to make  minimal  \textit{a priori} assumptions  about the underlying astrophysical population, e.g., a spline model. 
We elaborate on each approach in the following two subsections, and provide details for the models presented in this work falling in each category.

\subsection{Strongly Modeled Approach}

The \parametric~assumes that the underlying astrophysical distribution of binary properties can be described by a fixed functional form and associated hyperparameters. Consequently, this approach has far fewer hyperparameters as compared to the \nonparametric.
The parameterization may be motivated by theoretical expectations about the astrophysical population and/or could be selected because it seems to fit the observed data well.
This has the benefit of being simple and interpretable, as hyperparameters can be designed to correspond directly to physical features of interest.
Examples of such features are a distribution's minimum or maximum, the location parameter for an overdensity corresponding to e.g., pulsational pair-instability supernovae, etc.
On the other hand, these approaches can be overly restrictive: features present in the true astrophysical distribution that are not captured by our parameterization cannot be easily discovered. Additionally, the inferred parameters can be biased due to a mis-specified model~\citep[e.g.,][]{Romero-Shaw:2022ctb}.

The parameterized models selected for this work are generally simple extensions to those employed in the previous \ac{LVK}~astrophysical population studies paper from 
\gwtcthree~\citep{KAGRA:2021duu}. Updates to these models are motivated by features that have emerged due to new data in \gwtcfour and improved interpretations since \gwtcthree. 

Table~\ref{tab:summary_of_models} lists the \parametrics used in this paper, with details in Appendix~\ref{appendix:parametric_models_summary}. We also describe the parameterizations and priors for these models in Appendix~\ref{appendix:parametric_models_summary}, and our procedure for selecting the default \parametric in Appendix~\ref{appendix:model_comparison}.
For most models, we assume that masses, spins, and redshifts are all uncorrelated. 
We also study a selection of pairwise correlations between parameters in Section~\ref{sec:Population-level correlations between parameters}.

\subsection{Weakly Modeled Approach}

The \nonparametric adopts models that deliberately make few assumptions about the nature of the underlying compact-binary population.
Such approaches typically require a larger number of hyperparameters in order to effectively approximate a wide variety of distributions.
The philosophy of a \nonparametric is to discover unexpected features in the population, which may be unforeseen or difficult to parameterize. However, they could yield results that are more difficult to interpret astrophysically.
The difference between the \parametricAdj and \nonparametricAdj approaches can be understood in terms of the bias--variance tradeoff; the former has low variance with a risk of bias, whereas the latter has low bias but elevated variance.

\Nonparametrics must make some assumptions, however, and must be designed with different features in mind. 
For example, different approaches have sensitivity to astrophysical correlations, narrow structures, or gaps in the population. 
While a unified Bayesian approach to capture generic population features is still an open problem~\citep{Mandel:2016prl,Tiwari:2020vym,Rinaldi:2021bhm,Edelman:2022ydv,Golomb:2022bon,Payne:2022xan,Toubiana:2023egi,Callister:2023tgi,Ray:2023upk,Farah:2024xub,Heinzel:2024jlc}, we use two different \nonparametrics---\BSpline and \ac{BGP}---to verify results from our \parametrics (see Table~\ref{tab:summary_of_models}). 
We describe two additional \nonparametrics---\ac{AR} and \vamana---in Appendix~\ref{appendix:non_parametric_models_summary}, and compare these approaches in Appendix~\ref{appendix:non_parameteric_comparison}. 

%% file: summary_table.tex
\begin{deluxetable*}{lll}
\tablecaption{\label{tab:summary_of_models}Summary of Models}
\tablehead{
    \colhead{Model Type} & \colhead{Model Name} & \colhead{Description}
}
\startdata
\textbf{Strongly Modeled:} & \PDB & Models the mass spectrum of all \acp{CBC} simultaneously with \\
\textbf{Mass} & &  appropriate power-law and peak components. Also allows for a \\
& & gap between the most massive \ac{NS} and the least massive \ac{BH}. \\[6pt]
& \BrokenPLTwoPeaks $\bigstar$ & The primary mass distribution has a broken power law\\
&  & continuum between a minimum and maximum mass, plus \\
&  & two Gaussian peaks  around ${\sim} 10\,\Msun$ and ${\sim}35\,\Msun$. The   \\
& &  distribution of mass ratio $q$ is a power law between  \\
& & some minimum value and $1$. \\[6pt]
& \textsc{Extended Broken Power} & The mass-ratio power-law is allowed to differ between \\
& \textsc{Law + 2 Peaks} &  primary masses in the continuum and in the $35\,\Msun$ peak. \\
\hline
\textbf{Strongly Modeled:} & \default $\bigstar$ & The spin magnitude population is a Gaussian truncated over \\
\textbf{Component Spin} & & the physical range $\chi\in[0,1)$.  The distribution of cosine spin tilts \\
& &  relative to the orbital angular momentum  includes an isotropic \\
& &  (i.e., uniform) component and a truncated Gaussian component. \\
\hline
\textbf{Strongly Modeled:} & \effectiveSpinModel & The joint $\chieff$--$\chip$ effective spin distribution is a bivariate   \\
\textbf{Effective Spin} & & Gaussian allowing for correlations. \\[6pt]
& \skewnormalChiEff & The $\chieff$ marginal effective spin distribution is skew normal,\\
& &  truncated to $[-1,1]$. \\
\hline
\textbf{Strongly Modeled:} & \powerlawRedshift $\bigstar$ & The merger rate per unit comoving volume and source-frame \\
\textbf{Redshift} & & time evolves with redshift $z$ as a power law i.e., $\propto (1+z)^\kappa$. \\
\hline
\textbf{Strongly Modeled:} & \copulacorrelation & Truncated Gaussian distributions are assumed for $\chieff$ and  \\
\textbf{Correlations} & & $\chip$.  A Frank copula density function correlates two variables.\\
& &   Separate copula models correlate $(q,\chieff)$, $(z,\chieff)$, $(m_1,\chieff)$, \\
& &  and $(m_1,z)$. \\[6pt]
& \linearcorrelation & A truncated Gaussian distribution is assumed for $\chieff$ with \\
& & the mean and width linearly dependent on either $q$ or $z$. \\[6pt]
& \splinecorrelation & A truncated Gaussian distribution is assumed for $\chieff$ with \\
& & the mean and width dependent on either $q$ or $z$. This   \\
& & dependence is flexibly modeled with a cubic spline. \\
\hline
\textbf{Weakly Modeled:} & {\BSpline} & Fits the astrophysical distribution as a separable joint \\
\textbf{All Parameters} & & distribution with one-dimensional basis splines. Large numbers \\
& &  of basis functions allow for flexibility, with difference-based \\
& &  priors imposing smooth evolution \emph{a priori}. \\[6pt]
& {\textsc{Binned Gaussian Process}} & Assumes a fixed binning scheme and infers the event rate under \\
& &   the assumption of a constant rate within each bin, and a\\
& &   Gaussian process prior imposing smooth covariance across bins. \\
\enddata
\tablenotetext{}{
    \centering
    \parbox{\textwidth}{Strongly modeled and weakly modeled approaches used to study the mass, spin, and redshift distributions of merging compact binaries. We provide a brief description of each model here, with detailed descriptions in Appendix~\ref{appendix:parametric_models_summary} and Appendix~\ref{appendix:non_parametric_models_summary}. Models marked with a $\bigstar$ are treated as defaults and are used whenever no model is explicitly indicated for a certain parameter. All \parametrics mentioned below target \acp{BBH}, with the exception of \PDB that is used to model the entire \ac{CBC} population.}}
\end{deluxetable*}

%% file: data.tex
\section{Dataset}\label{sec:data}

\subsection{Data Collection Duration}
Analyses presented in this paper use selected data products from 
\gwtcfour~\citep{GWTC:Introduction, GWTC:Methods, GWTC:Results, OpenData}.
This section and Section ~\ref{subsec:event_selection} provide details on 
the selection criteria for events analyzed for this paper.
\gwtcfour includes \ac{GW} candidates and data from \ac{O1} through the end of 
\ac{O4a}, as well as a \ac{GW} candidate and data collected during an 
\ac{ER}~\citep{KAGRA:2013rdx} directly preceding the start of \ac{O4a}. 
\Acp{ER} are periods dedicated to final commissioning and configuration of the 
instruments prior to an observing run; the instruments may be in locked and 
low-noise configurations, but are not generally intended to perform astrophysical 
observations.
The \ac{ER} data included in \gwtcfour is deliberately chosen to contain 
a few days of data around a \ac{GW} event potentially originating from a 
\ac{NSBH} binary merger, GW230518\_125908 (henceforth, GW230518).
The analyses and results quoted in this paper exclude data obtained
during the \ac{ER}; the inclusion of this data would 
introduce human selection effects that cannot be easily incorporated
into $\xi(\PEhyperparam)$, and hence may bias our inferences. 

\subsection{Event Selection Criteria}
\label{subsec:event_selection}
\subsubsection{Significance Thresholds}
To ensure that the dataset we use in this paper has reduced contamination 
from noise events, we adopt a significance threshold of \ac{FAR} 
$<$ $1$ $\text{yr}^{-1}$ in at least one \ac{GW} search pipeline, 
which is consistent with the criterion adopted in~\citet{KAGRA:2021duu}. 
Based on the \ac{FAR} threshold, a total of \samplePurityEstimate[totalNumEvents]
\ac{CBC} candidates have been detected from \ac{O1} through 
\ac{O4a}~\citep{LIGOScientific:2018mvr, LIGOScientific:2020ibl, LIGOScientific:2021usb, KAGRA:2021vkt}
by the \ac{GW} search pipelines~\citep{GWTC:Results}, 
of which \samplePurityEstimate[totalNumEventsO4aOnly] were from \ac{O4a}.
This is a noteworthy increase in the number of observations reported in~\citet{KAGRA:2021vkt}, 
which contained \samplePurityEstimate[totalNumEventsGWTC3] events 
meeting the \ac{FAR} $<$ $1$ $\text{yr}^{-1}$ threshold. 
With this \ac{FAR} threshold and assuming noise signals are produced 
independently, we expect 
$\sum_k \mathrm{FAR} \times T_k$ ${\simeq}$ \samplePurityEstimate[expectedNumFalseTrialsFactor][rounded]
contaminant noise events in our results, where $T_k$ is an estimate of the time 
examined by the $k$th search.
The list of \ac{GW} events included in the analyses of this paper contains 
GW231123\_135430 (henceforth, GW231123), which has high probability for being 
the most massive \ac{BBH} with \ac{FAR} $<$ $1$ $\text{yr}^{-1}$ detected to 
date by the \acp{GW} with both component masses possibly in the upper mass 
gap~\citep{GW231123, Woosley:2016hmi, Mapelli:2019ipt, Farmer:2019jed, Farmer:2020xne, Woosley:2021xba, Hendriks:2023yrw}.
Not all events reported in the \acp{GWTC} 
papers~\citep{LIGOScientific:2018mvr, LIGOScientific:2020ibl, LIGOScientific:2021usb, KAGRA:2021vkt, GWTC:Results} 
are included in our analyses, as previous \ac{GWTC} papers thresholded event 
candidates using $p_{\rm astro}$ $\geq 0.5$ or \ac{FAR} $<$ $2$ $\text{yr}^{-1}$, 
whereas the analyses presented 
in our paper select events with a significance of \ac{FAR} $<$ $1$ $\text{yr}^{-1}$. 
Here, $p_{\rm astro}$ is the estimate of the probability of astrophysical
origin of the event candidates~\citep{GWTC:Methods, GWTC:Results}.
In addition, GW230630\_070659 is excluded from the analyses and the number of 
events reported in this paper, due to concerns of the data quality around the 
time of this event~\citep{GWTC:Results}.

\subsubsection{Mass and Significance Thresholds for Events with \acp{NS}}
To distinguish \ac{NS}-containing events from events containing only \acp{BH}, 
we first threshold events by checking whether the $1\%$ lower limit on the component 
mass is smaller or larger than $3\,\Msun$.
As far fewer \ac{GW} candidates with \acp{NS} have been observed compared to 
\acp{BBH}, this paper adopts a stricter \ac{FAR} threshold of $<0.25$ $\text{yr}^{-1}$ 
in at least one \ac{GW} search pipeline for \ac{GW} candidates with \acp{NS} 
to ensure a purer sample, as done in~\citet{KAGRA:2021duu}. 
This \ac{FAR} threshold excludes GW190917\_114630 and 
GW190426\_152155~\citep{LIGOScientific:2021usb}, 
which are consistent with originating from \acp{NSBH} but have \acp{FAR} of 
$>$ $0.25$ $\text{yr}^{-1}$. 
In addition, we exclude certain events whose category is ambiguous from 
dedicated \ac{BBH} and \ac{NSBH} analyses.
Specifically, we exclude GW190814 as its 
source's secondary mass is lower than the component masses of events classified 
as \acp{BBH} but higher than the inferred \ac{NS} mass range, leaving its 
classification 
ambiguous~\citep{LIGOScientific:2020zkf, Essick:2021vlx, KAGRA:2021duu}.
This event is included in Section~\ref{sec:all_masses} which considers binary-merger 
populations across all masses. 
Hence, the only \ac{NS}-containing event detected in \ac{O4a} considered in
this paper is GW230529~\citep{LIGOScientific:2024elc}, in addition to the  
previously reported \acp{NSBH}, GW200105\_162426 and GW200115\_042309 
(henceforth, GW200105 and GW200115, respectively).
The results are presented in Section~\ref{sec:ns}.

\subsubsection{Exclusion of Non-LVK Catalog Events}
In addition to the catalog of event candidates and its analyses conducted by the 
\ac{LVK}, independent teams have analyzed the public \ac{GW} data from \ac{O1} 
through \ac{O3b} using alternative algorithms and have identified additional 
\ac{GW} binary-merger event 
candidates~\citep{Venumadhav:2019tad, Venumadhav:2019lyq, Olsen:2022pin, Mehta:2023zlk, Wadekar:2023gea, Zackay:2019tzo, Nitz:2018imz, Nitz:2020oeq, Nitz:2021uxj, Nitz:2021zwj, Kumar:2024bfe, Mishra:2024zzs, Koloniari:2024kww}. 
We do not include these additional events in our analyses here due to subtleties 
with consistently combining sensitivity estimates from these independent catalogs.

\subsection{Sensitivity of \ac{GW} Searches}
A key ingredient in the estimation of population level properties 
is the sensitivity of our \ac{GW} searches $\xi(\PEhyperparam)$. 
The following four search pipelines analyzed detector data for \textit{real} 
\ac{GW} signals as well as simulated \ac{GW} signals, called \textit{injections}: 
\GSTLAL
~\citep{Messick:2016aqy, Sachdev:2019vvd, Hanna:2019ezx, Cannon:2020qnf, Ewing:2023qqe, Tsukada:2023edh, Sakon:2022ibh, Ray:2023nhx, Joshi:2025nty, Joshi:2025zdu}, 
\MBTA~\citep{Adams:2015ulm, Aubin:2020goo, Andres:2021vew, Allene:2025saz}, 
\PYCBC~\citep{Usman:2015kfa, Nitz:2017svb, Nitz:2018rgo, DalCanton:2020vpm}, 
and the \CWB analysis~\citep{Klimenko:2005xv, Klimenko:2008fu, Klimenko:2015ypf, Tiwari:2015bda, Drago:2020kic, Klimenko:2022nji, Mishra:2021tmu, Mishra:2022ott, Mishra:2024zzs}.
Injections were added to data at an artificially higher rate than observed \ac{GW} 
signals, and were used to quantify the pipelines' sensitivities to \ac{GW} 
signals~\citep{Essick:2025zed, GWTC:Methods}.
The distribution of injections was chosen to enable efficient and accurate 
resampling to a wide range of astrophysically plausible populations~\citep{Essick:2025zed}. 
For all the analyses in this paper, the detection efficiency $\xi(\PEhyperparam)$ 
is estimated using these injections through a Monte Carlo 
integral~\citep{GWTC:Methods, Tiwari:2017ndi, Farr:2019rap}.

\subsection{Source Properties}
\Ac{PE} pipelines use Bayesian inference to estimate the properties of GW events~\citep{GWTC:Methods}.
The hierarchical Bayesian inference framework described in Section~\ref{sec:methods} 
requires as input \ac{PE} samples from individual events. 
For events detected in \ac{O4a}, we use samples drawn from the posterior distribution using
the \SURSEVENDQFOUR{}~\citep{Varma:2019csw} waveform approximant if available. If these are not available e.g., because the signal duration is too long to be analyzed by \SURSEVENDQFOUR{}, we instead use a mixture of samples from the \IMRPhenomXPHMST{}~\citep{Pratten:2020ceb,Colleoni:2024knd} and \SEOBNRFIVEPHM{}~\citep{Ramos-Buades:2023ehm, Pompili:2023tna} approximants.
These are referred to as \textsc{Mixed} samples in the \ac{PE} data products. 
More details about various choices made in the \ac{PE} procedure can be found
in Section~5 of \citet{GWTC:Methods} and Section~3 of \citet{GWTC:Results}.

For all events detected before \ac{O4a}, we use the \textsc{Mixed} samples reported 
in the \gwtcthree \citep{KAGRA:2021vkt} and \gwtctwopone \citep{LIGOScientific:2021usb} data releases. 
One exception is the \ac{BNS} merger GW170817, for which we use samples obtained with the \IMRPhenomPTWONRTidal{} waveform approximant~\citep{Dietrich:2019kaq}  and a prior allowing for  large spin magnitudes~\citep{LIGOScientific:2018hze}.
We also reweight these samples to a distance prior that is uniform in comoving volume and source-frame time
following the prescription in Appendix C of \citet{LIGOScientific:2020ibl}.

While this work was in its final stages, a normalization error was discovered in the noise-weighted inner product used in the \ac{PE} likelihood function~\citep{GWTC:Methods, Talbot:2025vth}. While there is a version of the \ac{PE} samples that account for the correct likelihood via a reweighting prescription~\citep{GWTC:Methods, Talbot:2025vth}, we do not use these samples in this work. Further, for candidates detected during the first three observing runs, we discovered that incorrect priors were used when marginalizing over the uncertainty in the calibration of the LIGO detectors~\citep{GWTC:Methods}. Preliminary re-analysis indicates that for each candidate the impact of this error in marginalization is small, and we expect the impact on our population analyses to be negligible compared to other sources of systematic error.

%% file: full_cbc_population.tex
\section{Binary merger population across all masses} \label{sec:all_masses}

We begin our analysis of the astrophysical distribution of merging compact binaries with a joint analysis of all events discussed in Section~\ref{sec:data}\textemdash \acp{BNS}, \acp{NSBH}, and \acp{BBH}\textemdash without distinguishing between these different source classes.
This allows for a broad look at the complete population, self-consistent measurements of the merger rates in each binary source class, and an analysis of the population at the transition between \acp{NS} and \acp{BH}.
As mentioned in Section~\ref{sec:data}, we adopt a uniform detection threshold of FAR $<$ $0.25$ $\text{yr}^{-1}$ to ensure a high catalog purity of \ac{NS} systems, where we have fewer detections and so are more sensitive to non-astrophysical false-alarm contaminants.

\subsection{The Mass Spectrum of Compact Binaries}

\begin{figure*}
    \centering
    \includegraphics[width=0.98\linewidth]{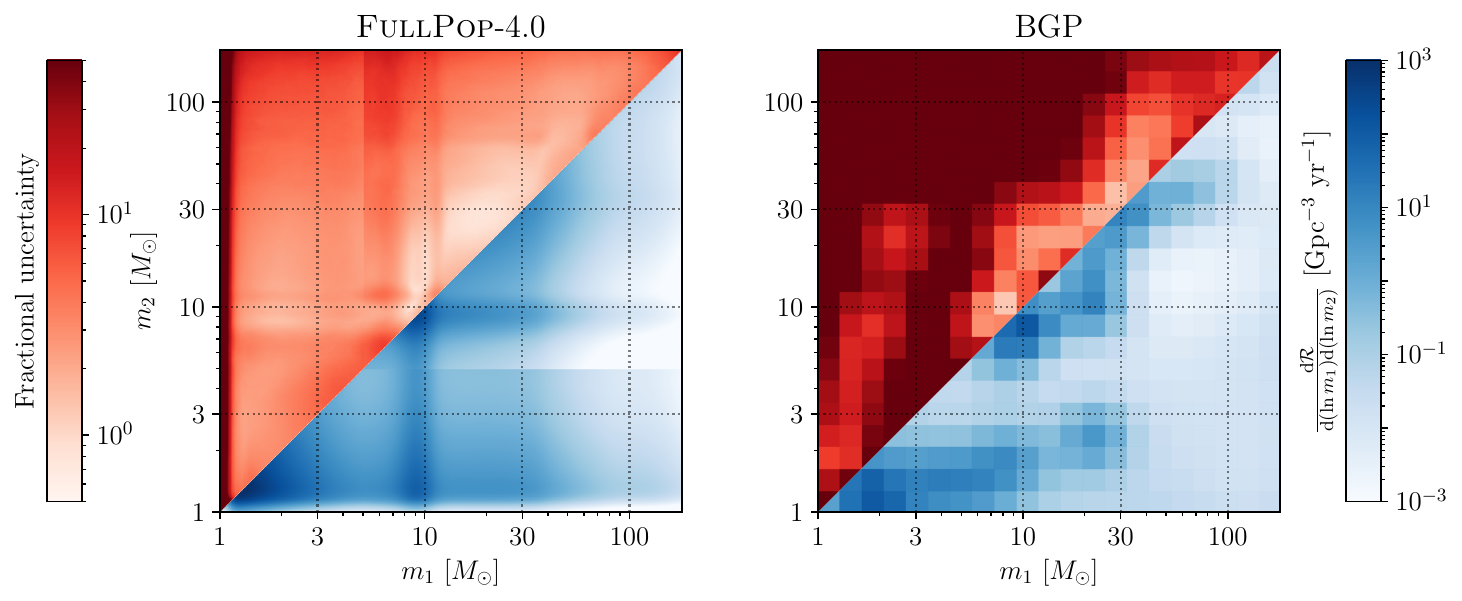}
    \caption{
    The complete mass spectrum, inferred using the \parametricAdj \PDB and \nonparametricAdj \ac{BGP} analyses. 
    In the lower triangle of the $m_1$--$m_2$ plane (where $m_1 > m_2$ by definition), we show median merger rate density across primary and secondary masses. At $m_2 = 5\,\Msun$, the \PDB model transitions to an alternative pairing function to allow for \acp{NSBH}, hence the discontinuity at $m_2 = 5\,\Msun$.
    In the upper triangle we show the fractional uncertainty in the merger rate reflected across the diagonal. The fractional uncertainty $\Delta\mathcal{R} /\mathcal{R}$ is difference of the 95th and 5th percentile values divided by the median merger rate.
    Both the \PDB and \ac{BGP} models show similar broad features: a population of \acp{BNS} at primary and secondary masses $\lesssim 2\,\Msun$, a population of \acp{NSBH} at primary mass $\sim 9\,\Msun$ and secondary mass $\lesssim 2\,\Msun$, and finally the population of \acp{BBH} with primary and secondary masses $\gtrsim 9\,\Msun$.
    The uncertainties are the smallest in the $\sim 9\,\Msun$ and $\sim 30\,\Msun$ \ac{BBH} peaks. 
    }
    \label{fig:inferred_m1_m2}
\end{figure*}

In Figure~\ref{fig:inferred_m1_m2} we show the joint primary and secondary-mass distributions, inferred using a \parametricAdj and a \nonparametric. 
Our \PDB \parametric is modified from the \textsc{Power law + Dip + Break} analysis of the previous \gwtcthree catalog \citep[][]{Fishbach:2020ryj, Farah:2021qom, KAGRA:2021duu, Mali:2024wpq}; see Appendix~\ref{appendix:PDB} for a model description. 
Our \nonparametricAdj analysis approximates the $m_1$--$m_2$ space with a \ac{BGP} \citep{Mandel:2016prl,KAGRA:2021duu,Ray:2023upk,Ray:2024hos}; see Appendix~\ref{appendix:BGPModel} for further discussion. 
The \PDB model uses the default models in Table~\ref{tab:summary_of_models} for the redshift and component spins, and for \ac{NS} masses ($m < 2.5\,\Msun$) the spin magnitude distribution is truncated over the range $\chi \in [0,0.4]$. 
The \ac{BGP} analysis fixes the \powerlawRedshift evolution to $\kappa = 3$ (see Appendix~\ref{appendix:redshiftModels}), and the spin distribution to be uniform in magnitude (again truncated over $\chi \in [0,0.4]$ for \ac{NS} masses) and isotropic in orientation.

\textbf{\boldmath We observe an enhanced merger rate around ${m_1 \sim m_2 \lesssim 2\,\Msun}$, representing \ac{BNS} systems, an additional subpopulation at unequal masses $m_1 \sim 9\,\Msun$ and $m_2 \lesssim 2 \,\Msun$ consistent with \acp{NSBH}, and a third \ac{BBH} subpopulation at $m_1, m_2 \gtrsim 9\,\Msun$.}
In the upper triangle of Figure~\ref{fig:inferred_m1_m2}, we show the fractional uncertainty as a function of mass, a unitless quantity $\Delta\mathcal{R} /\mathcal{R}$ defined as the 95th - 5th percentile uncertainty divided by the median merger rate. 
The rate is best constrained at equal masses, where the majority of mergers are observed, and in the \ac{BBH} range $10$--$40\,\Msun$. 
Due to fewer observations, the uncertainty is larger for \ac{BNS} and \ac{NSBH} systems.

\begin{figure*}
    \centering
    \includegraphics[width=0.98\linewidth]{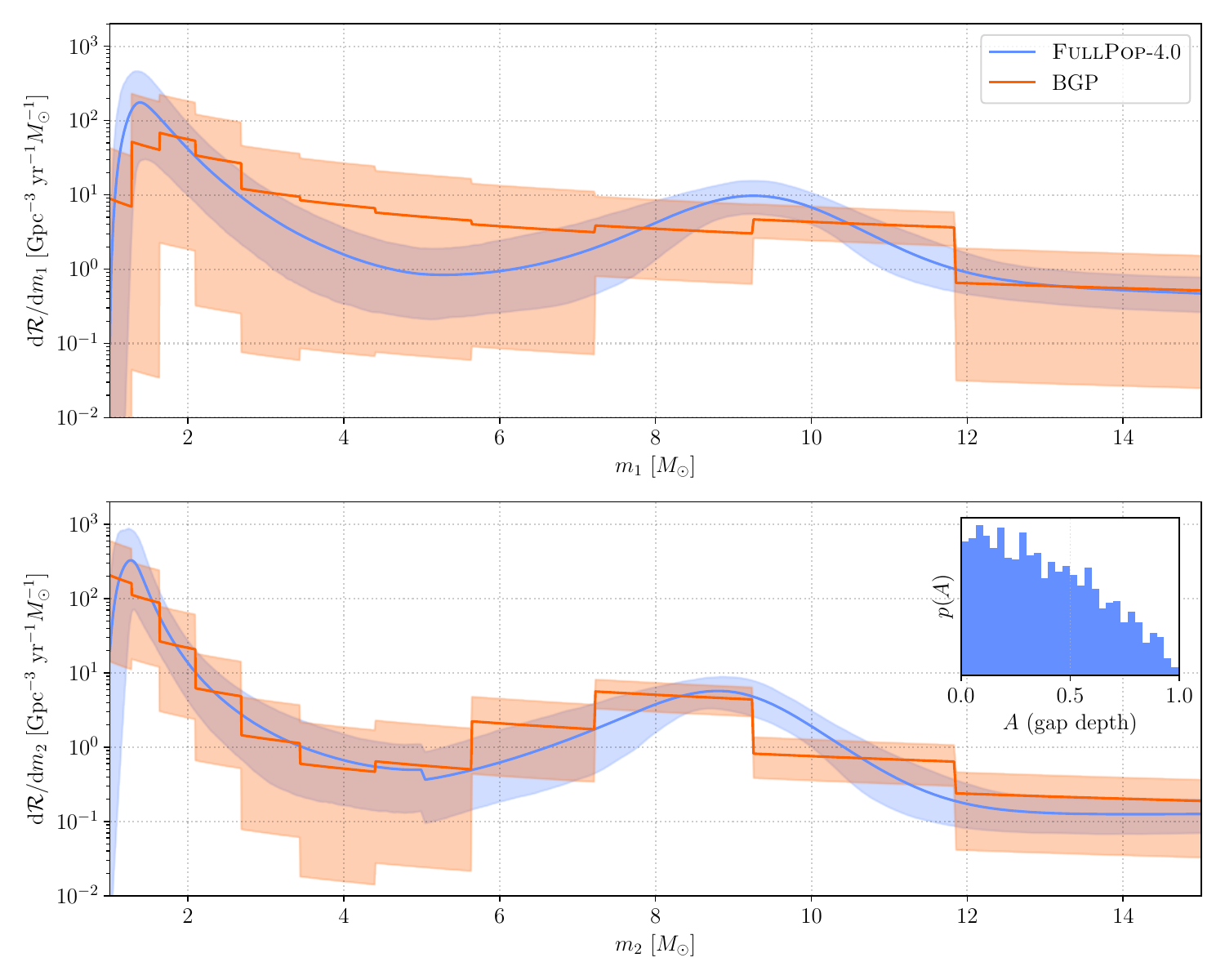}
    \caption{
    A comparison of the merger rate at redshift $z=0$, as a function of component mass for the \PDB and \ac{BGP} models. 
    In the upper panel, we show the merger rate as a function of primary mass, marginalized over secondary mass. In the lower panel, we show the merger rate as a function of the secondary mass, marginalized over the primary mass. At $m_2 = 5\,\Msun$, the \PDB model transitions to an alternative pairing function to allow for \acp{NSBH}, hence the discontinuity in the secondary mass.
    The median of the inferred merger rate is shown with a solid line, and the 90\% credible interval is shown in the shaded region.
    The \ac{BGP} \nonparametric has larger uncertainties due to increased flexibility. 
    Both the \parametricAdj and \nonparametricAdj approaches find local maxima in the merger rate at $m_2 \sim 1-2 \,\Msun$ and $m_2 \sim 6.5-9\,\Msun$ in the secondary-mass distribution. 
    However, the \ac{BGP} model does not confidently recover the low mass peaks in the primary-mass distribution.
    In the inset, we show the \PDB inferred gap depth parameter $A$ (see Appendix~\ref{appendix:PDB}), where $A=1$ (a completely empty gap) is disfavored.
    }
    \label{fig:full_mass_spectrum}
\end{figure*}

In Figure~\ref{fig:full_mass_spectrum}, we show the marginalized primary and secondary-mass distributions for our strongly and \nonparametricAdj reconstructions of the merger rate, focusing on the transition between \acp{NS} and \acp{BH}. 
The models are consistent within uncertainties, indicating that systematic error from model assumptions are smaller than the statistical uncertainties. 
The most notable difference between the \PDB and \ac{BGP} results is the large uncertainty in the \ac{BGP} primary-mass distribution relative to the secondary mass. 
There are significantly more observed \acp{CBC} with light secondary masses $m_2 \lesssim 15\,\Msun$ than primary masses $m_1 \lesssim 15\,\Msun$ (simply by the definition $m_2 < m_1$), and so the data-driven \ac{BGP} is more uncertain for small primary masses, while the \PDB results are more model driven.

At $m_2 = 5\,\Msun$, the \PDB model adopts an alternative pairing function to more naturally distinguish the pairing behavior of BNS and NSBH systems from that of BBH systems, introducing a discontinuity in the secondary-mass distribution (Figure~\ref{fig:full_mass_spectrum}) at the transition point.
We discuss the behavior of the population at the transition from \acp{NS} to \acp{BH} and the astrophysical insights below in Section~\ref{sec:lower_mass_gap}.

\begin{deluxetable*}{lcccccc}
\tablecaption{Merger rates in units $\perGpcyr$ in different mass ranges, according to the \PDB and \ac{BGP} models. }
\tablewidth{0pt}
\tablehead{
    \colhead{} & \colhead{BNS} & \colhead{NSBH} & \colhead{BBH} & \colhead{NS--Gap} & \colhead{BH--Gap} & \colhead{Full} \\
    \colhead{} & \colhead{$m_1 \in [1, 2.5]\,\Msun$} & \colhead{$m_1 > 2.5 \,\Msun$} & \colhead{$m_1 > 2.5\,\Msun$} & \colhead{$m_1 \in [2.5, 5]\,\Msun$} & \colhead{$m_1 > 2.5\,\Msun$} & \colhead{$m_1 > 1\,\Msun$} \\
    \colhead{} & \colhead{$m_2 \in [1, 2.5]\,\Msun$} & \colhead{$m_2 \in [1, 2.5]\,\Msun$} & \colhead{$m_2 > 2.5\,\Msun$} & \colhead{$m_2 \in [1, 2.5]\,\Msun$} & \colhead{$m_2 \in [2.5, 5]\,\Msun$} & \colhead{$m_2 > 1\,\Msun$}
}
\label{tab:inferred_merger_rate}
\startdata
\PDB & $\FullMassSpectrumPDB[rates][BNS][median]^{+\FullMassSpectrumPDB[rates][BNS][error plus]}_{-\FullMassSpectrumPDB[rates][BNS][error minus]}$ & $\FullMassSpectrumPDB[rates][NSBH][median]^{+\FullMassSpectrumPDB[rates][NSBH][error plus]}_{-\FullMassSpectrumPDB[rates][NSBH][error minus]}$ & $\FullMassSpectrumPDB[rates][BBH][median]^{+\FullMassSpectrumPDB[rates][BBH][error plus]}_{-\FullMassSpectrumPDB[rates][BBH][error minus]}$ & $\FullMassSpectrumPDB[rates][NS-Gap][median]^{+\FullMassSpectrumPDB[rates][NS-Gap][error plus]}_{-\FullMassSpectrumPDB[rates][NS-Gap][error minus]}$ & $\FullMassSpectrumPDB[rates][BH-Gap][median]^{+\FullMassSpectrumPDB[rates][BH-Gap][error plus]}_{-\FullMassSpectrumPDB[rates][BH-Gap][error minus]}$ & $\FullMassSpectrumPDB[rates][Full][median]^{+\FullMassSpectrumPDB[rates][Full][error plus]}_{-\FullMassSpectrumPDB[rates][Full][error minus]}$ \\
\ac{BGP} & $\FullMassSpectrumBGPRate[R_BNS][50th_percentile]^{+\FullMassSpectrumBGPRate[R_BNS][error_plus]}_{-\FullMassSpectrumBGPRate[R_BNS][error_minus]}$ & $\FullMassSpectrumBGPRate[R_NSBH][50th_percentile]^{+\FullMassSpectrumBGPRate[R_NSBH][error_plus]}_{-\FullMassSpectrumBGPRate[R_NSBH][error_minus]}$ & $\FullMassSpectrumBGPRate[R_BBH][50th_percentile]^{+\FullMassSpectrumBGPRate[R_BBH][error_plus]}_{-\FullMassSpectrumBGPRate[R_BBH][error_minus]}$ & $\FullMassSpectrumBGPRate[R_NS-Gap][50th_percentile]^{+\FullMassSpectrumBGPRate[R_NS-Gap][error_plus]}_{-\FullMassSpectrumBGPRate[R_NS-Gap][error_minus]}$ & $\FullMassSpectrumBGPRate[R_BH-Gap][50th_percentile]^{+\FullMassSpectrumBGPRate[R_BH-Gap][error_plus]}_{-\FullMassSpectrumBGPRate[R_BH-Gap][error_minus]}$ & $\FullMassSpectrumBGPRate[R_Full][50th_percentile]^{+\FullMassSpectrumBGPRate[R_Full][error_plus]}_{-\FullMassSpectrumBGPRate[R_Full][error_minus]}$ \\
Merged & $\FullMassSpectrumMerged[R_BNS][5th_percentile]$--$\FullMassSpectrumMerged[R_BNS][95th_percentile]$ & $\FullMassSpectrumMerged[R_NSBH][5th_percentile]$--$\FullMassSpectrumMerged[R_NSBH][95th_percentile]$ & $\FullMassSpectrumMerged[R_BBH][5th_percentile]$--$\FullMassSpectrumMerged[R_BBH][95th_percentile]$ & $\FullMassSpectrumMerged[R_NS-Gap][5th_percentile]$--$\FullMassSpectrumMerged[R_NS-Gap][95th_percentile]$ & $\FullMassSpectrumMerged[R_BH-Gap][5th_percentile]$--$\FullMassSpectrumMerged[R_BH-Gap][95th_percentile]$ & $\FullMassSpectrumMerged[R_Full][5th_percentile]$--$\FullMassSpectrumMerged[R_Full][95th_percentile]$ \\
\tableline
\textsc{Simple Uniform BNS} & $\CIBoundsDash{\simplerates[RBNS]}$ & -- & -- & -- & -- & -- \\
\enddata
\tablecomments{For \ac{BNS} systems, we also estimate the rate assuming a \textsc{Simple Uniform BNS} model. We show rates of \ac{BNS}, \ac{NSBH}, and \ac{BBH} assuming objects with mass $m \in [1, 2.5]\,\Msun$ are \acp{NS} and $m>2.5\,\Msun$ are \acp{BH}. We also show rates within the purported lower-mass gap between astrophysical \acp{NS} and \acp{BH}, according to these models. In the third row, we show the merged estimates, taking the union of the 90\% credible intervals for the \PDB and \ac{BGP} models, in order to account for model systematics. We quote merger rates at redshift $z=0$.}
\end{deluxetable*}

\subsection{Merger Rates}
\label{sec:merger_rates}

Since we self-consistently include all events in the catalog, our measurements of the binary merger rates are robust to events which straddle different source classes.
We calculate the rates of \ac{BNS}, \ac{NSBH}, and \ac{BBH} mergers by assuming any object with mass $1\,\Msun < m < 2.5\,\Msun$ is a \ac{NS}, and any object with mass $m > 2.5\,\Msun$ is a \ac{BH}. 
In all models, we assume the rate evolves with redshift in a manner that is uncorrelated with mass \citep{Fishbach:2018edt, KAGRA:2021duu}.
We quote merger rates in the local Universe, at redshift $z=0$. 

Our strongly-modeled \PDB and weakly-modeled \ac{BGP} approaches infer different rates over the $m_1$--$m_2$ space, and hence different rates in each binary source class. 
To marginalize over the systematic modeling uncertainty, we take the union of both 90\% credible intervals.
\textbf{\boldmath The rates thus obtained are ${\FullMassSpectrumMerged[R_BNS][5th_percentile]}$--${\FullMassSpectrumMerged[R_BNS][95th_percentile]\,{\perGpcyr}}$ for \ac{BNS} mergers, ${\FullMassSpectrumMerged[R_NSBH][5th_percentile]}$--${\FullMassSpectrumMerged[R_NSBH][95th_percentile]\,{\perGpcyr}}$ for \ac{NSBH} mergers, and ${\FullMassSpectrumMerged[R_BBH][5th_percentile]}$--${\FullMassSpectrumMerged[R_BBH][95th_percentile]\,{\perGpcyr}}$ for \ac{BBH} mergers.}
In Table~\ref{tab:inferred_merger_rate}, we show rates in these classes using different models and within purported mass gaps.

The \ac{BNS} merger rate measurements may be sensitive to assumptions about \ac{NS} pairing, and so we also estimate the \ac{BNS} merger rate assuming a simple, fixed population. 
We assume a uniform mass distribution between $1\,\Msun$ and $2.5\,\Msun$ for the component masses, isotropically distributed spins with uniform spin magnitudes below $0.4$, and a merger rate uniform in comoving volume up to $z=0.15$. 
Under this fiducial model (denoted \textsc{Simple Uniform BNS} in Table~\ref{tab:inferred_merger_rate}), we infer a \ac{BNS} merger rate of $\CIBoundsDash{\simplerates[RBNS]}\,\perGpcyr$. 

The estimates in Table~\ref{tab:inferred_merger_rate} are consistent with our previous analysis \citep{KAGRA:2021duu} and the uncertainties on the rate in each source class have generally decreased due to our larger catalog size. 
Although it is within uncertainties, our inferred merger rate for \ac{BNS} systems has notably decreased by a factor of $\sim 2$ (cf. PDB (pair) and \ac{BGP} in Table 2 of \citealt{KAGRA:2021duu}). 
This is a result of the improved detector range and observing time together with the lack of new \ac{BNS} detections. 

\subsection{The Neutron Star--Black Hole Transition}
\label{sec:lower_mass_gap}

\Ac{EM} observations have previously suggested the existence of a mass gap between \acp{NS} and \acp{BH} \citep{Bailyn:1997xt, Ozel:2010su, Farr:2010tu}. 
On the lower end of the gap, nonrotating \ac{NS} masses are bounded by a physical limit, the \ac{TOV} mass \citep[e.g.,][]{Kalogera:1996ci}. 
Astrophysical observations, heavy-ion collision experiments, and modeling of the dense matter \ac{EOS} at nuclear densities bound $M_{\rm max, TOV} \sim 2.2$--$2.5\,\Msun$ \citep{Landry:2020vaw, Dietrich:2020efo, Legred:2021hdx, Huth:2021bsp, Ai:2023ykc, Dittmann:2024mbo, Rutherford:2024srk, Koehn:2024set}, and
studies on the remnant in the \ac{BNS} merger GW170817 \citep{LIGOScientific:2017vwq} place limits in the range $\lesssim 2.3\,\Msun$ \citep{Margalit:2017dij,Rezzolla:2017aly,Ruiz:2017due,LIGOScientific:2019eut,Nathanail:2021tay}. 

On the upper end of the gap, \ac{EM} observations historically identified a dearth of \acp{BH} in the range $\sim 3$--$5\,\Msun$ \citep{Bailyn:1997xt, Ozel:2010su, Farr:2010tu}, hinting at an astrophysical mass gap between \acp{NS} and \acp{BH}, or perhaps a selection effect obscuring such objects. 
More recently, observations of noninteracting binary systems \citep{Thompson:2018ycv, Jayasinghe:2021uqb} and radio pulsar surveys \citep{Barr:2024wwl} suggest the presence of a population of compact objects within the gap.
Indeed, the \ac{GW} events GW190814 \citep{LIGOScientific:2020zkf} and GW230529 \citep{LIGOScientific:2024elc} are further evidence that the transition between \acp{NS} and \acp{BH} is populated, albeit sparsely. 

With the additional \gwtcfour data, the \ac{GW} picture of the purported lower-mass gap is becoming clearer and structures around the transition from \acp{NS} to \acp{BH} are emerging.
In both our \parametricAdj \PDB and \nonparametricAdj \ac{BGP} analyses, we find evidence for \textbf{a prominent pair of peaks at \ac{NS} masses $\bm{{\sim}1.5\Msun}$ and at \ac{BH} masses $\bm{{\sim}9\,\Msun}$ on each side of the lower-mass gap.
A completely empty gap between \acp{NS} and \acp{BH} is disfavored.}  
We cannot rule out the existence of extremely narrow gaps in the compact object spectrum, though such a feature requires fine tuning of the supernova explosion mechanism, fallback, binary interactions or other physical processes \citep[e.g.,][]{Fryer:2011cx,Belczynski:2011bn}.

Our \PDB model detects a peak in the \ac{BH} merger rate at $\FullPop[peak_locations][TenMsun][median]^{+\FullPop[peak_locations][TenMsun][error plus]}_{-\FullPop[peak_locations][TenMsun][error minus]}\,\Msun$.
We find that merging \acp{NS} represent the global maximum of the compact object mass spectrum at $\FullPop[peak_locations][OneMsun][median]^{+\FullPop[peak_locations][OneMsun][error plus]}_{-\FullPop[peak_locations][OneMsun][error minus]}\,\Msun$. 
At the transition from \acp{NS} to \acp{BH}, \PDB allows for an additional suppression in the merger rate, parameterized by a gap depth parameter $A$ where $A=1$ corresponds to an absolute gap with zero mergers and $A=0$ corresponds to no additional suppression.
We show the measurement on $A$ in the inset in Figure~\ref{fig:full_mass_spectrum}. $A$ is consistent with zero, and the lower and upper bounds of the gap (e.g., the maximum \ac{NS} mass and the minimum \ac{BH} mass) are not measured away from the prior. 

Earlier precursor analyses to \PDB showed that the merger rate between $\sim 3$--$5\,\Msun$ is likely suppressed relative to a power-law continuum \citep{Fishbach:2020ryj, LIGOScientific:2020kqk, Farah:2021qom}, using older datasets.
In \cite{KAGRA:2021duu}, we used a \parametricAdj analysis on \gwtcthree (compare the PDB model to \PDB) to argue that the transition from \acp{NS} to \acp{BH} is likely suppressed, but also may be partially filled.
\PDB measures a similar mass spectrum to these previous analyses, but reintreprets the data as evidence for a rise at $\sim 9\,\Msun$, and no \textit{additional} suppression between the \ac{NS} and \ac{BH} peaks (\citealt{Mali:2024wpq} make similar conclusions on \gwtcthree).

We corroborate our \parametricAdj \PDB conclusions with a \nonparametricAdj \ac{BGP} approach, and our constraints are improved relative to a similar \ac{BGP}-based model in \cite{KAGRA:2021duu}. 
Unlike the \PDB model, however, the \ac{BGP} analysis does not distinguish between peaks, gaps, or a continuum: as a \nonparametric, the shape of the population is inferred directly without an associated interpretation.

The \ac{BGP} model infers a nonzero merger rate between the \ac{NS} to \ac{BH} peaks consistent with an underlying continuum, and no evidence for a sharp gap. 
However, the smoothing kernel in the \ac{BGP} model makes it \textit{a priori} less sensitive to deep gaps, so we cannot rule them out either.
The \ac{BGP} model also observes \ac{NS} and \ac{BH} peaks. We quantify the significance as the fraction of hyperparameter samples where the merger rate density is higher than the merger rate in the neighboring bins. 
The merger rate maximizes for \acp{NS} in the range $1\,\Msun \leq m_2 \leq 2\,\Msun$ at \FullMassSpectrumBGPPeakSignificance[m2_peaks][(1, 2)]\% credibility and the low-mass \ac{BH} peak occurs in the secondary mass $7.5\,\Msun \leq m_2 \leq 9\,\Msun$ bin at \FullMassSpectrumBGPPeakSignificance[m2_peaks][(7.23021864, 9.2586008)]\% credibility.

As our knowledge improves about the compact object population at the boundary between \acp{NS} and \acp{BH}, we stand to learn about supernova physics \citep[e.g.,][]{Burrows:2020qrp} and the formation mechanisms of the heaviest \acp{NS} and lightest \acp{BH} \citep[][and references therein]{LIGOScientific:2024elc}.
If future catalogs continue to disfavor a completely empty gap, the standard picture of rapid core-collapse supernovae
---which features a sharp transition in remnant masses; successful explosions leave \acp{NS} with $m\lesssim 2\,\Msun$ and failed explosions promptly collapse to \acp{BH} with $m\gtrsim 5\,\Msun$ \citep[e.g.,][]{Fryer:1999ht,Fryer:2011cx}---
may require modifications to include fallback, slower instability growth, or stochasticity \citep[e.g.,][]{Fryer:1999ht,Fryer:2011cx,Belczynski:2011bn,Sukhbold:2015wba,Ertl:2019zks,Mandel:2020qwb}.
Fallback of stellar material can produce black holes from the maximum neutron star mass to the lightest \ac{BH} mass ($\sim 6\,\Msun$) in the rapid implosion scenario \citep[e.g.,][]{Ertl:2019zks}, or
a slower instability growth timescale could allow the proto-\ac{NS} to accrete enough mass before the explosion to populate the mass gap \citep[e.g.,][]{Belczynski:2011bn,Olejak:2022zee,Fryer:2022lla}.
Another possibility is that stochasticity in the stellar evolution and supernovae smooths out the remnant mass distribution and occupy the lower-mass gap \citep[e.g.,][]{Mandel:2020qwb,Mandel:2020cig}.
Alternatively, the gap between \acp{NS} and \acp{BH} may be populated by a pollution mechanism, such as the remnants of \ac{BNS} or white dwarf collisions which participate in further hierarchical mergers \citep[e.g.,][]{Gupta:2019nwj, Ye:2019xvf, Ye:2024wqj, Barr:2024wwl, Mahapatra:2025agb} or other exotic scenarios like primordial black holes \citep[e.g.,][]{Clesse:2020ghq} or gravitationally lensed events, which could be mistaken for mass gap objects \citep[e.g.,][]{Bianconi:2022etr,Janquart:2024ztv,Farah:2025ews}.

%% file: ns_results.tex
\section{Population Properties of Mergers Containing Neutron Stars}\label{sec:ns}

In O4a, GW230529~\citep{LIGOScientific:2024elc} is the only \ac{NS}-containing event identified with a ${\rm \ac{FAR} < 0.25\,{yr}^{-1}}$. We do not include the \ac{NSBH} candidate GW230518 identified during the \ac{ER} preceding \ac{O4a}. Additionally, no coincident \ac{EM} counterparts were identified for triggers (i.e., preliminary candidate GW signals flagged by the detection pipelines when their ranking statistic exceeded the alert threshold) during O4a based on follow-up efforts conducted by \ac{EM} telescopes. Thus, conclusions presented in previous papers~\citep{KAGRA:2021duu, LIGOScientific:2024elc} about the population properties of \ac{BNS} and \ac{NSBH} mergers remain largely unchanged.

We adopt models and methods consistent with those used in \gwtcthree~\citep{KAGRA:2021duu}. Specifically, we adopt two different models for the \ac{NS} mass distribution, one in which masses are assumed to be Gaussian distributed, and another in which they are assumed to follow a power law~\citep{KAGRA:2021duu}. These are referred to as the \textsc{Peak} model and the \textsc{Power} model respectively. We assume that the redshift evolution of the merger rate is fixed, and that the spins are distributed following the \ac{PE} prior~\citep{GWTC:Methods}. A uniform prior is used for hyperparameters of the population, with the condition \(m_\mathrm{min}\leq \mu \leq m_\mathrm{max}\) (where \(\mu\) is the mean of the Gaussian bump in the \textsc{Peak} model), assuming \(m_\mathrm{max}\) does not exceed \(M_\mathrm{max,TOV}\) (the maximum permissible \ac{NS} mass as expected from the TOV limit), as detailed in Appendix~\ref{appendix:parametric_models_summary}. 

When assumed to follow the \textsc{Power} model, \ac{NS} masses favor a mass distribution with a power-law slope constrained to $\alpha=\CIPlusMinus{\powersodapop[woGW190814][param][alpha]}$. The inference from the \textsc{Peak} model is broad, with largely unconstrained peak width $\sigma=\CIPlusMinus{\peakcutsodapop[woGW190814][param][sigma]}\,\Msun$ and location $\mu=\CIPlusMinus{\peakcutsodapop[woGW190814][param][mu]}\,\Msun$. While these results may hint at a peak emerging near \(1.4\,\Msun\), it is much broader than the relatively sharp peak in the mass distribution of Galactic \ac{NS} systems~\citep{Farrow:2019xnc,2024OJAp....7E..58E}. Our inferred \ac{NS} mass distribution remains broad, with greater support for high-mass \acp{NS}.

As no new \acp{NSBH} beyond GW230529 were confidently observed in O4a, the population-level results in \citet{LIGOScientific:2024elc} produced using the \textsc{NSBHPop} model~\citep{Biscoveanu:2022iue} remain unchanged. Specifically, our inferred minimum \ac{BH} mass in \ac{NSBH} systems remains {$3.4^{+1.0}_{-1.2}\,\Msun$}~\citep{LIGOScientific:2024elc}.

%% file: bbh_results.tex
\begin{figure*}[ht!]
	\centering
	\includegraphics[width=\textwidth]{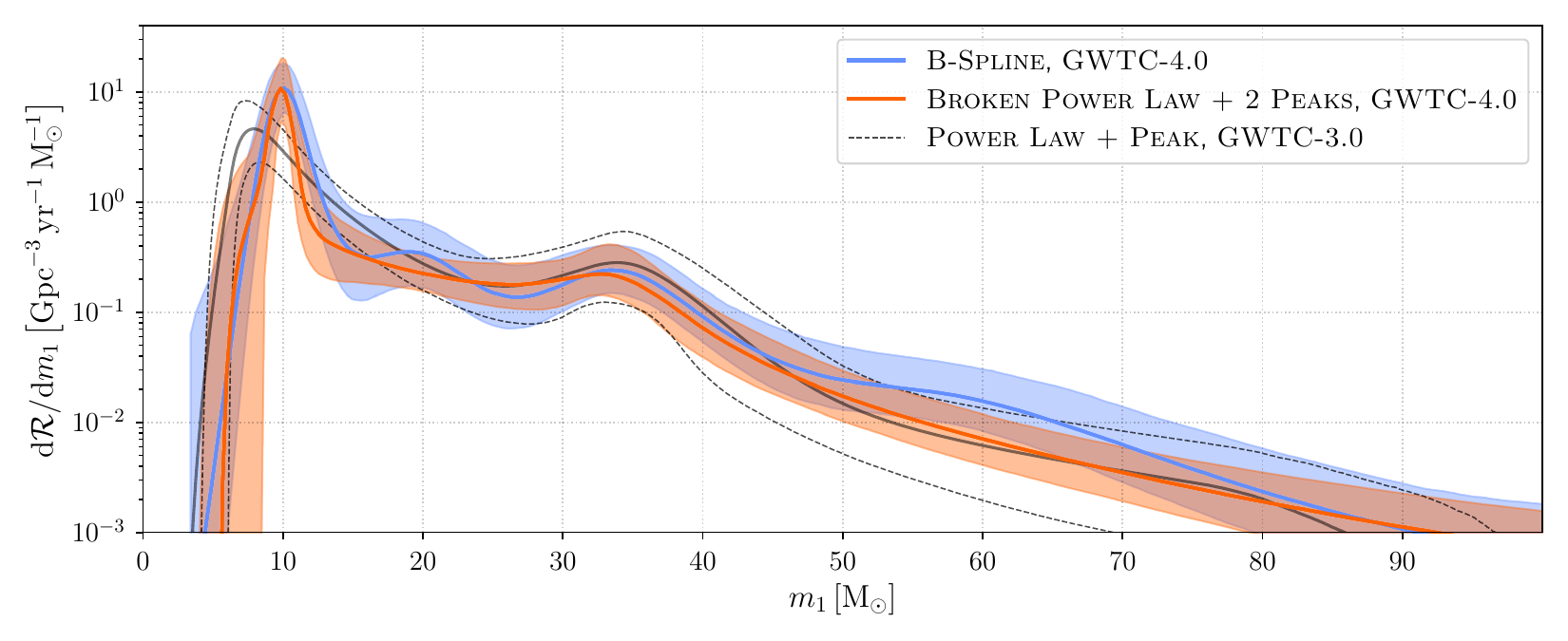}
	\caption{Differential merger rate as a function of primary mass (evaluated at $z=0.2$) of the \BrokenPLTwoPeaks model (orange) and \BSpline model (blue) compared to the \PLPeak model from \gwtcthree \citep{KAGRA:2021duu}. The solid lines indicate the posterior medians and the shaded regions show the 90\% credible interval of each model. Comparing these results, it is clear that a single power law is a poor description of the the low-mass end of the spectrum.}
	\label{fig:primary_mass_default}
  \end{figure*}
  
\section{Binary Black Hole Population}\label{sec:bbh_results}
In this section, we analyze the astrophysical population of \acp{BBH} using data from \ac{GW} events with a \ac{FAR} $< 1 \,\text{yr}^{-1}$. We only include events whose $1\%$ lower limit on both component mass posteriors (under the \ac{PE} priors) is larger than $3 \, \Msun$.
There are {$\counts[FAROnePerYear][O4aBBH]$} events from \ac{O4a} that meet this criterion in addition to the $69$ \ac{BBH} events from \gwtcthree, which brings the total number of \ac{BBH} events passing our \ac{FAR} threshold to {$\counts[FAROnePerYear][totalBBH]$}. Of the events considered in \ac{O4a}, only GW230529 does not meet this criterion and is not considered in the analyses below (see Section~\ref{sec:data}).

\input{bbh_masses.tex}
\input{bbh_spins.tex}
\input{bbh_redshift.tex}
\input{bbh_correlations.tex}

%% file: bbh_masses.tex
\subsection{Primary Mass}
\label{subsec:bbh_masses}

In this section, we illustrate the main findings of the strongly and weakly modeled approaches using results from the \BrokenPLTwoPeaks (see Appendix \ref{appendix:BrokenPLTwoPeaks}) and \BSpline (see Appendix \ref{appendix:BSpline}) models, respectively. We chose the \BrokenPLTwoPeaks model as our fiducial mass model because it performed the best in our model comparison study. It was first introduced by \citet{Callister:2023tgi} to describe the \gwtcthree mass distribution. Results from a selection of other models considered in the model comparison study can be found in Appendix \ref{appendix:model_comparison_mass}, while models that employ a \nonparametric can be found in Appendix \ref{appendix:non_parameteric_comparison}, both of which include comparisons to the fiducial \BrokenPLTwoPeaks and \BSpline models. The \parametric employs the default models listed in Table \ref{tab:summary_of_models} for the other source parameters, namely the \powerlawRedshift and \default models.

\textbf{We identify a global peak at $\bm{{\sim}10 \,\Msun}$ and find that it is robust to model variations}. A Gaussian-like peak at ${\sim}10 \,\Msun$ was first identified by \citet{Tiwari:2020otp} and \citet{Edelman:2021zkw} using \gwtctwo and later by the \nonparametrics in \citet{KAGRA:2021duu} using \gwtcthree. Figure \ref{fig:primary_mass_default} shows the rate ${\rm d}\mathcal{R}/{\rm d} m_1$ as a function of primary mass at $z=0.2$ inferred by these models compared to the \PLPeak (see Appendix B of \citealt{KAGRA:2021duu} for a description of this model) inference with \gwtcthree from \citet{KAGRA:2021duu}.\footnote{We quote \ac{BBH} merger rates at $z=0.2$ to be consistent with \citet{KAGRA:2021duu} and because $z\sim0.2$ is where we best constrain the merger rate.} The low-mass Gaussian component of the \BrokenPLTwoPeaks model infers a global peak in the primary mass spectrum at $m_1 = \CIPlusMinus{\defaultbbh[mass_1][peak_1_location]} \,\Msun$ relative to the underlying broken power law continuum. The \BSpline model also recovers a global peak at $m_1 = \CIPlusMinus{\BsplineIID[mass_1][peak_1_location]} \,\Msun$. Both values are consistent with \gwtcthree, which inferred a global peak at $m_1 = $ \muOOT by modeling the mass spectrum as a power law modulated by a spline.

\begin{figure*}[t]
	\centering
	\includegraphics[width=\textwidth]{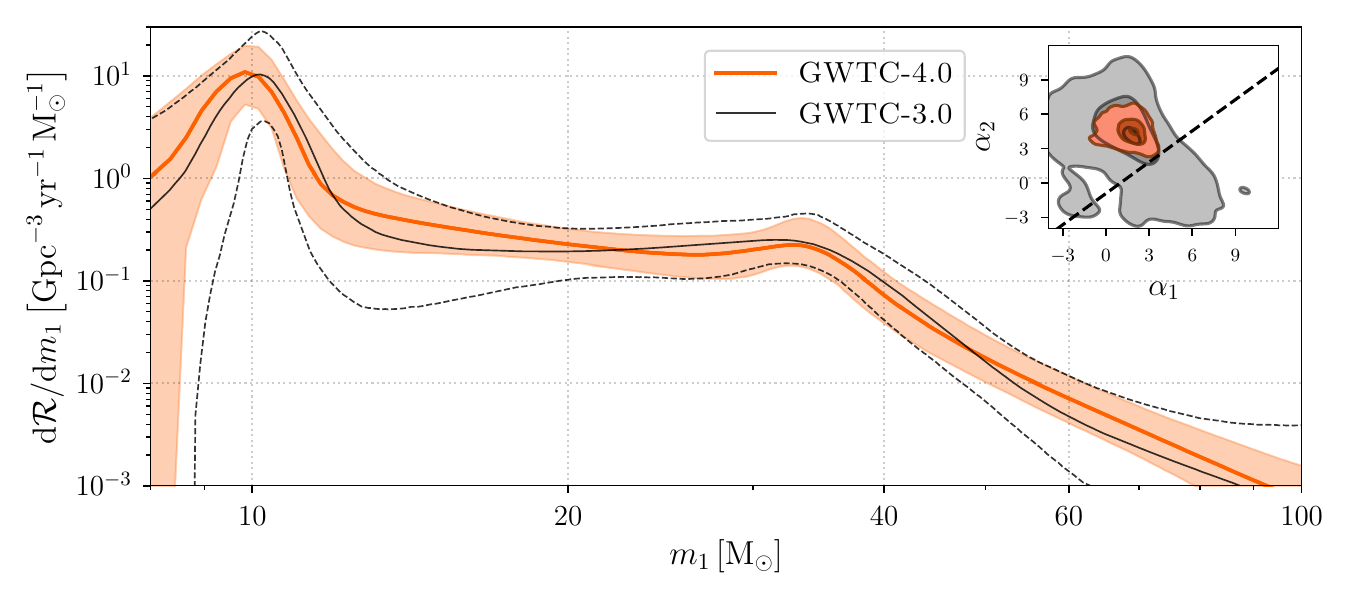}
	\caption{Differential merger rate as a function of primary mass (evaluated at $z=0.2$) of the \BrokenPLTwoPeaks model inferred with \gwtcfour compared to the same model applied to only \gwtcthree \ac{BBH} events. The orange shaded region shows the 90\% credible interval for \gwtcfour and the solid orange curve shows the posterior median, while the black dashed curves bound the $90\%$ credible region for \gwtcthree and the solid black curve shows the posterior median. The inferred distribution is similar between catalogs, which highlights that the low-mass structure identified in \gwtcfour was present in \gwtcthree. The inset figure shows the joint posterior of the broken power law index parameters $\alpha_1$ and $\alpha_2$ for \gwtcthree (gray) and \gwtcfour (orange), with the contours showing the $5^{\rm th}, 50^{\rm th}$, and $95^{\rm th}$ percentiles. The black dashed curve in the inset indicates where $\alpha_1 = \alpha_2$.}
	\label{fig:primary_mass_o3b_comparison}
  \end{figure*}

\textbf{A broken power law is necessary to describe the continuum structure above $\bm{{\sim}15 \,\Msun}$} (see Figure \ref{fig:primary_mass_default}). Below ${\sim}35 \, \Msun$, the continuum  of \ac{BH} masses is well described by a power law with spectral index $\alpha_1 = \CIPlusMinus{\defaultbbh[alpha_1]}$. Above ${\sim}35 \, \Msun$, the continuum steepens, with a spectral index of $\alpha_2 = \CIPlusMinus{\defaultbbh[alpha_2]}$ that is consistent with the \PLPeak result from \gwtcthree ($\alpha = $ \alphaOT). We find that $\alpha_2 > \alpha_1$ at $\defaultbbh[alpha_diff_perc] \%$ credibility. Though this continuum structure was identified by \citet{Callister:2023tgi}, it was not identified by the \parametrics in \citet{KAGRA:2021duu}, in part due to the limited flexibility of models employed in \citet{KAGRA:2021duu}. To illustrate this, in Figure \ref{fig:primary_mass_o3b_comparison} we present a re-analysis of \gwtcthree using the \BrokenPLTwoPeaks model alongside the \gwtcfour result, which shows improved constraints across the full mass spectrum. 

\textbf{We identify a feature at $\bm{{\sim}35\, \Msun}$}. Using the \parametric, we find that this feature is consistent with either: (i) an over-density that peaks at $m_1 = \CIPlusMinus{\defaultbbh[mass_1][peak_2_location]} \, \Msun$ relative to an underlying broken power law (i.e., the \BrokenPLTwoPeaks result), or (ii) a broken power law with break mass $m_\text{break} = \CIPlusMinus{\PowerLawTwoPeakOneBBH[break_mass]} \,\Msun$ (i.e., a broken power law without a second peak near the break). See Appendix \ref{appendix:model_comparison_mass} for a more detailed discussion of this result. The former conclusion is supported by the \BSpline model, which exhibits a local peak at $m_1 = \CIPlusMinus{\BsplineIID[mass_1][peak_2_location]} \,\Msun$, consistent with other models that employ a \nonparametric in Appendix \ref{appendix:non_parameteric_comparison}. 

\textbf{Additional structure may be present in the mass spectrum}. A bump near ${\sim}20 \,\Msun$ is present in some of the \nonparametrics (see Figure \ref{fig:bbh_mass_non_parametrics} in Appendix \ref{appendix:non_parameteric_comparison}). We cannot conclude with the \parametric whether adding a third Gaussian component to the \BrokenPLTwoPeaks model in this region is required by the data or not (as quantified by the log Bayes factor $\log_{10}\mathcal{B} = \BBHMassModelBayesFactors[pl2pk3]$ between the default model and one including a third peak). This feature was first reported in an analysis of \gwtctwo \citep{Tiwari:2020otp} and was present in several analyses of \gwtcthree \citep{KAGRA:2021duu,Edelman:2022ydv, Tiwari:2023xff,Godfrey:2023oxb}. Additionally, the \BSpline and other \nonparametrics show a rise in the merger rate relative to the \BrokenPLTwoPeaks result in the ${\sim}60 \,\Msun$ region, which can be seen in Figure \ref{fig:primary_mass_default} and Figure \ref{fig:bbh_mass_non_parametrics}. A previous study of \gwtcthree found evidence for a similar feature in this region \citep{MaganaHernandez:2024qkz}.

We do not place informative constraints on the individual parameters that govern low-mass smoothing for the \BrokenPLTwoPeaks model. We caution against astrophysically interpreting the \ac{BBH} primary mass distribution below ${\sim} 8 \,\Msun$ because of the bias that may be introduced by removing the probable \ac{NS}-containing events in the manner described in Section~\ref{sec:data}. Removing such events is responsible for the discrepancy below ${\sim} 8 \,\Msun$ between Figure \ref{fig:full_mass_spectrum} and Figure \ref{fig:primary_mass_default}.

\gwtcfour includes an exceptional high-mass BBH, GW231123~\citep{GW231123}, whose inferred component masses lie at the extreme upper end of those in our dataset. To check if GW231123 is an outlier with respect to the mass distribution of \acp{BBH}, we construct mock catalogs containing $153$ detected events following the inferred population without GW231123, and calculate the distribution of maximum detectable \ac{BH} masses. The total mass of GW231123 lies at the $\CIPlusMinus{\OutlierTestMassiveEvent[percentilemmax300][total_mass_source]}$th percentile of this distribution. This shows that while GW231123 lies in the tail of the distribution, its total mass is consistent with the inferred mass spectrum; the degree of consistency is more than was the case with \gwtcthree data alone~\citep{GW231123}.

The main compact binary formation scenarios (\citealt{{Mandel:2018hfr}} and references therein)---isolated binary evolution and dynamical assembly in dense stellar environments---have both been shown to produce populations consistent with current observations \citep{Mandel:2021smh}. A peak near $m_1 \sim 10 \, \Msun$ is often predicted by isolated binary evolution models \citep{Dominik:2014yma, Belczynski:2017gds, Giacobbo:2018etu, Wiktorowicz:2019dil, Neijssel:2019irh}, while mass distributions from dynamical formation, such as in young and globular clusters, typically peak above $10\, \Msun$ \citep{Rodriguez:2016kxx, Hong:2018bqs, Rodriguez:2019huv, Banerjee:2020qub, Antonini:2020xnd}. However, predictions from isolated and dynamical formation often overlap and can vary significantly based on the assumptions and methodologies used, making it difficult to conclude the origin of the observed catalog or constrain formation physics based on features in the marginal mass distributions. 

A feature that may provide distinguishing power between the isolated and dynamical channels is the existence of an upper mass gap, in the range $45\,\Msun \lesssim m \lesssim 120 \,\Msun$. Such a dearth would be consistent with the theorized \textit{pair-instability mass gap}, arising from the complete disruption of massive stars due to runaway electron--positron pair production \citep{Woosley:2016hmi,Mapelli:2019ipt,Farmer:2019jed}. While the precise locations of the lower and upper edges of the pair-instability mass gap are sensitive to physical assumptions \citep{Renzo:2020rzx, vanSon:2020zbk, Woosley:2021xba, Shen:2023rco, Winch:2024xdt}, the feature tends to be a robust prediction of most stellar evolution models \citep{Marchant:2018kun,Marchant:2020haw,Renzo:2020rzx,Woosley:2021xba}. The gap may not be completely empty due to overmassive stellar envelope fallback or stellar mergers \citep{DiCarlo:2019pmf, DiCarlo:2019fcq, Mapelli:2019ipt, Kremer:2020wtp}, hierarchical BBH mergers in clusters or in AGN environments \citep{Mckernan:2017ssq, Rodriguez:2019huv, McKernan:2019beu, Yang:2019okq, Mapelli:2020xeq, Antonini:2018auk, Fragione:2020nib, Liu:2020gif, Martinez:2020lzt, Sedda:2020jvg, Mahapatra:2021hme, Mahapatra:2022ngs, Mahapatra:2024qsy}, or even due to primordial \acp{BH} \citep{Postnov:2014tza, Bird:2016dcv, Clesse:2020ghq}. The decrease in the merger rate above $m_1 \sim35 \, \Msun$ seen in all models and the tentative rise near $m_1\sim60 \, \Msun$ seen in the \nonparametrics may hint toward a polluted mass gap.

\begin{figure}
	\centering
	\includegraphics[width=\columnwidth]{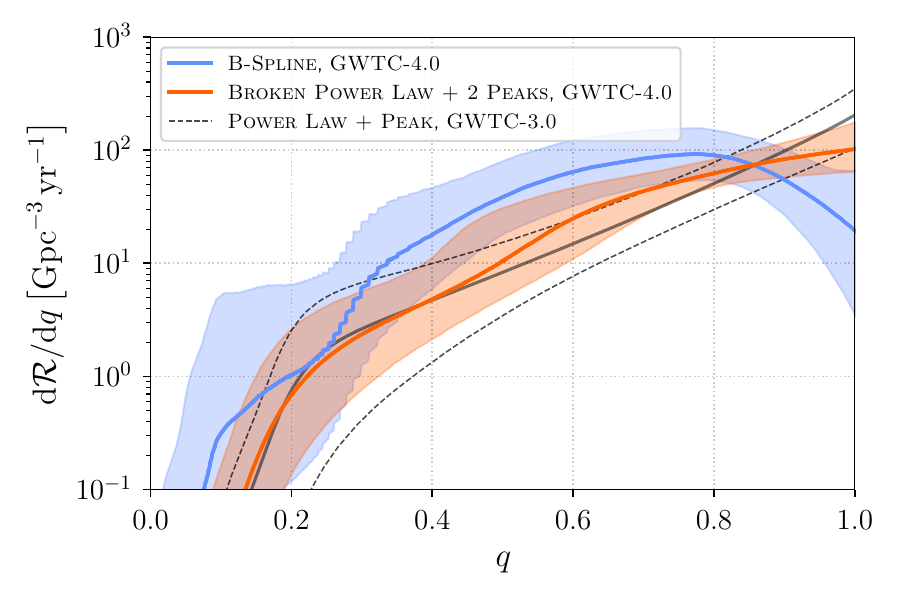}
	\caption{Differential merger rate as a function of mass-ratio (evaluated at $z=0.2$) of the \BrokenPLTwoPeaks model (orange) and \BSpline model (blue). The fiducial \textsc{Power Law + Peak} model from \citet{KAGRA:2021duu} is included for comparison. Solid curves indicate posterior medians and the shaded (dashed) regions show 90\% credible intervals.}
	\label{fig:mass_ratio_default}
  \end{figure}

\subsection{Mass-Ratio}
\label{subsec:bbh_ratio}

Figure \ref{fig:mass_ratio_default} shows the differential merger rate ${\rm d}\mathcal{R}/{\rm d}q$ evaluated at $z = 0.2$ inferred by the \BrokenPLTwoPeaks and \BSpline models compared to the \PLPeak result from \gwtcthree.\footnote{Models that utilize B-Splines infer the separable distribution $p(q)$ but the results of such models shown in this section are actually the conditional marginal distribution $p(q|m_2{>}3\Msun)$. See Appendix \ref{appendix:BSpline} for further details.}

\textbf{We find that the mass-ratio distribution can be described by a power-law with an index $\bm{\beta_q = \CIPlusMinus{\defaultbbh[beta]}}$} that is consistent with \gwtcthree ($\beta_q = 1.1^{+1.7}_{-1.3}$) with a reduction in uncertainty. The \BSpline model shows a reduced rate above $q \gtrsim 0.8$ relative to the \BrokenPLTwoPeaks result. More flexible parametrizations in the \parametric were explored (see also the other \nonparametrics in Figure \ref{fig:bbh_mass_non_parametrics}, which do peak near unity), but model comparisons were inconclusive. Similarly, in \gwtcthree, \citet{Godfrey:2023oxb} inferred a mass-ratio distribution peaked away from unity using a \nonparametric, while \citet{Rinaldi:2025emt} found equal evidence for two \parametrics with very distinct behavior above $q \gtrsim 0.7$. 

Models that incorporate correlations between source parameters have a greater potential to distinguish between formation channels than uncorrelated ones. We next present results from two different models that correlate features of the primary mass spectrum with different mass-ratio distributions. 

\begin{figure}[t]
	\centering
	  \includegraphics[width=\columnwidth]{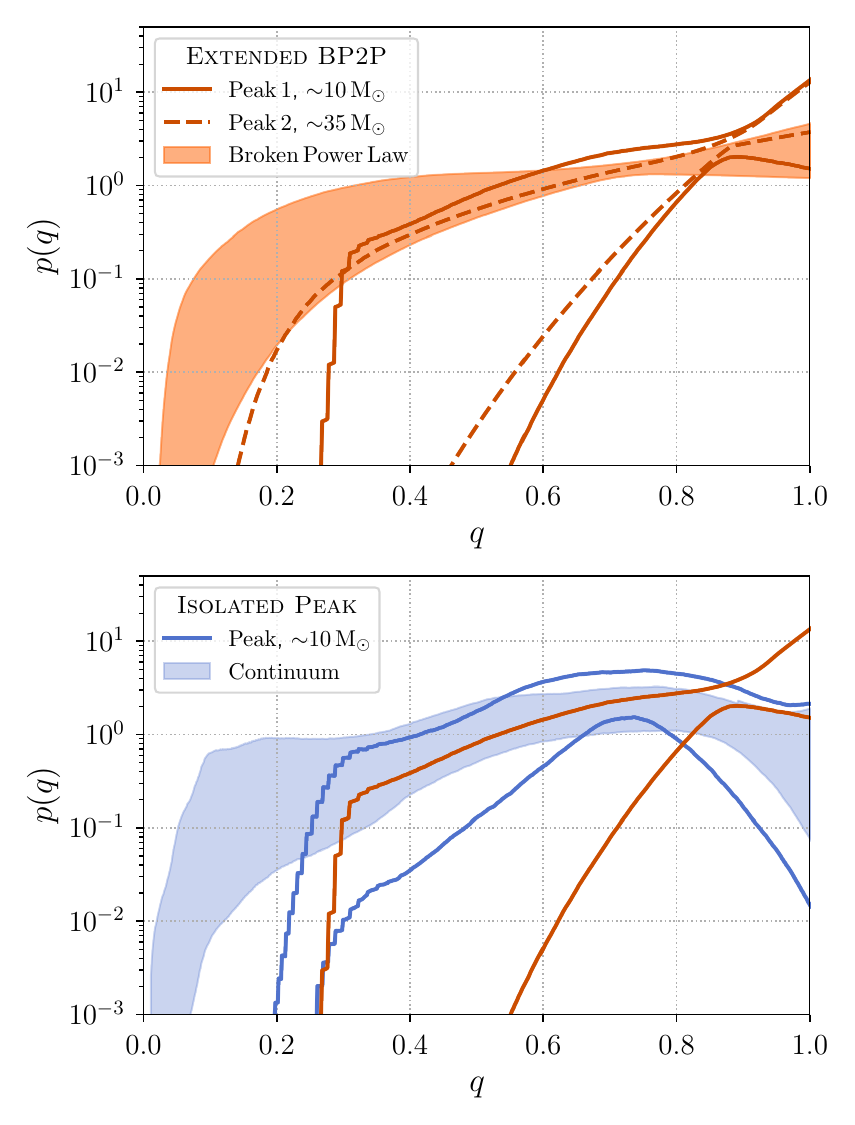}
	  \caption{Top Panel: The $90\%$ credible regions for the mass-ratio distribution of the ${\sim}10 \,\Msun$ peak (solid orange curves), ${\sim}35 \,\Msun$ peak (dashed curves), and continuum (orange shaded region) components of the \ExtendedBrokenPLTwoPeaks model. Bottom Panel: The $90\%$ credible regions for mass-ratio distribution of the ${\sim}10 \,\Msun$ peak (solid blue curves) and the rest of the mass spectrum (blue shaded region) inferred with the \textsc{Isolated Peak} model. The ${\sim}10\, \Msun$ peak mass-ratio distribution from the \ExtendedBrokenPLTwoPeaks is included for comparison.}
	  \label{fig:mass_ratio_peak}
  \end{figure}

\textbf{\acp{BH} with masses $\bm{{\sim} 35\, \Msun}$ preferentially merge with other \acp{BH} of more equal mass relative to those in the underlying mass continuum.} The \ExtendedBrokenPLTwoPeaks model modifies the \BrokenPLTwoPeaks model by allowing each primary mass mixture component (i.e., the broken power law and two Gaussian components) to be associated with a different power-law-mass-ratio distribution. Figure \ref{fig:mass_ratio_peak} shows the mass-ratio distribution for each mixture component. Specifically, each component infers a different power-law index $\beta_q$, with $\beta_q^{\rm BP} = \CIPlusMinus{\VaryingBetaQsTwoModes[param][beta_q_power_law]}$ for the broken power law, $\beta_q^{\rm peak1} = \CIPlusMinus{\VaryingBetaQsTwoModes[param][beta_q_low_mass_peak]}$ for the Gaussian component that captures the ${\sim}10\, \Msun$ peak, and $\beta_q^{\rm peak2} = \CIPlusMinus{\VaryingBetaQsTwoModes[param][beta_q_high_mass_peak]}$ for the second Gaussian component that captures the ${\sim}35 \, \Msun$ feature. Critically, $\beta_q^{\rm peak2} > \beta_q^{\rm BP}$ at $\VaryingBetaQsTwoModesBetaDiffPerc[beta_peak_pl_diff_perc] \%$ credibility.  Other studies of \gwtcthree have drawn similar conclusions about the pairing preferences of high mass \acp{BH} \citep[e.g., ][]{Li:2022jge, Baibhav:2022qxm, Sadiq:2023zee, Galaudage:2024meo, Roy:2025ktr}. 

\textbf{\acp{BH} with masses $\bm{{\sim}10\, \Msun}$ may prefentially merge with lighter \acp{BH}}. The region bounded by the solid blue curves in the bottom panel of Figure \ref{fig:mass_ratio_peak} shows the mass-ratio distribution of the ${\sim}10 \,\Msun$ peak inferred with the \textsc{Isolated Peak} model. This model is a mixture of a Gaussian peak and a B-Spline (continuum) in primary mass, and the mass ratio and spin distributions are inferred separately for each mixture component with B-Splines (see Appendix \ref{appendix:non_parametric_models_summary} for further details). The Gaussain peak is inferred at ${\sim}10 \,\Msun$ and its mass-ratio distribution exhibits a peak at $q = \CIPlusMinus{\IsoPeakIID[mass_ratio][peak][peak_location]}$, a feature that a single power law is unable to reproduce. This could explain the large uncertainty in $\beta_q^{\rm peak1}$ from the \ExtendedBrokenPLTwoPeaks model. The solid blue curve in Figure \ref{fig:mass_ratio_peak} shows the mass-ratio distribution of masses outside of the ${\sim}10 \,\Msun$ peak inferred by the B-Spline mixture component. Unlike the full population mass-ratio distribution inferred by the \BSpline model in Figure \ref{fig:mass_ratio_default}, this result does not possess a peak near $q \sim 0.8$, indicating that the peak seen in the full population is due largely to the events around ${\sim}10 \,\Msun$. This mass feature was identified in \gwtcthree by \citet{Godfrey:2023oxb}. 

Most formation channels generally favor equal mass systems. For example, dynamical formation can produce systems with a wide range of mass ratios, but predicted distributions typically peak at unity \citep{Rodriguez:2016kxx, Torniamenti:2024uxl}. Certain hierarchical mergers may not necessarily follow this trend, in particular mergers between first generation and second generation BHs have been shown to produce a mass-ratio distribution peaked near $q \sim 0.5$ \citep{Rodriguez:2019huv}. Mass transfer during the contact phase of binaries formed via chemically homogeneous evolution~\citep{deMink:2016vkw, Marchant:2016wow} leads to a strong preference for equal mass-ratio systems, but this mechanism is thought to be important for binaries with $\massone \gtrsim 10\,\Msun$~\citep{duBuisson:2020asn, Zevin:2020gbd, Riley:2020btf}. Mass-ratio reversal within the stable mass transfer channel can lead to a peak in the mass-ratio distribution between $q\sim$ 0.6-0.8 \citep{Neijssel:2019irh,vanSon:2021zpk}, which is qualitatively consistent with the mass-ratio distribution of the ${\sim}10 \,\Msun$ inferred with the \textsc{Isolated Peak} model. Stable mass transfer also predicts a peak near ${\sim}10 \,\Msun$ that is robust to uncertainties in the metallicity-dependent star formation history \citep{vanSon:2022ylf} and other physical uncertainties of the channel \citep{vanSon:2022myr}.

%% file: bbh_spins.tex
\subsection{Spin}
\label{subsec:bbh_spins}

We next present the spin distribution of the \ac{BBH} population through O4a.
We model spins using two different parameterizations: the magnitudes $\chi_i$ and tilt angles $\theta_i$ (Section~\ref{subsubsec:bbh_spin_mag_and_tilts}), and the effective inspiral spin $\chieff$ and effective precessing spin $\chip$ (Section~\ref{subsubsec:bbh_effective_spins}).
The effective inspiral spin $\chieff$ is the mass-weighted average of the component spins \textit{aligned} with the binary's orbital angular momentum~\citep{Racine:2008qv,Ajith:2009bn}.
The effective precessing spin $\chip$ characterizes the degree of relativistic precession caused by spin--orbit misalignment, capturing the effect of the \textit{in-plane} spin components~\citep{Schmidt:2010it,Schmidt:2012rh,Schmidt:2014iyl,Gerosa:2020aiw}; these are defined in Equations 15 and 16 of \cite{GWTC:Introduction}.
More details about spin parameterization are given in Appendix~\ref{appendix:spinModels}.

Previous analyses of the \acp{BBH} observed through O3 found that \ac{BH} spin vectors tend to be small in magnitude ($\chi\lesssim 0.3$) and preferentially---but not exclusively---lie above the orbital plane \citep[$\cos\theta>0$;][]{LIGOScientific:2018jsj,LIGOScientific:2020kqk,KAGRA:2021duu}. 
The new observations in \gwtcfour further support these conclusions and additionally suggest more structure in the component and effective spin distributions, enabled by changes to the models used in previous population analyses:
(i) \textbf{the spin magnitude distribution has support at $\bm{\chi\approx 0}$}, (ii) \textbf{the spin tilt distribution may peak away from perfect alignment with the orbital angular momentum}, and (iii) \textbf{the $\bm{\chieff}$ distribution is asymmetric about its peak}.
We elaborate upon these new features in the following subsections.

\subsubsection{Spin Magnitudes and Tilts}
\label{subsubsec:bbh_spin_mag_and_tilts}

\begin{figure*}
  \centering
  \includegraphics[width=\linewidth]{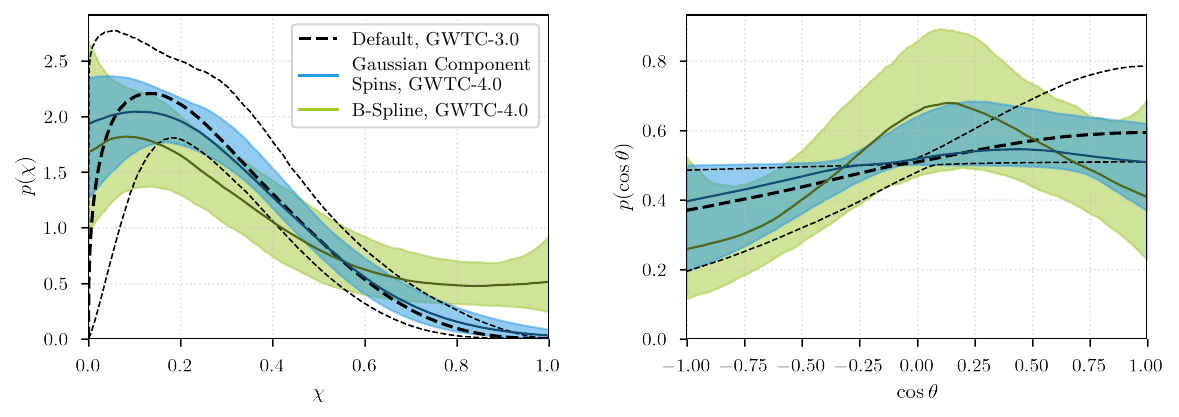}
  \caption{Marginal spin magnitude $\chi$ (left) and cosine tilt angle $\cos\theta$ (right) distributions under the \default (blue) and \BSpline (green) models.
  The results from the \gwtcthree~{\sc Default} are shown for comparison (black dashed); note that this model is different from what is used for \gwtcfour.
  The \gwtcthree analysis employed a Beta distribution for spin magnitudes rather than a truncated normal, and fixed the location of the Gaussian component of the spin tilt angle distribution to $\cos\theta=1$ rather than letting it vary freely in inference. 
  The solid lines show the median of the inferred distributions from \gwtcfour, and the shaded regions show their 90\% credible regions.
  The thick dashed lines show the median from \gwtcthree, while the thin dashed lines show its 90\% credible region.}
  \label{fig:spin_mag_tilt_pdf}
\end{figure*}

Spin magnitudes and tilt angles provide insight about \ac{BBH} formation and evolution~\citep[e.g.,][]{Mandel:2018hfr,Mapelli:2018uds}; we begin with spin magnitudes.
If angular momentum transport in stars is efficient, stellar cores rotate slowly, resulting in small spin magnitudes for isolated \acp{BH}~\citep{Fuller:2019sxi,Ma:2019cpr,Fuller:2019ckz}.
However, binary interactions can significantly influence \ac{BH} spins through mechanisms such as tides~\citep{Hut:1981,Packet:1981,Zaldarriaga:2017qkw,Qin:2018vaa,Bavera:2020inc,Mandel:2020lhv} and accretion~\citep{Hut:1981,Packet:1981,vandenHeuvel:2017pwp,Neijssel:2019irh,Steinle:2020xej,Stegmann:2020kgb}. 
If a \ac{BBH} forms from a pre-existing isolated stellar binary, tidal interactions can spin up the progenitor of the second-born \ac{BH}~\citep{Bavera:2020inc,Qin:2018sxk}, yielding spin magnitudes of $\chi\approx 0.2{-}0.4$~\citep{Ma:2023nrf}, although these can be reduced by stellar winds~\citep{Tout:1992}.
Chemically homogeneous evolution---involving tidally locked, high-mass, low-metallicity, close binaries---may produce even larger spins~\citep{Mandel:2015qlu,deMink:2016vkw}. 
Alternatively, large spin magnitudes may point toward a hierarchical merger origin, where one or both \acp{BH} are remnants of previous \ac{BBH} mergers~\citep{Rodriguez:2019huv,Zhang:2023fpp,Doctor:2019ruh,Kimball:2020opk,Payne:2024ywe,Gerosa:2021mno,Fishbach:2017dwv,Gerosa:2017kvu,Mould:2022ccw,Mahapatra:2021hme,Mahapatra:2022ngs,Mahapatra:2024qsy}, which are predicted to have $\chi \sim 0.7$~\citep{Lousto:2009mf}.

With this astrophysical context, we present our measurement of the spin magnitude distribution through O4a. 
The left column of Figure~\ref{fig:spin_mag_tilt_pdf} shows the marginal distributions of $\chi$ using the \parametric~(\default; blue) and \nonparametric~(\BSpline; green).
The \default~model (Equation \ref{eqn:spin_mag_model}) describes the $\chi$ population as a truncated Gaussian distribution.
This is a departure from the {\sc Default} spin model of \gwtcthree.
There, a non-singular Beta distribution was used~\citep{KAGRA:2021duu}, which forces $p(\chi)=0$ at $\chi=0,1$ and thus cannot measure contributions to the population at near-minimal or near-maximal spins.
Allowing for more model flexibility at the $\chi$ distribution's boundaries is crucial, as there exists an ongoing discussion in the literature about whether or not there is an over-density of \acp{BBH} with $\chi \lesssim 0.01$~\citep{Kimball:2020opk,Galaudage:2021rkt,Callister:2022qwb,Tong:2022iws,Mould:2022xeu,Hussain:2024qzl}.
We find that the \default~model is preferred over the {\sc Default} spin model of \gwtcthree by $\log_{10} {\cal B} = \spinModelComparisonLogTenBayesFactors[MagTruncnormIidTiltIsotropicTruncnormNid][log10_bayes_factor_over_old_default]$; additional parametric spin magnitude (and tilt) models are discussed in Appendix~\ref{appendix:model_comparison_spins}.
The widening of the 90\% credible regions for the $\chi$ and $\cos\theta$ distributions at their boundaries under the \BSpline~model seen in Figure \ref{fig:spin_mag_tilt_pdf} is a prior-driven effect common in spline modeling \citep[e.g.,][]{Golomb:2022bon,Edelman:2022ydv}.

\textbf{In \gwtcfour, we constrain $\bm{p(\chi\approx 0) > 0}$ under both the strongly and weakly modeled approaches.}
At 90\% credibility, our recovered spin magnitude distribution peaks between $\chi=\CIBoundsDash{\MagTruncnormIidTiltIsotropicTruncnormNid[param][mu_chi]}$, as measured by the $\mu_{\chi}$ location parameter.
Broadly, \ac{BH} spins are inferred to be predominantly non-extremal, with the \default~model finding that $90\%$ of \acp{BH} having $\chi<\MagTruncnormIidTiltIsotropicTruncnormNid[ppd][a][90th percentile]$.
The comparative dearth of observed \acp{BBH} with large spins disfavors a population dominated by second-generation \acp{BH}. 
However, the precise fraction of systems with large spin magnitudes is model dependent: the \default and \BSpline models only disagree at 90\% credibility for $\chi > \defaultVersusBSplineSpinMagnitudes[chi at discrepancy=90]$.
At $\chi = 0.8$, $0.9$, and $1.0$, their $p(\chi)$ distributions differ at the $\defaultVersusBSplineSpinMagnitudes[discrepancy percentile for chi=0.8]$\%, $\defaultVersusBSplineSpinMagnitudes[discrepancy percentile for chi=0.9]$\%, and $\defaultVersusBSplineSpinMagnitudes[discrepancy percentile for chi=1.0]$\% levels, respectively.
While the \default model approaches $p(\chi)\sim 0$ at $\chi=1$,
the \BSpline~model infers a larger, nearly flat distribution from $\chi=0.6$ to $1$. 
Similar behavior was found in \gwtcthree~\citep{Godfrey:2023oxb}, and attributed to a subpopulation with near-uniform spin magnitudes.
The discrepancy between our two models may arise from limited flexibility in the \parametric; a single truncated Gaussian cannot increase the probability at $\chi \gtrsim 0.8$ without also increasing it at $\chi\sim 0.2{-}0.5$. 
In \gwtcfour, a handful of events do have preferentially large spin magnitudes, including GW231123 which has primary spin $\chi_1 > 0.5$ with high confidence~\citep{GW231123}.

Under both the \default and \BSpline models, the results presented in Figure~\ref{fig:spin_mag_tilt_pdf} assume that the spin magnitudes are \ac{IID}, meaning that the function describing the population distribution is factorizable in terms of $\chi_1$ and $\chi_2$, and the two are described by the same set of hyperparameters.
The data prefer identically distributed spin magnitudes over those which are non-identically distributed, cf.~Table~\ref{tab:modelComparisonSpins}.
However, mathematically, the primary and secondary spins cannot be \ac{IID} if the purported correlation between $q$ and $\chieff$ is true~\citep{Farr:2025}. 
We probe this correlation in Section~\ref{subsubsec:mass_spin_correlations}, and find support for its existence, albeit with evidence that has diminished since \gwtcthree. 
Thus, we interpret the Bayes factor in favor of \ac{IID} spins as a statement that, under the \default~model, we cannot yet say with statistical certainty that the spins are not identically distributed. 
This statement is likely driven by the fact that individual-event spin magnitude posterior distributions are typically wide, making $\chi_1$ and $\chi_2$ hard to distinguish. 
In general, $\chi_1$ and $\chi_2$ are not expected to be identically distributed in nature; if $q$ and $\chieff$ are indeed correlated, more informative $\chi_i$ measurements may eventually reveal their non-identical nature.
Figure~\ref{fig:spin_mags_tilts_comparison_cornerplot} in Appendix~\ref{appendix:model_comparison_spins} presents posteriors on the \default hyperparameters under the assumption that spin magnitudes (and tilts) are identically versus non-identically distributed.

\begin{figure}
  \centering
  \includegraphics[width=\linewidth]{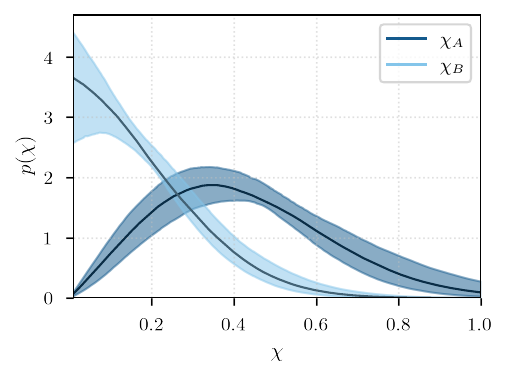}
  \caption{Larger ($\chi_A$, darker blue) and smaller ($\chi_B$, lighter blue) spin magnitude distributions, derived from imposing order statistics on the \default $\chi$ distribution shown in Figure~\ref{fig:spin_mag_tilt_pdf}.
  The solid lines show the median of each inferred distribution, and the shaded regions show the 90\% credible intervals.}
  \label{fig:spin_sorting_pdf}
\end{figure}

Next, Figure~\ref{fig:spin_sorting_pdf} shows the inferred spin magnitude distributions if, rather than sorting by the more versus less massive \ac{BH}, we instead sort by the \ac{BH} with the larger (subscript A) and smaller (subscript B) spin magnitude~\citep{Biscoveanu:2020are}. 
The magnitudes $\chi_A$ and $\chi_B$ are derived quantities and are not fit independently:
to generate their distributions, we take results from the \default~model and impose order statistics, assuming that $\chi_A$ is the larger of two draws from the $p(\chi)$ distribution in the left panel of Figure~\ref{fig:spin_mag_tilt_pdf} and $\chi_B$ is the smaller~\citep{KAGRA:2021duu, Biscoveanu:2020are}.
This analysis does not assume any sort of pairing function between spins.
Spin sorting offers an alternative way to visualize and interpret the spin information from the \ac{IID} model.
The more rapidly spinning component has a wide spin magnitude distribution, with a \ac{PPD} peaking at $\chi_A = \MagTruncnormIidTiltIsotropicTruncnormSpinSorting[ppd][chi_A][spin at peak]$, and support up to $\chi_{A,99\%}= \CIPlusMinus{\MagTruncnormIidTiltIsotropicTruncnormSpinSorting[param][chi_A_99th]}$ (value of $\chi_A$ at which each $p(\chi_A)$ trace reaches its 99th percentile, serving as a proxy for the distribution's maximum).
The vanishing probability at $\chi_A=0$ is a Jacobian effect of the order statistics.
However, if both \acp{BH} had $\chi\approx 0$, the $\chi_A$ distribution would be much more strongly peaked at small values~\citep{Szemraj:2025fmm}, meaning that spin-sorting results on \gwtcfour indicate that \textbf{at least one BH per binary has $\bm{\chi\gtrsim 0}$}.
The more slowly spinning component, on the other hand, is consistent with a narrower distribution peaking at $\chi_B = 0$, and only has support up to $\chi_{B,99\%} = \CIPlusMinus{\MagTruncnormIidTiltIsotropicTruncnormSpinSorting[param][chi_B_99th]}$.
The observation that spin $\chi_B$ magnitudes are preferentially small could indicate small natal spins for at least one of the two \acp{BH}, while the fact that the population is consistent with only one \ac{BH} per binary having large spin supports the existence of some variety of spin-up mechanism in \ac{BBH} evolution.

We next turn to the spin tilt distribution. 
Many authors argue that if \acp{BBH} are formed primarily in the isolated binary scenario, large tilt angles are hard to explain without invoking large supernovae kicks and inefficient tides~\citep{Kalogera:1999tq,Gerosa:2018wbw, Steinle:2020xej,Wysocki:2017isg,Stevenson:2022hmi,Callister:2020vyz}, although others claim that, depending on the specifics of poorly understood supernovae physics, even small kicks can misalign a binary~\citep{Baibhav:2024rkn,Tauris:2022ggv}.
Complete anti-alignment ($\cos\theta = -1$) is a possible result of mass transfer in the isolated channel~\citep{Stegmann:2020kgb}. 
On the other hand, \acp{BBH} formed dynamically in stellar clusters are predicted to have istropically distributed spin orientations, as there is no \textit{a priori} preferential spin direction in these environments~\citep{Rodriguez:2015oxa,Rodriguez:2016kxx,Rodriguez:2017pec,Farr:2017uvj}.
However, recent studies indicate that mechanisms could possibly exist for dynamically formed \acp{BBH} to have slight preference for $\cos\theta>0$~\citep{Trani:2021lcv,Banerjee:2023ycw,Kiroglu:2025bbp}, especially for \acp{BBH} formed dynamically in the disks of active galactic nuclei~\citep{Wang:2021yjf,McKernan:2021nwk}.

The right-hand column of Figure~\ref{fig:spin_mag_tilt_pdf} shows the marginal distribution of the cosine of the tilt angle, $\cos\theta$.
The $\cos\theta$ population under the \default~model is a mixture between isotropic and truncated Gaussian sub-populations (Equation~\ref{eqn:spin_tilt_model}).
Following \gwtcthree and earlier population analyses, we assume that spin tilt angles are \ac{NID} under the \default~model, meaning that while $\cos\theta_1$ and $\cos\theta_2$ are described by the same hyperparameters, the population distribution is \textit{not} separable in terms of the two: we require that both \acp{BH} in a binary are drawn from the same sub-population (either the isotropic or the truncated Gaussian).
\ac{NID} tilts are favored by the data over non-identical distribution, cf.~Table~\ref{tab:modelComparisonSpins}.
The \BSpline model naturally assumes spin tilt angles are \ac{IID}, as it does not probe separate tilt sub-populations.

The Default model of \gwtcthree fixed the location of the Gaussian sub-population at exact spin--orbit alignment~\citep{Vitale:2015tea,Talbot:2017yur,KAGRA:2021duu}, making it a half-Gaussian peaking at $\cos\theta=1$.
In \gwtcfour, the strongly and weakly modeled approaches both find that \textbf{the spin tilt distribution may peak away from exact spin--orbit alignment}. 
The \default~model finds that the $\cos\theta$ distribution reaches its maximum between $\defaultbbh[mu_spin][5th percentile]$ and $\defaultbbh[mu_spin][95th percentile]$ at 90\% credibility; consistently, the \BSpline~model finds the peak to lie between $\BsplineIID[cos_tilt][peak][5th percentile]$ and $\BsplineIID[cos_tilt][peak][95th percentile]$.
The possibility that the tilt angle distribution peaks away from alignment is also found in \gwtcthree under various models which permit such a feature~\citep{Vitale:2022dpa,Edelman:2022ydv,Golomb:2022bon}.
It is not impossible, however, for the peak of a $\cos\theta$ distribution to be inferred away from alignment even when the true underlying population does peak at $\cos\theta=1$~\citep{Vitale:2025lms}.
The inferred peak of the \gwtcfour $\cos\theta$ distribution is more strongly constrained away from unity than nearly all of those spuriously found with simulated catalogs of the same size~\citep[cf.~Figure~\ref{fig:spin_mags_tilts_comparison_cornerplot} with Figure 4 of][]{Vitale:2025lms}.

Under both the strongly and weakly modeled approaches, the $\cos\theta$ distribution shown in Figure~\ref{fig:spin_mag_tilt_pdf} has support across a wide range of spin tilts, with a slightly larger fraction of positive $\cos\theta$ compared to negative.
Given this broad support, we next ask the question: what is the lowest spin tilt angle that is absolutely required to fit \gwtcfour reasonably?
To probe this, we use a model which is similar to the \default~model but with $p(\cos\theta < t_{\rm min}) = 0$ for an inferred value $t_{\rm min}$, where $t\equiv \cos\theta$~\citep{Tong:2022iws,Callister:2022qwb}; see Equation~\eqref{eqn:spin_tilt_model_with_min}.
We find $t_{\rm min} = \CIPlusMinus{\MinimumTiltModel[param][tmin]}$ at \confidenceLevel~credibility.
That the posterior for the minimum required $\cos\theta$ is inconsistent with $-1$ but remains confidently negative is a somewhat unexpected astrophysical result.
If most \acp{BBH} form in isolation, a minimum tilt cutoff is plausible but would be expected to be closer to zero, potentially even positive. 
Conversely, dynamically formed \acp{BBH} should exhibit isotropic tilts extending down to $\cos\theta = -1$. 
Our observed tilt distribution therefore does not preclude either broad scenario of isolated or dynamical formation---or a mixture of both.
However, the fact that the inferred minimum required tilt is significantly negative suggests the contribution of a dynamical formation channel to the \ac{BBH} population.
We further discuss the astrophysical interpretation of negative spin tilts and potential sub-populations in Section~\ref{subsubsec:bbh_effective_spins} in the context of effective spins.

\subsubsection{Effective Spins}
\label{subsubsec:bbh_effective_spins}

\begin{figure*}
  \centering
  \includegraphics[width=\linewidth]{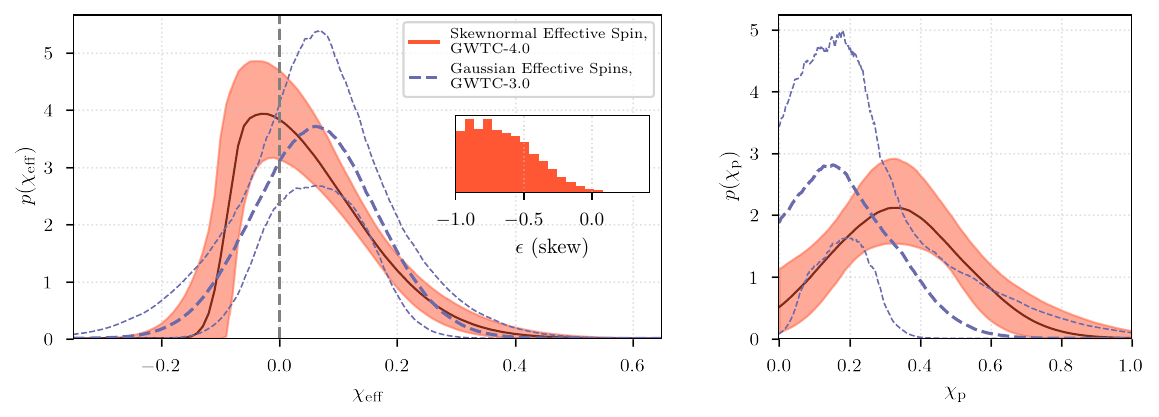}
  \caption{Marginal $\chieff$ (left panel) and $\chip$ (right panel) distributions.
  The solid lines show the median of each inferred distribution, and the shaded regions show the 90\% credible intervals.
  The \gwtcfour results under the \skewnormalChiEff model are shown in red; the \gwtcthree results under the \effectiveSpinModel~model are shown in purple dashed.
  The histogram inset in the left panel shows the skew parameter $\epsilon$ for the \skewnormalChiEff~model. 
  There is significant preference for a skewed $\chieff$ distribution, indicated by $\epsilon < 0$ with $\EpsSkewNormalChiEff[param][percent of epsilon less than zero]$\% credibility.
  }
  \label{fig:chi_eff_chi_p_pdf}
\end{figure*}

We next turn to the inferred distributions of the effective spins $\chieff$ and $\chip$. 
Figure~\ref{fig:chi_eff_chi_p_pdf} shows the marginal distributions of $\chieff$ (left) and $\chip$ (right) under the \skewnormalChiEff model (red solid; Equation~\ref{eqn:chi_eff_skewnormal_model}), compared to the \effectiveSpinModel model result from \gwtcthree (purple dashed; Equation~\ref{eqn:chi_eff_chi_p_gaussian_model}).
The \skewnormalChiEff model~\citep{Banagiri:2025dxo} differs from the previously-used \effectiveSpinModel in two ways. 
First, it allows for asymmetry in the $\chieff$ marginal distribution.
Second, the marginal $\chip$ distribution is a truncated Gaussian which is not correlated with $\chieff$.
On \gwtcfour data, the \effectiveSpinModel model finds that $\chieff$ and $\chip$ are preferentially uncorrelated, although this conclusion depends on analysis settings; see Figure~\ref{fig:effective_spins_comparison_dists} in Appendix \ref{appendix:model_comparison_effective_spins} and the discussion therein.

\textbf{We find that the $\bm{\chieff}$ distribution is skewed and asymmetric about zero with more support for positive values.}
Asymmetry can be directly probed with the $\epsilon$ skewness parameter of the \skewnormalChiEff~model, defined in Equation~\eqref{eqn:chi_eff_skewnormal_model}.
We find that the skewness $\epsilon$ is less than 0 (symmetry) with $\EpsSkewNormalChiEff[param][percent of epsilon less than zero]$\% credibility, as can be seen in the inset of the left panel of Figure~\ref{fig:chi_eff_chi_p_pdf}.
This corresponds to a wider $\chieff$ distribution to the right of the peak, and a narrower distribution to the left, and is known as \textit{positive skew}.
As discussed in Section~\ref{subsubsec:mass_spin_correlations}, asymmetry is also found when allowing for a linear or spline correlation between $\chieff$ and mass ratio.
Results for the \gwtcfour $\chieff$ distribution measured with more models, including the \effectiveSpinModel model and with the \nonparametric~are presented in Figures~\ref{fig:effective_spins_comparison_dists} and~\ref{fig:effective_spins_nonparametric} in  Appendix~\ref{appendix:model_comparison}.

\begin{deluxetable*}{c | c | c | c}
\tablecaption{
    Summary of probes of spin misalignment, as measured by $\chieff$.
}
\tablehead{
    \colhead{Model} & \colhead{$\chi_{\mathrm{eff}, 1\%}$} & \colhead{Fraction $\chieff<0$} & \colhead{HM Fraction}
}
\label{tab:min_chieff}
\startdata
\effectiveSpinModel~(\gwtcthree) & $\chiEffMinGaussianGTWCThree$ & $\fNegChieffGaussianGTWCThree$ & $< 3.1\times 10^{-2}$ \\
\effectiveSpinModel~(\gwtcfour) & $\CIPlusMinus{\gaussianChiEffChiP[param][min_eff]}$ & $\CIPlusMinus{\gaussianChiEffChiP[param][f_neg_eff]}$ & $< 1.9\times 10^{-2}$ \\
\skewnormalChiEff~(\gwtcfour) & $\CIPlusMinus{\EpsSkewNormalChiEff[param][min_chi_eff]}$ & $\CIPlusMinus{\EpsSkewNormalChiEff[param][f_neg_chi_eff]}$ & $< 1.3\times 10^{-4}$ \\
\enddata
\tablecomments{
    Summary of probes of spin misalignment, as measured by $\chieff$.
    We give 90\% credible intervals for the $\chieff$ value of the first percentile of the distribution ($\chi_{\mathrm{eff}, 1\%}$), serving as a proxy for the minimum $\chieff$, and the fraction of $\chieff<0$.
    The HM fraction provides an upper limit to the fraction of \acp{BBH} of hierarchical merger origin, equal to $0.16$ times the fraction of systems with $\chieff < -0.3$~\citep{Fishbach:2022lzq,Baibhav:2020xdf}; we provide its 90\% upper limit.
    Posteriors for these quantities for these and other models are plotted in Figure~\ref{fig:effective_spins_nonparametric}.
}
\end{deluxetable*}

We now discuss \acp{BBH} with spin tilts lying below the orbital plane. 
These interesting probes of dynamical~\ac{BBH} formation have $\cos\theta<0$ and thus negative $\chieff$.
In Table \ref{tab:min_chieff}, we present \confidenceLevel~bounds for three quantities which probe negative $\chieff$ in the population, using both the \skewnormalChiEff and \effectiveSpinModel models for a straightforward comparison to \gwtcthree.
First, we report the first percentile of $\chieff$ distribution, which serves as a proxy for the distribution's minimum without necessitating the inclusion of sharp features in the distribution itself~\citep{Callister:2022qwb}, constraining it to fall between $\gaussianChiEffChiP[param][min_eff][5th percentile]$ and $\EpsSkewNormalChiEff[param][min_chi_eff][95th percentile]$.
The \effectiveSpinModel~model finds more extremal minimum spins than the \skewnormalChiEff~model, likely due to its required symmetry about its peak; fitting the positive side of the distribution forces a longer tail into the negative region.
Second, we report the fraction of the population with negative $\chieff$ to be $\gaussianChiEffChiP[param][f_neg_eff][5th percentile]$--$\EpsSkewNormalChiEff[param][f_neg_chi_eff][95th percentile]$, with the two models finding nearly identical results. 
Assuming isotropy, an upper bound on the fraction of \acp{BBH} formed dynamically in gas-free environments can be placed by doubling the fraction of negative $\chieff$~\citep[Equation 7 of][]{LIGOScientific:2020kqk}; we thus find that at most $84\%$ of \acp{BBH} form dynamically. 
Third, we give the 90\% upper limit on the fraction of \acp{BBH} coming from the hierarchical merger (HM) scenario. 
The HM fraction is bounded by the consideration that $\sim 16\%$ of \acp{BBH} coming from the HM formation channel will have $\chieff < -0.3$~\citep{Baibhav:2020xdf,Fishbach:2022lzq}; we limit this fraction to $\lesssim 3\%$.
Figure~\ref{fig:effective_spins_nonparametric} in Appendix~\ref{appendix:non_parameteric_comparison} shows posterior distributions on the quantities in Table \ref{tab:min_chieff} for the models presented here and those using the \nonparametric.

It is possible that the support for $\chieff < 0$ may be a byproduct of the models fitting for a peak at $\chieff\sim 0$ without having the flexibility for a sharp decline beyond $\chieff < 0$.
This type of sharp feature could occur if component spin magnitudes cluster around $\chi\sim 0$ but their tilts seldom reach below $\cos\theta=0$. 
In Section~\ref{subsubsec:bbh_spin_mag_and_tilts}, we address this by directly fitting for the minimum required $\cos\theta$ which is found to be confidently negative, and corresponds to a fraction $\CIPlusMinus{\MinimumTiltModel[fraction negative cos theta]}$ of systems with spins laying below the orbital plane.
These probes of negative spin indicate that \textbf{between $\sim$20--40\% the BBH population has spins which are more than 90 degrees misaligned with the orbital angular momentum}.

It is unlikely that the entire observed \ac{BBH} population originates from a single formation channel~\citep{Zevin:2020gbd,Mandel:2021smh,Cheng:2023ddt,Afroz:2024fzp,Colloms:2025hib}.
Features in our spin distributions indeed suggest the presence of multiple sub-populations.
A purely random spin channel would produce a symmetric distribution around $\chieff=0$~\citep{Rodriguez:2016kxx,Rodriguez:2016vmx,Farr:2017uvj}.
However, for all models, the observed $\chieff$ distribution is asymmetric about zero, c.f., Figure \ref{fig:effective_spins_nonparametric}.
This asymmetry suggests contribution from a preferentially aligned subpopulation~\citep{Gerosa:2018wbw,Sedda:2023mlv,Banagiri:2025dxo}.
The skew observed in the \skewnormalChiEff, as well as the \qchiefflinearcorrelation and \qchieffsplinecorrelation models~(see Section~\ref{subsubsec:mass_spin_correlations} and Figure~\ref{fig:effective_spins_nonparametric}) further supports the presence of an aligned component, either as a sub-dominant sub-population or a dominant sub-population with small spin magnitudes. 
We find a preference for small spin magnitude (Figure~\ref{fig:spin_mag_tilt_pdf}), perhaps favoring the latter.

Finally, we discuss the effective precessing spin $\chip$, which can provide additional insight about spin precession in the population.
The right panel of Figure~\ref{fig:chi_eff_chi_p_pdf} shows the marginal inferred $\chip$ distribution (red).
We find that precession exists on a population level, as our models do not support $(\mu_p, \sigma_p) = (0,0)$ at high credibility.
The \gwtcfour results support larger $\chip$ than \gwtcthree, shown for comparison in Figure~\ref{fig:chi_eff_chi_p_pdf} (purple dashed).
However, different likelihood convergence criteria (Appendix~\ref{sec:likelihood-estimator-variance}) were used between \gwtcfour and \gwtcthree, meaning these results are not directly comparable. 
Under the less-stringent \gwtcthree convergence criterion, the $\chip$ distribution inferred from \gwtcfour data is more consistent with that found in \gwtcthree, albeit still with less support for low $\mu_{\rm p}$ and low $\sigma_{\rm p}$. 
The inferred $\chip$ distribution is more dependent on analysis settings than other parameters, largely because the individual-event prior has no support at $\chip=0$ making it technically difficult to reweight individual-event posteriors to the population.
We discuss the sensitivity of the $\chip$ distribution to analysis settings in detail in Appendix~\ref{appendix:model_comparison_effective_spins}; see Figures~\ref{fig:effective_spins_comparison_cornerplot} and \ref{fig:effective_spins_comparison_dists}.

%% file: bbh_redshift.tex
\subsection{Merger Rate and Redshift Evolution}
\label{subsec:bbh_redshift}

Improvements in the sensitivity of current \ac{GW} observatories \citep{LIGOO4Detector:2023wmz, Capote:2024rmo, GWTC:Introduction} not only provide more \ac{BBH} detections, but also observations of increasingly faint sources at higher redshifts.
These observations allow us to improve our population-level constraints on the evolution of the merger rate across redshift.

\begin{figure}
  \centering
  \includegraphics[width=0.98\columnwidth]{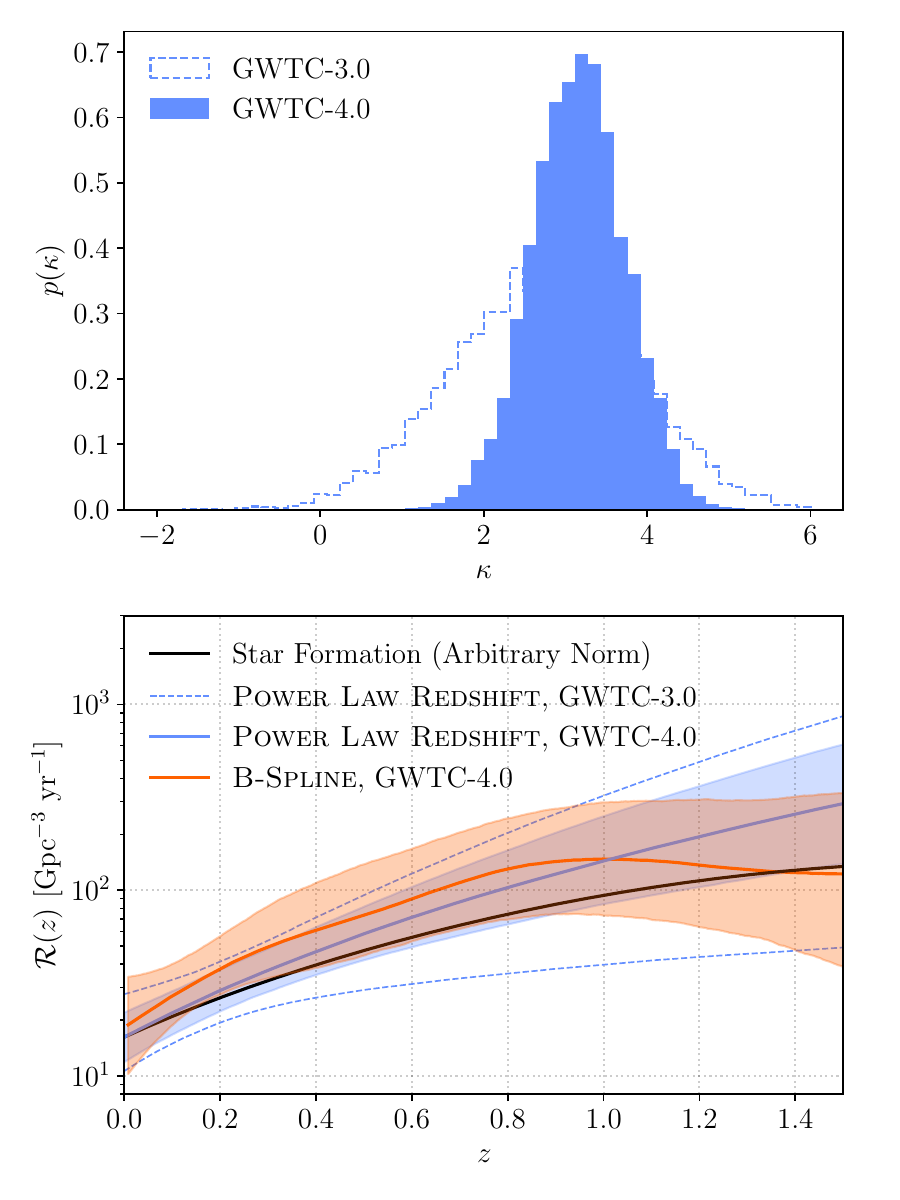}
  \caption{
  Comparison of redshift models between \gwtcthree and \gwtcfour. 
  \textit{Top:} Posterior on the $\kappa$ parameter for the \powerlawRedshift model. \gwtcfour shows increased support for positive values.
  \textit{Bottom:} Median and 90\% credible regions for the comoving source frame rate in the \powerlawRedshift model. 
  We also show a comparison of the \powerlawRedshift model and the \nonparametricAdj \BSpline model and a scaled cosmic star formation rate density \citep{Madau:2014bja}, which are consistent within uncertainties ($\kappa_{\rm SFR} = 2.7$).
  }
  \label{fig:powerlaw_redshift_gwtc3_vs_gwtc4}
\end{figure}

Following \citet{KAGRA:2021duu}, we repeat the \powerlawRedshift \parametric. We assume that the merger rate evolves as $\mathcal{R}(z) \propto (1+z)^\kappa$, and we infer the proportionality constant and the power-law index $\kappa$.
{\textbf{\boldmath We find that the \ac{BBH} merger rate at $\bm{z=0.2}$ is $\defaultbbh[redshift][rate_at_z_0-2][median]^{+\defaultbbh[redshift][rate_at_z_0-2][error plus]}_{-\defaultbbh[redshift][rate_at_z_0-2][error minus]}\,\bm{\perGpcyr}$, and the power-law exponent is constrained to $\bm{\kappa = }\defaultbbh[lamb][median]^{+\defaultbbh[lamb][error plus]}_{-\defaultbbh[lamb][error minus]}$,}}
which represent consistent and improved constraints over our \gwtcthree analysis. We infer that 99\% of detectable \acp{BBH} fall below $z=1.5_{-0.2}^{+0.2}$ (cf. the maximum observable redshift inferred as $z=1.1_{-0.2}^{+0.2}$ in \citealt{Fishbach:2023pqs}, from \gwtcthree).
We also use the \nonparametricAdj \BSpline approach to explore if the data support behavior beyond a power-law evolution, and compare our results in Figure~\ref{fig:powerlaw_redshift_gwtc3_vs_gwtc4}.
While the results are consistent---indicating that the \powerlawRedshift model is sufficient---the \BSpline approach infers a larger merger rate of $\BsplineIID[rate_at_z_0-2][median]^{+\BsplineIID[rate_at_z_0-2][error plus]}_{-\BsplineIID[rate_at_z_0-2][error minus]}\,\perGpcyr$ at $z=0.2$. 
Our results are nominally consistent with the cosmic star formation rate density, with $\kappa_{\rm SFR} = 2.7$ \citep{Madau:2014bja}.

Our constraints on the merger rate evolution informs our understanding of the \ac{BBH} progenitor formation rate and the delay time distribution between formation and merger \citep{Vitale:2018yhm,Rodriguez:2018rmd, Fragione:2018vty,Fishbach:2018edt, Baibhav:2019gxm, Romero-Shaw:2020siz,Broekgaarden:2021efa,Mapelli:2021gyv,Fishbach:2021mhp,Chruslinska:2022ovf,Fishbach:2023xws,Boesky:2024msm}. 
For \acp{BBH} formed from isolated binary evolution, the merger rate evolution is typically well approximated as a power law at small redshift, though the value of the power-law index $\kappa$ is sensitive to the assumed population synthesis parameters \citep{Neijssel:2019irh, Broekgaarden:2021efa, Gallegos-Garcia:2021hti, deSa:2024otm}.
Nevertheless, models typically prefer values around $\kappa\sim 1$ \citep{Dominik:2013tma, Baibhav:2019gxm}. 
\acp{BBH} originating from dense star clusters predict local merger rates of $\mathcal{R}\sim 10\,\perGpcyr$ \citep{Sedda:2023qlx} and steeper values of $\kappa \sim 2$ \citep{Antonini:2020xnd}, albeit with large theoretical uncertainties that can account for a similar evolution to the isolated binary evolution predictions.
{\textbf{\boldmath Our observations suggest a steeper evolution, with $\bm{\kappa\sim 2}$--$\bm{4}$.}}
While this does not rule out either class of \ac{BBH} formation, it may be suggestive of (i) progenitor formation rates that peak at an earlier redshift as compared to the cosmic star formation rate density, e.g., due to a preference for low-metallicity progenitors, and/or (ii) shorter delay times, perhaps with a tail toward long delay times~\citep{Fishbach:2021mhp, Karathanasis:2022rtr,Fishbach:2023pqs, Turbang:2023tjk,Vijaykumar:2023bgs,Schiebelbein-Zwack:2024roj}.

Models of isolated binary evolution and dense star clusters often predict that the merger rate evolves differently across the mass spectrum \citep{vanSon:2021zpk, Mapelli:2021gyv, Ye:2024ypm};
however, analyses of the previous catalog have not found evidence for or against differential rate evolution \citep{Fishbach:2021yvy,Sadiq:2021fin, vanSon:2021zpk, Ray:2023upk, Heinzel:2024hva, Sadiq:2025aog}.
We discuss potential mass-redshift correlations in Section~\ref{subsubsec:redshift_mass_correlations}.

%% file: bbh_correlations.tex
\subsection{Population-level Correlations between Parameters}
\label{sec:Population-level correlations between parameters}

While much can be learned by studying the distributions of individual \ac{BBH} parameters, we can glean additional information on the population by considering how parameters are correlated with one another across systems.
In this subsection, we provide an overview of how parameters in the \ac{BBH} population appear to be broadly structured in various two-dimensional slices of parameter space, briefly discussing the astrophysical implications of our findings.

\subsubsection{Mass Ratio and Spin Correlations}
\label{subsubsec:mass_ratio_spin_correlations}
We begin by following up on the purported anti-correlation between \ac{BBH} mass ratio $q$ and effective inspiral spin $\chieff$.
\textbf{Although we find less support for the specific case of anti-correlation between $\bm{q}$ and $\bm{\chieff}$ relative to \gwtcthree, we find compelling evidence for some correlated structure in $\bm{(q,\chieff)}$.}
Namely, larger positive values of $\chieff$ appear to be favored as $q$ decreases, but it is unclear if this is accompanied by a preference for larger negative values of $\chieff$ as well.
The $(q,\chieff)$~\linearcorrelation~model imposes a linear functional dependence between \ac{BBH} mass ratio, and the mean and (natural log) width of the $\chieff$ distribution (see Appendix~\ref{appendix:LinearCorrelationModels}).
Fitting this model to \gwtcthree data suggests that a linear correlation coefficient of $\delta \mu_{\mathrm{eff}|q} < 0$ (i.e., the case of a negative $q$-dependence on the mean of the $\chieff$ distribution) with 98\% credibility \citep{KAGRA:2021duu}.
Updating this analysis to include data obtained over \ac{O4a} softens the evidence for an anti-correlation between $q$ and the mean of the $\chieff$ distribution, with a value of $\delta \mu_{\mathrm{eff}|q} < 0$ now inferred at $\qchieffLinearCorrelationModel[param][mu_chieff_1][percentile exclude zero]\%$ credibility.
However, we now see notable evidence for a linear increase in the log width of the $\chieff$ distribution as mass ratios become more unequal -- with $\delta \ln \sigma_{\mathrm{eff}|q} < 0$ at $\qchieffLinearCorrelationModel[param][ln_sigma_chieff_1][percentile exclude zero]\%$ credibility.

\begin{figure}
  \centering
  \includegraphics[width=0.98\columnwidth]{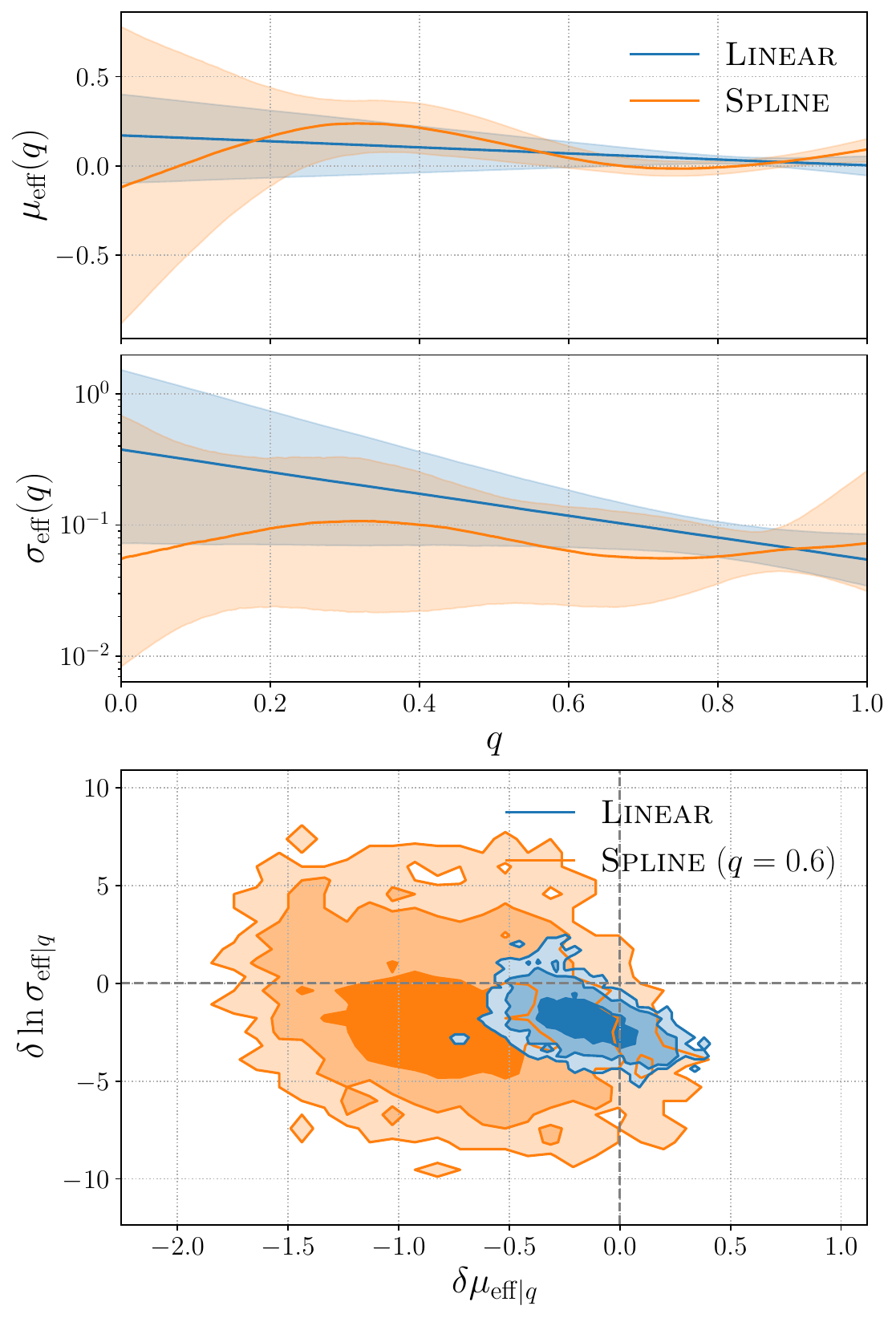}
  \caption{
  The inferred peak (top) and width (middle) of the $\chieff$ distribution as a function of mass ratio for the $(q,\chieff)$~\linearcorrelation~model (blue) and \splinecorrelation~model (orange).
  The shaded regions in these panels give the \confidenceLevel~credible intervals.
  The bottom panel gives the posterior distribution for the gradient of the $\chieff$ distribution's peak ($\delta \mu_{\mathrm{eff}|q}$) and natural log width ($\delta \ln \sigma_{\mathrm{eff}|q}$) dependent on mass ratio.
  From dark to light, the shaded regions represent the 50\%, 90\% and 99\% credible intervals.
  Again, blue gives the result of the \linearcorrelation~model, while orange shows the result of the \splinecorrelation~model sliced through $q=0.6$ (the approximate point at which the gradients are largest).
  It appears that mass ratio and $\chieff$ exhibit some kind of correlation, but the exact nature is unclear.
  }
  \label{fig:q_chieff_mu_sigma}
\end{figure}

We plot the mass-ratio dependent mean and width of the $\chieff$ distribution inferred using the \linearcorrelation~model in Figure~\ref{fig:q_chieff_mu_sigma}.
In the bottom panel of Figure~\ref{fig:q_chieff_mu_sigma}, we include the two-dimensional posterior distribution of $\delta \mu_{\mathrm{eff}|q}$ and $\delta \ln \sigma_{\mathrm{eff}|q}$.
Inspecting this plot, we see an anti-correlation in the posterior of the two hyperparameters, where larger negative values of $\delta \mu_{\mathrm{eff}|q}$ imply values of $\delta \ln \sigma_{\mathrm{eff}|q}$ closer to zero, and larger negative values of $\delta \ln \sigma_{\mathrm{eff}|q}$ imply values of $\delta \mu_{\mathrm{eff}|q}$ closer to zero.
The case in which both parameters are zero (no correlation between $q$ and $\chieff$ of any kind) appears to be ruled out at $> 99\%$ credibility.
In practice, this implies that the \linearcorrelation~model finds support for larger positive values of $\chieff$ at more unequal mass ratios, but cannot yet conclude whether these are accompanied by larger negative values of $\chieff$ as well.

Next, we probe for more intricate correlations between $q$ and $\chieff$ using the \splinecorrelation~model.
Similar to the \linearcorrelation~model, the \splinecorrelation~model allows for the mean and width of the $\chieff$ distribution to evolve with mass ratio.
However, these mass ratio dependences, rather than being linear, are modeled flexibly with cubic splines \citep[see Appendix~\ref{appendix:LinearCorrelationModels} and][]{Heinzel:2023hlb}.
We plot the $q$-dependent means and widths inferred from the \splinecorrelation~model alongside those from the \linearcorrelation~model in Figure~\ref{fig:q_chieff_mu_sigma}.
Despite fluctuations emerging in the \splinecorrelation~model, the two models are, within \confidenceLevel~credible bounds, consistent.
The preference for broadening in the $\chieff$ distribution as $q$ decreases found with the \linearcorrelation~model, does not clearly appear in the \splinecorrelation~model.
In the bottom panel of Figure~\ref{fig:q_chieff_mu_sigma}, we also plot the inferred gradient of the $\chieff$ distribution's mean and (natural log) width relative to mass ratio at $q = 0.6$ (roughly the value at which covariance appears most pronounced).
Here, we see a similar structure to that of the \linearcorrelation~model's $(\delta \mu_{\mathrm{eff}|q}, \delta \ln \sigma_{\mathrm{eff}|q})$ posterior, albeit with much more uncertainty.
As a final observation from the \splinecorrelation~model in Figure~\ref{fig:q_chieff_mu_sigma}, we see that the inferred $\chieff$ distribution at low mass ratios ($q \lesssim 0.2$) grows in uncertainty, roughly recovering the prior.
This implies that any correlation inferred by the \linearcorrelation~model is likely driven by observations with mass ratios $q \gtrsim 0.2$.
As such, the inferences from this model should be interpreted with caution as $q \rightarrow 0$, where the trend is effectively being extrapolated from the more populated $q \gtrsim 0.2$ region due to a lack of model flexibility.

\begin{figure}
  \centering
  \includegraphics[width=0.98\columnwidth]{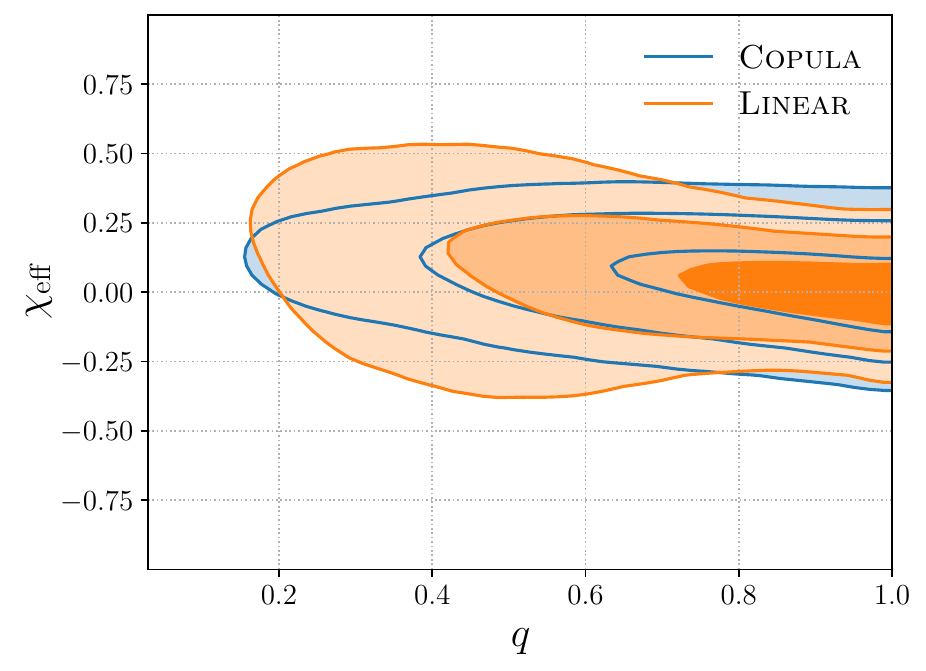}
  \caption{
  Mass ratio and $\chieff$ \acp{PPD} for the \copulacorrelation~model (blue) and \linearcorrelation~model (orange).
  The contours, from dark to light, mark 50\%, 90\%, and 99\% of the volume.
  We see a subtle preference for an anti-correlation, although this feature is far less prevalent than it appeared in \gwtcthree.
  We also see evidence for a broadening in the distribution as mass ratios become more unequal in the \linearcorrelation~model.
  The \copulacorrelation~model is not flexible in a way that it can capture a broadening.
  }
  \label{fig:q_chieff_ppd}
\end{figure}

We now move to the \copulacorrelation~model, which allows for a correlation between $q$ and $\chieff$ with a variable strength $\kappa_{q, \mathrm{eff}}$ (see Appendix~\ref{appendix:copulaCorrelationModels}).
This framework has the added advantage that the level of correlation is decoupled from the shape of the marginal distribution \citep[e.g.,][]{Adamcewicz:2022hce}, with less flexibility for covariant structure relative to other correlated population models.
Fitting the \copulacorrelation~model to \gwtcfour suggests that $q$ and $\chieff$ are anti-correlated ($\kappa_{q, \mathrm{eff}} < 0$) with $\qchieffCopulaCorrelationModel[param][kappa][correlated percentile]\%$ credibility.
Specifically, we infer $\kappa_{q, \mathrm{eff}} = \CIPlusMinus{\qchieffCopulaCorrelationModel[param][kappa]}$.
\cite{Adamcewicz:2023mov} analyzed \gwtcthree data with a copula model to find that $(q, \chieff)$ are anti-correlated with $>99\%$ credibility, suggesting greater evidence for an anti-correlation than is measured here.
However, these results are not directly comparable to those presented in this work, due to different modeling assumptions and different convergence criteria in the population likelihood.
Qualitatively, we see that the \copulacorrelation~model and \linearcorrelation~model exhibit a subtle anti-correlation in $(q, \chieff)$, as seen in the respective two-dimensional \acp{PPD} in Figure~\ref{fig:q_chieff_ppd}.

There are a number of potential astrophysical implications if the anti-correlation in $(q,\chieff)$ is real.
If isolated binaries make up a substantial fraction of the population and undergo tidal spin up, stable mass transfer and the resulting mass ratio reversal of systems \citep{Broekgaarden:2022nst, Zevin:2022wrw, Olejak:2024qxr} may produce an such an anti-correlation in the population.
This feature could also be explained by binaries undergoing a common-envelope phase provided common envelope efficiencies are sufficiently high \citep{Bavera:2020uch}.

Covariance in $(q,\chieff)$ could also be a result of hierarchical mergers contributing to a considerable fraction of the \ac{BBH} merger rate \citep{Antonini:2024het, Antonini:2025zzw}.
Hierarchical mergers should exhibit mass ratios that are more unequal and spin magnitudes that are larger than isolated, or first-generation dynamical mergers.
If hierarchical mergers occur in typical dynamical environments, the spins of the \acp{BH} will be isotropically distributed, thus producing a broadening in the $\chieff$ distribution as mass ratios become unequal.
Meanwhile, \acp{BH} in \ac{AGN} may have spins preferentially aligned with one another, meaning hierarchical mergers in these environments can produce a correlation between mass ratios and spins that is asymmetrical about $\chieff = 0$ \citep{McKernan:2021nwk, Santini:2023ukl}.
Stronger evidence for unequal mass binaries preferring positive values of $\chieff$ could then indicate that binaries merging in \ac{AGN} are predominantly coaligned with the rotation of the \ac{AGN} disk \citep{Santini:2023ukl}.

\subsubsection{Mass and Spin Correlations}
\label{subsubsec:mass_spin_correlations}
\begin{figure}
	\centering
	\includegraphics[width=0.98\columnwidth]{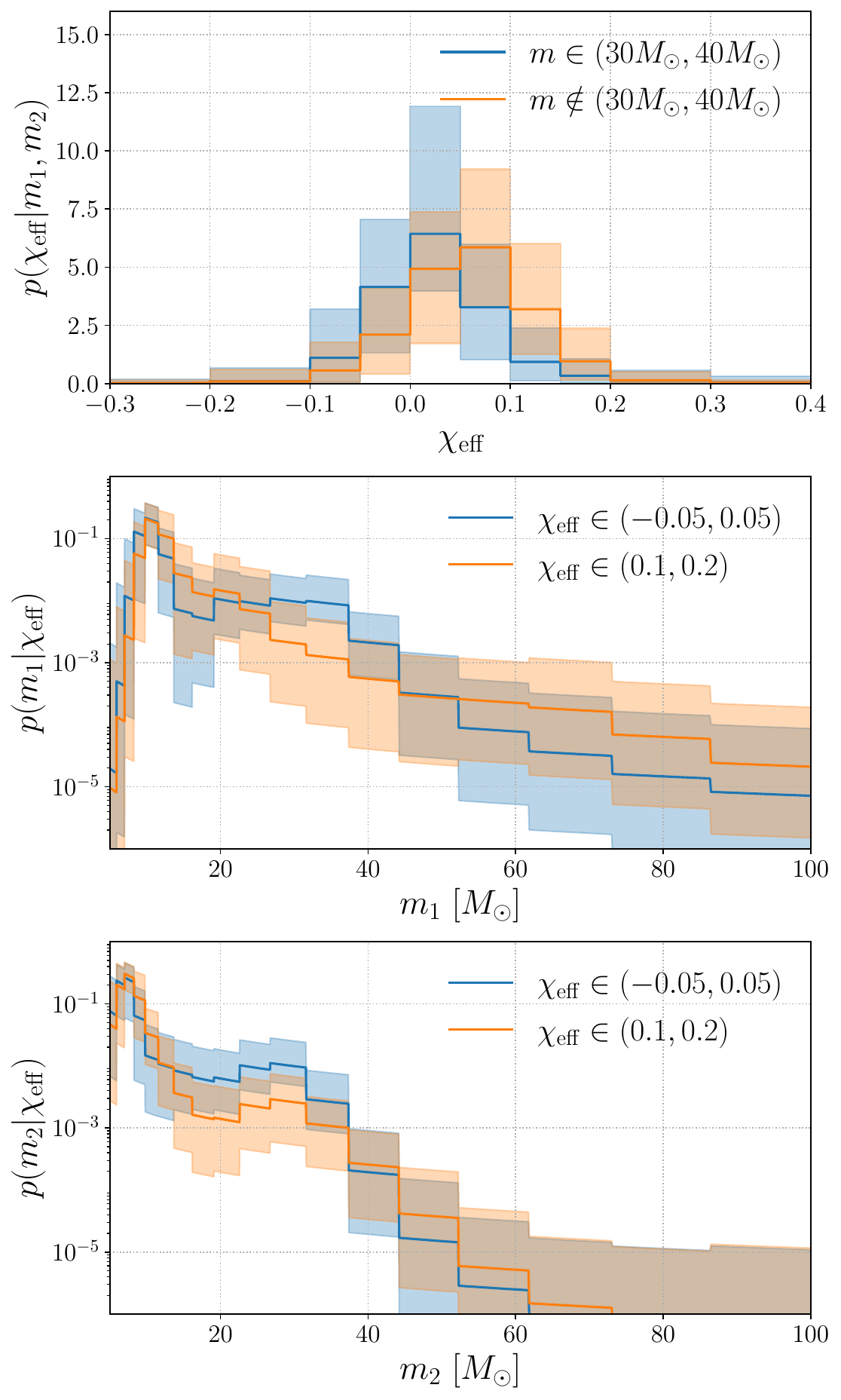}
	\caption{
  \textit{Top:} inferred distributions of effective inspiral spin $\chieff$ in the correlated mass--spin \ac{BGP}~analysis.
  The solid lines give the median, while the shaded regions indicate the \confidenceLevel~credible intervals.
  In blue, we have the $\chieff$ distribution of \ac{BBH} systems with at least one mass inside the $30{-}40 \,\Msun$ peak range.
	In orange, we have the $\chieff$ distribution of \ac{BBH} systems masses outside the $30{-}40 \,\Msun$ peak range.
  While the two distributions are consistent within \confidenceLevel~credible intervals, systems outside the peak range appear to favour a distribution of $\chieff$ skewed toward more positive values.
	Meanwhile, systems inside the peak range appear to favour a more symmetrical distribution about $\chieff = 0$.
  \textit{Middle:} inferred distributions of primary mass from the correlated mass--spin \ac{BGP}~analysis.
  \textit{Bottom:} inferred distributions of secondary mass from the correlated mass--spin \ac{BGP}~analysis.
	In both mass plots, blue shows the distribution for systems with small spins, $\chieff$ in the range $(-0.05,0.05)$, while orange shows the distribution for systems with larger spins, $\chieff$ in the range $(0.1,0.2)$.
	Note the preference for a higher proportion of mergers with masses $\approx 30{-}40 \,\Msun$ for low spin systems (blue).
	}
	\label{fig:mass_spin_correlated_bgp_combined}
\end{figure}

\textbf{We find model-dependent evidence for correlations between $\bm{m_1}$ and $\bm{\chieff}$.}
Motivated by the feature in the mass distribution around the ${\sim} 30{-}40 \,\Msun$ range, we use a \ac{BGP}~analysis allowing for correlations in $m_1$ and $\chieff$ (see Appendix~\ref{appendix:BGPModel}) to see if this feature in the distribution of masses is accompanied by a deviation in the \ac{BBH} spin distribution.
Using \gwtcfour, these studies find weak evidence that \ac{BBH} systems with at least one mass in the ${\sim} 30{-}40 \,\Msun$ peak tend to have spins that are symmetrically distributed about $\chieff = 0$, while binaries outside of this mass range have spins that are skewed toward positive (aligned) values of $\chieff$.
Quantitatively, this \ac{BGP}~analysis infers that for every merger in the ${\sim} 30{-}40 \,\Msun$ peak with a negative value of $\chieff$, there are $\CIPlusMinus{\MassSpinCorrelatedBGPModel[param][positive-to-negative chi_eff ratio for m in 30-40]}$ with a positive value of $\chieff$.
Meanwhile, for every event outside of this mass range with a negative value of $\chieff$, there are $\CIPlusMinus{\MassSpinCorrelatedBGPModel[param][positive-to-negative chi_eff ratio for m outside 30-40]}$ with a positive value of $\chieff$.
We illustrate this result in the top panel of Figure~\ref{fig:mass_spin_correlated_bgp_combined}.
Furthermore, in the \ac{BGP}~analysis the inferred distribution of masses for \ac{BBH} systems with effective inspiral spins ($|\chieff| \lesssim 0.1$) exhibits a preference for a larger proportion of mergers with $m_1 \sim 30{-}40 \,\Msun$, compared to systems with $\chieff \gtrsim 0.1$.
These distributions, however, remain consistent within \confidenceLevel~credible intervals.
This is illustrated in the bottom two panels of Figure~\ref{fig:mass_spin_correlated_bgp_combined}.
These features were also recovered with same analyses applied to \gwtcthree data, albeit with less certainty \citep{Ray:2024hos}.
However, more recent \gwtcthree analyses find conflicting evidence, suggesting that the $\chieff$ distribution of $\approx 30 \,\Msun$ \acp{BH} is positively skewed \citep{Sadiq:2025vly, Roy:2025ktr}.

The \textsc{Isolated Peak}~model \citep[see also][]{Godfrey:2023oxb} fits the data to a model containing multiple mass subpopulations with independent spin distributions.
One subpopulation consists of a single peak in the mass distribution (which is inferred to center on ${\sim} 10 \,\Msun$), while other subpopulations are flexibly fit with the \BSpline~method.
Applying these analyses to \gwtcfour suggests that \acp{BH} within the ${\sim} 10 \,\Msun$ peak have a spin magnitude distribution consistent (within 90\% credibile intervals) with the rest of the \ac{BBH} population.
Meanwhile, although there is some overlap in the 90\% credibile regions of the cosine tilt distributions for both subpopulations, the distribution of tilts for \acp{BH} in the ${\sim} 10 \,\Msun$ peak favors alignment and is inconsistent with isotropy.
The distribution of cosine tilts for \acp{BH} outside this mass peak appears more symmetrical about $\cos \theta = 0$ and does not rule out isotropy.
Roughly speaking, this corresponds to a distribution of $\chieff$ that may be symmetrical about zero for the \ac{BBH} population outside of the ${\sim} 10 \,\Msun$ peak.
Inside this mass peak, however, the implied $\chieff$ distribution skews toward positive values.

We also model a correlation between primary mass and $\chieff$ using a copula model (see Appendix~\ref{appendix:copulaCorrelationModels}).
Fitting this \copulacorrelation~model with \gwtcfour data, we infer a correlation of $\kappa_{m_1, \mathrm{eff}} = \CIPlusMinus{\mchieffCopulaCorrelationModel[param][kappa]}$ between $m_1$ and $\chieff$.
Hence, the \copulacorrelation~analysis finds no evidence for a single, smooth, population-wide correlation between primary mass and $\chieff$.

Broadly, our analyses of the joint mass and spin distribution recover similar features to recent works that probe for features in \gwtcthree.
First, analyses of \gwtcthree do not find evidence for a smoothly correlated distribution in primary mass and $\chieff$ \citep{Safarzadeh:2020mlb, Biscoveanu:2022qac, Fishbach:2022lzq, Heinzel:2023hlb, Heinzel:2024hva, Antonini:2024het}, as is the case in this work.
Meanwhile, a number of analyses find evidence for separate subpopulations in mass and spin consistent with predictions of dynamical and field mergers.
More specifically, these works tend to find a subpopulation with lower masses and smaller, mostly aligned spins, and a second subpopulation of heavier binaries with larger, isotropically distributed spins, where the subpopulations transition around ${\sim} 40 \,\Msun$ \citep{Wang:2022gnx, Mould:2022ccw, Godfrey:2023oxb, Li:2023yyt, Antonini:2024het, Pierra:2024fbl, Guo:2024wwv, Li:2025rhu, Sadiq:2025vly}.
While the analyses presented in this work do not recover this exact feature (nor do they probe directly for it), the \textsc{Isolated Peak}~model's preference for aligned spins at ${\sim} 10 \,\Msun$ and isotropically distributed spins at higher masses may be related.

\subsubsection{Redshift and Spin Correlations}
\label{subsubsec:redshift_spin_correlations}
\begin{figure}
  \centering
  \includegraphics[width=0.98\columnwidth]{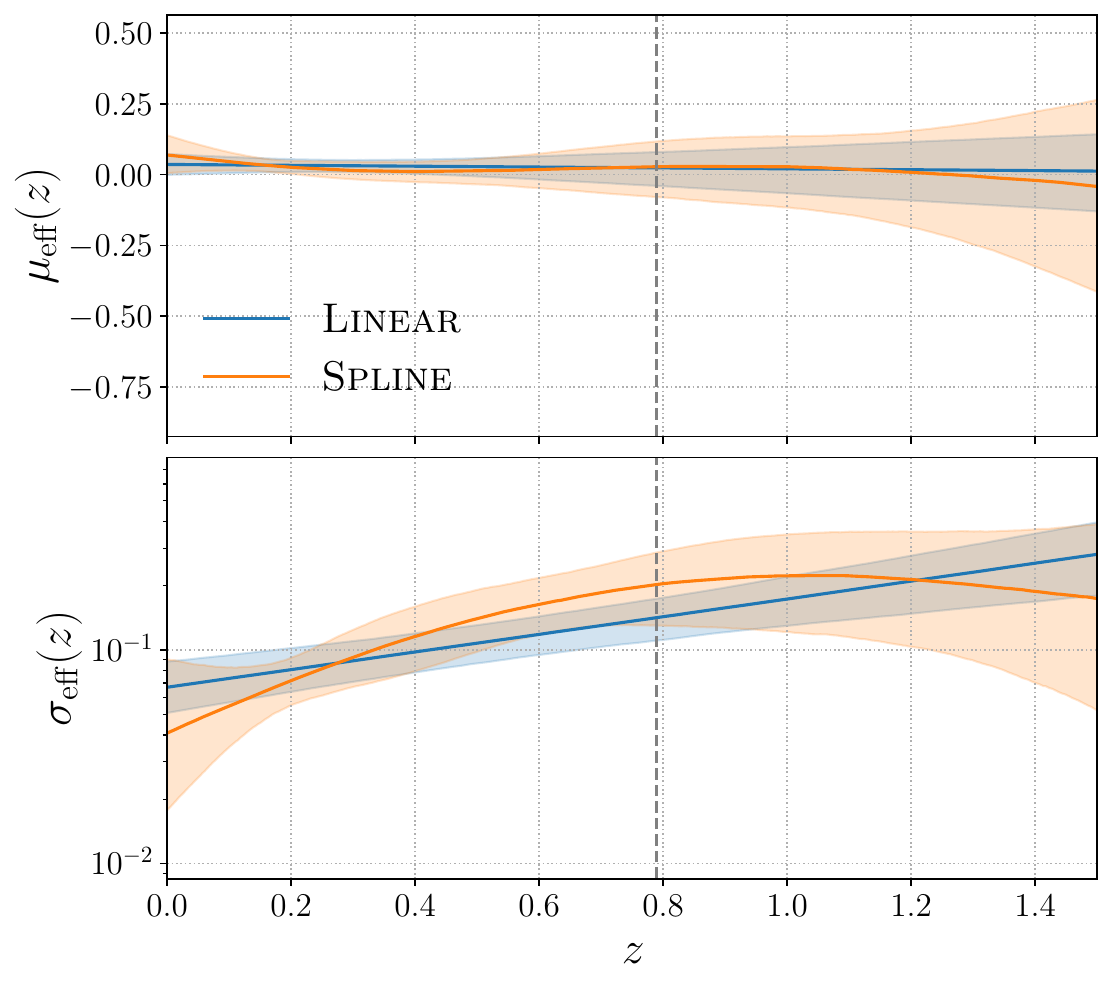}
  \caption{
  The inferred mean (top) and width (bottom) of the $\chieff$ distribution as a function of redshift for the \linearcorrelation~model (blue) and the \splinecorrelation~model (orange).
  The shaded regions give the \confidenceLevel~credible intervals.
  The gray dashed line indicates the redshift under which 90\% of the cumulative posterior probability lies across all events in \gwtcfour.
  Therefore, inferences above this redshift are dominated by the prior or population model.
  Note the logarithmic scale on the width (bottom panel).
  In either model, we find evidence that the width of the effective inspiral spin distribution increases with redshift up to $z \sim 1$.
  }
  \label{fig:z_chieff_mu_sigma}
\end{figure}

\textbf{We find evidence that the $\bm{\chieff}$ distribution broadens as redshift increases up to $\bm{z \sim 1}$}.
We again look for correlations between redshift and $\chieff$ by modeling the mean and width of the $\chieff$ distribution with a linear dependence on redshift (see Appendix~\ref{appendix:LinearCorrelationModels}).
\cite{Biscoveanu:2022qac} employed a \linearcorrelation~model for $(z,\chieff)$ to analyze \gwtcthree data, finding no evidence for redshift dependence in the mean of the $\chieff$ distribution, but suggesting that the $\chieff$ distribution broadens with increasing redshift ($\delta \ln \sigma_{\mathrm{eff}|z} > 0$ at 99\% credibility).
Updating this analysis to include \ac{O4a} data, we find more evidence for a broadening in the $\chieff$ distribution with redshift, with $\delta \ln \sigma_{\mathrm{eff}|z} > 0$ now inferred at $> \zchieffLinearCorrelationModel[param][ln_sigma_chieff_1][percentile exclude zero]\%$ credibility.
The mean and width of the $\chieff$ distribution as functions of redshift are shown in Figure~\ref{fig:z_chieff_mu_sigma}.

The \splinecorrelation~model, which models the redshift dependence on the mean and width of the $\chieff$ distribution flexibly with cubic splines (see Appendix~\ref{appendix:SplineCorrelationModels}), is broadly consistent with the \linearcorrelation~model.
The exception is that the \splinecorrelation~model tends to recover the prior beyond a redshift of $z \gtrsim 1$.
This likely indicates that the \linearcorrelation~model is fitting a trend at low redshifts, then extending this trend to redshifts $z \gtrsim 1$ due to a lack of flexibility.
To further illustrate this point, the gray dashed line in Figure~\ref{fig:z_chieff_mu_sigma} indicates the redshift under which 90\% of the catalog's cumulative posterior support is contained.
Roughly speaking, this means our inferences above this redshift are informed by only ${\sim} 10\%$ of the data.
Therefore, while we find evidence that the $\chieff$ distribution broadens as binaries approach $z \sim 1$, we are unable to determine if this trend continues at higher redshifts.

We also model a correlation between redshift and $\chieff$ using a copula model, with variable correlation $\kappa_{z, \mathrm{eff}}$ (see Appendix~\ref{appendix:copulaCorrelationModels}).
In contrast to the above analyses, the \copulacorrelation~model finds evidence for a positive correlation in $(z,\chieff)$, inferring a value of $\kappa_{z, \mathrm{eff}} = \CIPlusMinus{\zchieffCopulaCorrelationModel[param][kappa]}$ (or $\kappa_{z, \mathrm{eff}} > 0$ with $\zchieffCopulaCorrelationModel[param][kappa][correlated percentile]\%$ credibility).
While the \copulacorrelation~model lacks the flexibility to fit a broadening directly, the long (albeit shallow) posterior tails reaching into both large-negative and large-positive values for $\kappa_{z, \mathrm{eff}}$ may relate to the broadening in the $\chieff$ distribution recovered by the \linearcorrelation~and~\splinecorrelation~models (see the subtle mode at negative values of $\kappa_{z, \mathrm{eff}}$ in Appendix~\ref{appendix:supplementary_correlations}).
If the $z$-dependent broadening and constant mean of the $\chieff$ distribution from the \linearcorrelation~and~\splinecorrelation~models is to be believed, it is unclear why the posterior on $\kappa_{z, \mathrm{eff}}$ skews toward positive values, rather than being symmetric about zero.

It is unclear what this broadening in the $\chieff$ distribution at greater redshifts might imply astrophysically.
Provided \ac{BH} progenitors experience significant spin up due to tidal torques, smaller orbital separations and periods will result in more efficient spin up of \ac{BBH} components \citep{Zaldarriaga:2017qkw, Mapelli:2018uds, Bavera:2020uch, Bavera:2022mef, Fuller:2022ysb}.
Of course, smaller orbital separations correspond to shorter delay-times to merger.
This means that these systems may contribute more to the merger rate in the earlier Universe (higher $z$), where systems with wider orbits have not had time to merge.
Furthermore, given that higher metallicity systems are expected to evolve to longer orbital periods due to experiencing more mass loss during core helium burning \citep{Qin:2018vaa, Fuller:2022ysb}, one might predict that larger spin magnitudes should be observed in regions of lower metallicity and thus also at higher redshifts.
However, the correlation in the \linearcorrelation~and~\splinecorrelation~models is between redshift and the width of the $\chieff$ distribution rather than the mean.
This includes an increasing number of systems with both large-positive and large-negative values of $\chieff$.
An increase in systems with large-negative $\chieff$ may be somewhat difficult to square with the above hypothesis, given that tidal spin up is only relevant in isolated systems, and that large supernovae kicks are required to misalign the \ac{BH} spins so significantly \citep{Kalogera:1999tq,  Wysocki:2017isg, Gerosa:2018wbw, Callister:2020vyz, Steinle:2020xej, Stevenson:2022hmi, Tauris:2022ggv, Baibhav:2024rkn}.
On the other hand, if the \copulacorrelation~results are to be taken at face-value, the preference for a positive correlation in $(z,\chieff)$ fits more neatly with the aforementioned hypothesis.

Another possibility, following discussion from Section~\ref{subsec:bbh_spins}, is that multiple separate subpopulations of \ac{BBH} mergers are being observed, where the more dominant (in terms of redshift-dependent merger rate) subpopulation changes at some nearby redshift.
Hierarchical mergers becoming more dominant at higher redshifts for example could potentially explain such a broadening in the effective spin distribution.
This interpretation would also imply some level of correlation between mass and redshift, which we do not find evidence for below in Section~\ref{subsubsec:redshift_mass_correlations}.

As a final caveat, using \gwtcthree data, \cite{Biscoveanu:2022qac} fit the \ac{BBH} population to a model in which the $\chieff$ distribution is linearly dependent on both primary mass and redshift.
In doing so, the authors find degeneracies between the $(m_1,\chieff)$ and $(z,\chieff)$ correlations.
While we do not explore correlations beyond two dimensions in this work, we acknowledge that the assumption of independence between other pairs of parameters may have notable effects on our inferences.

\subsubsection{Redshift and Mass Correlations}
\label{subsubsec:redshift_mass_correlations}
Finally, we consider potential correlations between BBH mass and redshift.
\textbf{We do not find evidence for evolution in the mass distribution with redshift.}
We emphasize that these inferences are constrained to the nearby Universe, with relatively little data beyond $z \sim 1$ (19 out of 153 \ac{BBH} events having posteriors consistent with $z>1$ to \confidenceLevel~credibility).
We model correlations between primary mass and redshift using a copula with correlation $\kappa_{m_1, z}$ (see Appendix~\ref{appendix:copulaCorrelationModels}).
Using this \copulacorrelation~model, we infer a correlation of $\kappa_{m_1, z} = \CIPlusMinus{\mzCopulaCorrelationModel[param][kappa]}$.
We also search for correlations between mass and redshift using a \ac{BGP}~analysis that allows for covariance in $m_1$ and $z$ (see Appendix~\ref{appendix:BGPModel}).
Similarly, this analysis finds that the mass distribution does not show a distinguishable evolution with redshift (see Appendix~\ref{appendix:supplementary_correlations}).
This conclusion is mostly in line with studies using \gwtcthree, which are also unable to infer any correlation between mass and redshift with confidence \citep{Fishbach:2021yvy, Sadiq:2021fin, vanSon:2021zpk, Karathanasis:2022rtr, Ray:2023upk, Heinzel:2023hlb, Heinzel:2024hva, Lalleman:2025xcs, Sadiq:2025aog}.
This is with the exception of \cite{Rinaldi:2023bbd} finding a positive correlation between \ac{BBH} primary mass and redshift, although a novel method is used to account for selection effects.

%% file: conclusion.tex
\section{Conclusion}\label{sec:conclusion}
In this paper, we present population-level analyses of events included in the fourth Gravitational-Wave Transient Catalog \gwtcfour. This dataset more than doubles the number of events analyzed compared to the previous catalog. Our main findings are:

\begin{enumerate}

	\item Features identified in the third catalog \gwtcthree persist, including clear overabudances in the mass distribution at $1$\textendash$2\,\Msun$ and around $ 10\, \Msun$, and a feature near $35\,\Msun$. There is no conclusive evidence to either support or refute a suppression of the merger rate between these features.

	\item We estimate the merger rates at redshift $z=0$ to be $\FullMassSpectrumMerged[R_BNS][5th_percentile]\textendash\FullMassSpectrumMerged[R_BNS][95th_percentile]\,\perGpcyr$ for \aclp{BNS}, $\FullMassSpectrumMerged[R_NSBH][5th_percentile]\textendash \FullMassSpectrumMerged[R_NSBH][95th_percentile]\,\perGpcyr$ for \aclp{NSBH}, $\FullMassSpectrumMerged[R_BBH][5th_percentile] \textendash\FullMassSpectrumMerged[R_BBH][95th_percentile]\,\perGpcyr$ for \aclp{BBH}.

	\item The \acl{BBH} primary mass distribution is well described by a broken power law, shallow at low masses and steep at high masses, modulated by overdensities near $10\, \Msun$ and $35\,\Msun$. A \nonparametric finds evidence of overdensity around $ 20\, \Msun$.

	\item \Aclp{BH} in the $35\,\Msun$ feature tend to pair with companions of similar mass more frequently than lower-mass \aclp{BH} do. 

	\item The distribution of effective inspiral spins is asymmetric about $\chieff=0$ and is skewed toward positive $\chieff$ values. Spin magnitudes span a broad range from $0$ to $1$, although ${\sim} 90\%$ of BHs have $\chi < \MagTruncnormIidTiltIsotropicTruncnormNid[ppd][a][90th percentile]$. 

	\item The redshift evolution of the \acl{BBH} merger rate $\mathcal{R}(z)$ remains consistent with the cosmic star formation rate density. A merger rate uniform in comoving volume and source-frame time is ruled out. 
	
	\item We find that \aclp{BH} outside the $30$\textendash$40\,\Msun$ range prefer an asymmetric effective inspiral spin distribution skewed toward higher values, while those within this range show no such preference. Compared to \gwtcthree, we observe stronger evidence that the width of the effective spin distribution increases with redshift and weaker evidence for an anti-correlation between mass ratio and effective spin. No redshift evolution is observed in the mass distribution.

	\item The \acl{NS} mass distribution remain consistent with previous results, favoring a broad distribution of \acl{NS} masses between $1 \, \Msun$ and $3 \, \Msun$.

\end{enumerate}

The analysis of this expanded dataset refines the statistical significance of previously reported trends and reveals new features in the population of compact binary mergers. Several of these findings, such as the persistent  ${\sim} 10\, \Msun$ feature, pose challenges to current models of supernova physics, binary mass transfer, and dynamical formation in dense stellar environments. Future data from the remainder of the fourth observing run O4 will further enhance our understanding and may uncover additional structure in the population.

%% file: gwtc-common-files__LVKack.tex
This material is based upon work supported by NSF's LIGO Laboratory, which is a
major facility fully funded by the National Science Foundation.
The authors also gratefully acknowledge the support of
the Science and Technology Facilities Council (STFC) of the
United Kingdom, the Max-Planck-Society (MPS), and the State of
Niedersachsen/Germany for support of the construction of Advanced LIGO 
and construction and operation of the GEO\,600 detector. 
Additional support for Advanced LIGO was provided by the Australian Research Council.
The authors gratefully acknowledge the Italian Istituto Nazionale di Fisica Nucleare (INFN),  
the French Centre National de la Recherche Scientifique (CNRS) and
the Netherlands Organization for Scientific Research (NWO)
for the construction and operation of the Virgo detector
and the creation and support  of the EGO consortium. 
The authors also gratefully acknowledge research support from these agencies as well as by 
the Council of Scientific and Industrial Research of India, 
the Department of Science and Technology, India,
the Science \& Engineering Research Board (SERB), India,
the Ministry of Human Resource Development, India,
the Spanish Agencia Estatal de Investigaci\'on (AEI),
the Spanish Ministerio de Ciencia, Innovaci\'on y Universidades,
the European Union NextGenerationEU/PRTR (PRTR-C17.I1),
the ICSC - CentroNazionale di Ricerca in High Performance Computing, Big Data
and Quantum Computing, funded by the European Union NextGenerationEU,
the Comunitat Auton\`oma de les Illes Balears through the Conselleria d'Educaci\'o i Universitats,
the Conselleria d'Innovaci\'o, Universitats, Ci\`encia i Societat Digital de la Generalitat Valenciana and
the CERCA Programme Generalitat de Catalunya, Spain,
the Polish National Agency for Academic Exchange,
the National Science Centre of Poland and the European Union - European Regional
Development Fund;
the Foundation for Polish Science (FNP),
the Polish Ministry of Science and Higher Education,
the Swiss National Science Foundation (SNSF),
the Russian Science Foundation,
the European Commission,
the European Social Funds (ESF),
the European Regional Development Funds (ERDF),
the Royal Society, 
the Scottish Funding Council, 
the Scottish Universities Physics Alliance, 
the Hungarian Scientific Research Fund (OTKA),
the French Lyon Institute of Origins (LIO),
the Belgian Fonds de la Recherche Scientifique (FRS-FNRS), 
Actions de Recherche Concert\'ees (ARC) and
Fonds Wetenschappelijk Onderzoek - Vlaanderen (FWO), Belgium,
the Paris \^{I}le-de-France Region, 
the National Research, Development and Innovation Office of Hungary (NKFIH), 
the National Research Foundation of Korea,
the Natural Sciences and Engineering Research Council of Canada (NSERC),
the Canadian Foundation for Innovation (CFI),
the Brazilian Ministry of Science, Technology, and Innovations,
the International Center for Theoretical Physics South American Institute for Fundamental Research (ICTP-SAIFR), 
the Research Grants Council of Hong Kong,
the National Natural Science Foundation of China (NSFC),
the Israel Science Foundation (ISF),
the US-Israel Binational Science Fund (BSF),
the Leverhulme Trust, 
the Research Corporation,
the National Science and Technology Council (NSTC), Taiwan,
the United States Department of Energy,
and
the Kavli Foundation.
The authors gratefully acknowledge the support of the NSF, STFC, INFN and CNRS for provision of computational resources.

This work was supported by MEXT,
the JSPS Leading-edge Research Infrastructure Program,
JSPS Grant-in-Aid for Specially Promoted Research 26000005,
JSPS Grant-in-Aid for Scientific Research on Innovative Areas 2402: 24103006,
24103005, and 2905: JP17H06358, JP17H06361 and JP17H06364,
JSPS Core-to-Core Program A.\ Advanced Research Networks,
JSPS Grants-in-Aid for Scientific Research (S) 17H06133 and 20H05639,
JSPS Grant-in-Aid for Transformative Research Areas (A) 20A203: JP20H05854,
the joint research program of the Institute for Cosmic Ray Research,
University of Tokyo,
the National Research Foundation (NRF),
the Computing Infrastructure Project of the Global Science experimental Data hub
Center (GSDC) at KISTI,
the Korea Astronomy and Space Science Institute (KASI),
the Ministry of Science and ICT (MSIT) in Korea,
Academia Sinica (AS),
the AS Grid Center (ASGC) and the National Science and Technology Council (NSTC)
in Taiwan under grants including the Science Vanguard Research Program,
the Advanced Technology Center (ATC) of NAOJ,
and the Mechanical Engineering Center of KEK.

Additional acknowledgements for support of individual authors may be found in the following document: \\
\url{https://dcc.ligo.org/LIGO-M2300033/public}.
For the purpose of open access, the authors have applied a Creative Commons Attribution (CC BY)
license to any Author Accepted Manuscript version arising.
We request that citations to this article use 'A. G. Abac {\it et al.} (LIGO-Virgo-KAGRA Collaboration), ...' or similar phrasing, depending on journal convention.

%% file: appendix.tex
\section{Hierarchical Inference Details}\label{appendix:technical_details}
\input{appendix_hier_inference.tex}

\section{Summary of models used in the strongly modeled approach}\label{appendix:parametric_models_summary}
\input{appendix_parametric_models.tex}

\section{Summary of models used in the weakly modeled approach}\label{appendix:non_parametric_models_summary}
\input{appendix_non_parametric_models.tex}

\section{Result Validation Studies}\label{appendix:model_comparison}
\input{appendix_validation_studies.tex}

%% file: appendix_hier_inference.tex
\subsection{Likelihood Estimation}
\label{sec:likelihood-estimator-variance}

The analytic integrals in Equation~\eqref{eq:hierarchical_likelihood} are not tractable, and so we estimate the integrals using Monte Carlo estimation \citep{Tiwari:2017ndi, Farr:2019rap, Essick:2022ojx, Talbot:2023pex}. 
For example, the estimator for the likelihood $\hat{\mathcal{L}}(d_i|\PEhyperparam)$ is
\begin{equation}
    \hat{\mathcal{L}}(d_i|\PEhyperparam) \propto \frac{1}{\Npe} \sum_{j=1}^{\Npe} \frac{\pi(\PEparameter_{ij}|\PEhyperparam)}{p(\PEparameter_{ij})},
    \label{eq:single_event_estimators}
\end{equation}
where $\{\PEparameter_{ij}\}_{j=1}^{\Npe}$ are a collection of $\Npe$ samples from the posterior on the \ac{GW} parameters of the $i{\rm th}$ event $d_i$. 
We divide out by the parameter estimation prior $p(\PEparameter)$, and so
the Monte Carlo estimator in Equation~\eqref{eq:single_event_estimators} converges to the $i{\rm th}$ integral inside the product of Equation~\eqref{eq:hierarchical_likelihood} in the limit of $\Npe \to \infty$.

Similarly, the estimator for the selection efficiency $\hat{\xi}$ is \citep{Tiwari:2017ndi, Farr:2019rap, Essick:2022ojx}
\begin{equation}
    \hat{\xi}(\PEhyperparam) \propto \frac{1}{\Ndraw} \sum_{j=1}^{\Nfound} \frac{\pi(\PEparameter_{j}|\PEhyperparam)}{\pi(\PEparameter_{j}|\PEhyperparam_{\rm draw})},
    \label{eq:selection_estimator}
\end{equation}
where $\Ndraw$ events with parameters $\PEparameter_i$ are injected into representative noise from the detectors. 
The search pipelines then search these synthetically generated data and recover some subset $\Nfound$ of the events with the detection statistic exceeding some threshold.
In the limit of $\Ndraw \to \infty$, this approaches the true integral in Equation~\eqref{eq:selection_efficiency}.

Because we only have a finite number of samples from each event and finite $\Ndraw$, we must be careful to account for the intrinsic variance in the estimation of the likelihood. 
To be sure our Monte Carlo estimators for the likelihood are trustworthy, we study the impact of Monte Carlo uncertainty in every inference we perform. 
Specifically, we compute the variance in the log-likelihood estimator, which varies across parameter space due to our resampling techniques in Equation~\eqref{eq:single_event_estimators} and Equation~\eqref{eq:selection_estimator}.
Propagating the uncertainty in the log-likelihood along independent degrees of freedom, the variance in the log-likelihood estimator $\sigma^2_{\ln \hat{\mathcal{L}}}$ in combining Equation~\eqref{eq:hierarchical_likelihood}, Equation~\eqref{eq:selection_efficiency}, and Equation~\eqref{eq:single_event_estimators} can be estimated as~(e.g., \citealt{Essick:2022ojx})
\begin{equation}
	\sigma^2_{\ln \hat{\mathcal{L}}}(\PEhyperparam) = \sum_{i=1}^{N_{\rm det}}\frac{\sigma^2_{\hat{\mathcal{L}}_i}(\PEhyperparam)}{\hat{\mathcal{L}}_i(\PEhyperparam)^2} + N_{\rm det}^2 \sigma^2_{\xi}(\PEhyperparam) ,
	\label{eq:variance_likelihood}
\end{equation}
where
\begin{equation}
	\sigma^2_{\hat{\mathcal{L}}_i}(\PEhyperparam) = \frac{1}{N_{\rm PE}}\left[\frac{1}{N_{\rm PE} -1 }\sum_{j=1}^{N_{\rm PE}}\left(\frac{\pi(\PEparameter_{ij}|\PEhyperparam)}{p(\PEparameter_{ij})}\right)^2 - \hat{\mathcal{L}}_i(\PEhyperparam)^2 \right]
\end{equation}
is the Monte Carlo variance in the single event Monte Carlo integrals of Equation~\eqref{eq:single_event_estimators} and 
\begin{equation}
	\sigma^2_{\xi}(\PEhyperparam) = \frac{1}{N_{\rm draw}}\left[\frac{1}{N_{\rm draw} -1}\sum_{j=1}^{N_{\rm found}}\left(\frac{\pi(\PEparameter_{j}|\PEhyperparam)}{p(\PEparameter_{j}|\PEhyperparam_{\rm draw})}\right)^2 - \hat{\xi}(\PEhyperparam)^2\right] 
\end{equation}
is the variance in the detection efficiency Monte Carlo integral of Equation~\eqref{eq:selection_estimator}.
When the rate-marginalized likelihood of Equation~\eqref{eq:rate_marginalized_hierarchical_likelihood} is used, $\sigma^2_{\ln \hat{\mathcal{L}}}$ takes a slightly different form
\begin{equation}
	\sigma^2_{\ln \hat{\mathcal{L}}}(\PEhyperparam) = \sum_{i=1}^{N_{\rm det}}\frac{\sigma^2_{\hat{\mathcal{L}}_i}(\PEhyperparam)}{\hat{\mathcal{L}}_i(\PEhyperparam)^2} + N_{\rm det}^2 \frac{\sigma^2_{\xi}(\PEhyperparam)}{\hat{\xi}(\PEhyperparam)^2}.
	\label{eq:variance_rate_marginalized_likelihood}
\end{equation}

It has been shown that \ac{GW} population inference can be biased when the variance in log-likelihood estimator exceeds $1$.
Consequently, we adopt a threshold on $\sigma^2_{\ln \hat{\mathcal{L}}}$ of $1$ to manage the bias of the posterior. Above this threshold the likelihood estimate may not be converged, and thus we ignore posterior samples with variances above this chosen threshold. Equation~\eqref{eq:variance_likelihood} describes the pointwise variance in the estimation of the log-likelihood. However, for accurate sampling of the posterior we only require that the difference of log-likelihoods to be small. A small pointwise log-likelihood variance is \textit{sufficient} for the variance of the difference of log-likelihoods to be small but not \textit{necessary}, rendering our threshold conservative~\citep{Farr:2019rap, Essick:2022ojx}. Indeed, for some models, a large region of hyperparameter space is removed by this threshold, limiting the range of potential populations that can be explored; for an example, see Appendix~\ref{appendix:model_comparison_effective_spins}.
Improvements to likelihood estimation are an active area of ongoing research~\citep{Wysocki:2018mpo,Doctor:2019ruh,Delfavero:2021qsc,Golomb:2021tll,Mould:2023eca,Hussain:2024qzl,Mancarella:2025uat}.

\subsection{Sampling Techniques}

In each model, we draw samples from the posterior to study the population distributions consistent with the data and the population model.
We draw samples using a variety of stochastic sampling algorithms, where the exact approach depends on the model. 
For most \parametrics, we use the nested sampler \DYNESTY\citep{Speagle:2019ivv} wrapper in \BILBY\citep{Ashton:2018jfp}, and the \GWPOPULATION implementation of the hierarchical likelihood \citep{Talbot:2024yqw}.

However, our \nonparametrics tend to have a large number of hyperparameters and so have a high dimensional posterior.
Nested sampling struggles with high dimensional distributions, so we use the \ac{HMC} adaptive \ac{NUTS} implementation in \NUMPYRO\citep{Phan:2019elc, Bingham:2019}. 
The \NUMPYRO adaptive \ac{NUTS} requires an autodifferentiable implementation of the likelihood, which we write in \JAX\citep{Bradbury:2018}.
This gradient information allows the \ac{NUTS} algorithm to efficiently explore high dimensional posteriors.

%% file: appendix_parametric_models.tex
\subsection{Mass Model for the Full CBC Population}\label{appendix:PDB}

Our \parametricAdj \PDB mass model is based on \textsc{Power Law+Dip}~\citep{Fishbach:2020ryj}, \textsc{Broken Power Law + Dip}~\citep{Farah:2021qom}, and \textsc{MultiPDB}~\citep{Mali:2024wpq}.
It is also similar to the \textsc{Power Law+Dip+Break} model~\citep{KAGRA:2021duu}, but in our \PDB model we now include additional structure.
We allow for the possibility of an upper mass gap, as well as separate pairing functions for \ac{NS}-containing binaries and \acp{BBH}.
Additionally, we include explicit peaks at low and mid-range \ac{BH} masses to capture the $\sim 9$--$10\,\Msun$ and $\sim 30$--$40\, \Msun$ features found by the \nonparametrics.
As discussed in Section~\ref{sec:all_masses}, we find that the addition of a low-mass \ac{BH} peak eliminates the need to explicitly model a gap between \acp{NS} and \acp{BH} masses.

Our mass model is parameterized as
\begin{equation}
    \pi(m_1, m_2|\PEhyperparam) \propto \pi_m(m_1|\PEhyperparam)\pi_m(m_2|\PEhyperparam)f(m_1, m_2)\Theta(m_2 < m_1)
\label{eq: pdb joint mass model}
\end{equation}
where the one-dimensional mass distribution $\pi_m(m|\PEhyperparam)$ is given by
\begin{align}
\pi_m(m|\PEhyperparam) &=& [1 + c_1 \mathcal{N}_{[m_{\text{min}}, m_{\text{max}}]}(m|\mu_1, \sigma_1) + c_2 \mathcal{N}_{[m_{\text{min}}, m_{\text{max}}]}(m|\mu_2, \sigma_2)]
n_1(m\,|\,m_{\text{NSmax}}, m_{\text{BHmin}}, \eta_{\text{NSmax}}, \eta_{\text{BHmin}}, A) \nonumber \\
&& \times n_2(m\,|\,m_{\text{UMGmin}}, m_{\text{UMGmax}}, \eta_{\text{UMGmin}}, \eta_{\text{UMGmax}}, A_2) h(m\,|\,m_{\text{NSmin}}, \eta_{\text{NSmin}}) l(m\,|\,m_{\text{BHmax}}, \eta_{\text{BHmax}}) \nonumber \\
&& \times
\begin{cases}
m^{\alpha_1} & \text{if } m < m_{\text{NSmax}} \\
m^{\alpha_{\text{dip}}} m_{\text{NSmax}}^{\alpha_1 - \alpha_{\text{dip}}} & \text{if } m_{\text{NSmax}} \leq m < m_{\text{BHmin}} \\
m^{\alpha_2} m_{\text{NSmax}}^{\alpha_1 - \alpha_{\text{dip}}} m_{\text{BHmin}}^{\alpha_{\text{dip}} - \alpha_2} & \text{if } m \geq m_{\text{BHmin}}.
\end{cases}
\label{eq: pdb mass model}
\end{align}

This $\pi_m(m|\PEhyperparam)$ represents a universal mass function to describe the primary and secondary mass distributions. Note that the marginal mass distribution is different from the universal mass distribution due to the pairing formalism.
$\mathcal{N}_{[a,b]}(\mu, \sigma)$ represents a truncated normal distribution over $[a,b]$ with location and width parameters $\mu$ and $\sigma$. 
The high-pass, low-pass and notch functions are defined as follows:

\begin{align}
    l(m|m_{\text{BHmax}}, \eta_{\text{BHmax}}) &= \left[1+\left(\frac{m}{m_{\text{BHmax}}}\right)^{\eta_{\text{BHmax}}}\right]^{-1},\\
    h(m|m_{\text{NSmin}}, \eta_{\text{NSmin}}) &= 1 - l(m|m_{\text{NSmin}}, \eta_{\text{NSmin}}),\\
    n_1(m|m_{\text{NSmax}}, m_{\text{BHmin}}, \eta_{\text{NSmax}}, \eta_{\text{BHmin}}, A) &= 1 - A l(m|m_{\text{NSmax}}, \eta_{\text{NSmax}}) h(m|m_{\text{BHmin}}, \eta_{\text{BHmin}}),\\
	n_2(m|m_{\text{UMGmin}}, m_{\text{UMGmax}}, \eta_{\text{UMGmin}}, \eta_{\text{UMGmax}}, A_2) &= 1 - A_2 l(m|m_{\text{UMGmin}}, \eta_{\text{UMGmin}}) h(m|m_{\text{UMGmax}}, \eta_{\text{UMGmax}}).\\
\end{align}

The pairing function
\begin{equation}
f(m_1,m_2|\beta_{\rm BH}, \beta_{\rm NS}) = \begin{cases}
    \left(\dfrac{m_2}{m_1}\right)^{\beta_{1}} & {\rm if} \; m_2 < 5\,\Msun \\
    \left(\dfrac{m_2}{m_1}\right)^{\beta_{2}} & {\rm if} \; m_2 > 5\,\Msun \\
\end{cases}
\end{equation}
controls how much merging binaries favor/disfavor equal masses. We allow for alternative pairing for binaries with very light secondary masses (\acp{NSBH} or \ac{BBH} with the secondary component in the lower-mass gap, e.g., GW190814 \citealt{LIGOScientific:2020zkf}).
We show the priors and describe the parameters of the \PDB model in Table~\ref{tab:parameters_pdb}.

\begin{deluxetable}{ccccc}
\tablecaption{Summary of \PDB model parameters and priors.}
\tablehead{
\colhead{Category} & \colhead{Parameter} & \colhead{Unit} & \colhead{Description} & \colhead{Prior}
}
\label{tab:parameters_pdb}
\startdata
Pairing Function & $\beta_1$ & -- & Spectral index below $5 \, \Msun$ & U($-2, 3$) \\
& $\beta_2$ & -- & Spectral index above $5\, \Msun$ & U($-2, 7$) \\
\tableline
Broken Power-Law & $\alpha_1$ & -- & Powerlaw below $m_{\text{NS max}}$ & U($-10, 2$) \\
& $\alpha_{\text{dip}}$ & -- & Powerlaw between $m_{\text{NS max}}$ and $m_{\text{BH min}}$ & U($-3, 2$) \\
& $\alpha_2$ & -- & Powerlaw above $m_{\text{BH min}}$ & U($-3, 2$) \\
& $m_{\text{brk}}$ & $\Msun$ & Break point between $\alpha_1$ and $\alpha_2$ & $5$ \\
\tableline
Highpass Filter & $m_{\text{NS min}}$ & $\Msun$ & Low-mass roll-off & U($1, 1.4$) \\
& $\eta_{\text{min}}$ & -- & Sharpness at $m_{\text{NS min}}$ & $50$ \\
\tableline
Lowpass Filter & $m_{\text{BHmax}}$ & $\Msun$ & High-mass roll-off & U($60, 200$) \\
& $\eta_{\text{max}}$ & -- & Sharpness at $m_{\text{BHmax}}$ & U($-4, 12$) \\
\tableline
Low-Mass Notch & $m_{\text{NS max}}$ & $\Msun$ & Lower notch edge & U($1.4, 5$) \\
& $\eta^{\text{low}}_1$ & -- & Sharpness at $m_{\text{NS max}}$ & $50$ \\
& $m_{\text{BH min}}$ & $\Msun$ & Upper notch edge & U($5, 9$) \\
& $\eta^{\text{high}}_1$ & -- & Sharpness at $m_{\text{BH min}}$ & $50$ \\
& $A_1$ & -- & Notch depth & $0$ \\
\tableline
High-Mass Notch & $m_{\text{UMGmin}}$ & $\Msun$ & Lower notch edge & U($30, 90$) \\
& $\eta^{\text{low}}_2$ & -- & Sharpness at $m_{\text{UMGmin}}$ & $30$ \\
& $m_{\text{UMGmax}}$ & $\Msun$ & Upper notch edge & U($60, 150$) \\
& $\eta^{\text{high}}_2$ & -- & Sharpness at $m_{\text{UMGmax}}$ & $30$ \\
& $A_2$ & -- & Depth of high-mass notch & U($0, 1$) \\
\tableline
Low-Mass Peak & $\mu^{\text{peak}}_2$ & $\Msun$ & Peak location & U($6, 12$) \\
& $\sigma^{\text{peak}}_2$ & $\Msun$ & Peak width & U($0, 5$) \\
& $c_2$ & -- & Peak height & U($0, 500$) \\
\tableline
High-Mass Peak & $\mu^{\text{peak}}_1$ & $\Msun$ & Peak location & U($17, 50$) \\
& $\sigma^{\text{peak}}_1$ & $\Msun$ & Peak width & U($4, 20$) \\
& $c_1$ & -- & Peak height & U($0, 1000$) \\
\enddata
\tablecomments{U($x$, $y$) denotes a Uniform prior between $x$ and $y$.}
\end{deluxetable}

\subsection{Neutron Star Mass Models}\label{appendix:NSmodels}

\begin{deluxetable}{cccc}
\tablecaption{Summary of \textsc{Power} and \textsc{Peak} \ac{NS} mass model parameters.}
\tablehead{
\colhead{Parameter} & \colhead{Unit} & \colhead{Description} & \colhead{Prior}
}
\label{tab:parameters_ns}
\startdata
$\alpha$ & -- & Spectral index for the power-law in the \textsc{Power} NS mass distribution. & U($-15$, $5$) \\
$m_{\rm min}$ & $\Msun$ & Minimum mass of the \ac{NS} mass distribution. & U($1.0$, $1.5$) \\
$m_{\rm max}$ & $\Msun$ & Maximum mass of the \ac{NS} mass distribution. & U($1.5$, $3.0$) \\
$\mu$ & $\Msun$ & Location of the Gaussian peak in the \textsc{Peak} \ac{NS} mass distribution. & U($1.0$, $3.0$) \\
$\sigma$ & $\Msun$ & Width of the Gaussian peak in the \textsc{Peak} \ac{NS} mass distribution. & U($0.01$, $2.00$) \\
\enddata
\tablecomments{U($x$, $y$) denotes a Uniform prior between $x$ and $y$.}
\end{deluxetable}

\noindent Following previous work~\citep{Landry:2021hvl,KAGRA:2021duu},
the mass distribution of \ac{NS}-containing events is modeled as
\begin{equation}
\pi(m_1, m_2 | \PEhyperparam) \propto
\begin{cases}
\pi(m_1| \PEhyperparam) \, \pi(m_2| \PEhyperparam) & \text{if BNS,} \\
\mathrm{U}(3\,\Msun, 60\,\Msun) \, \pi(m_2 | \PEhyperparam) & \text{if NSBH},
\end{cases}
\end{equation}
To model $\pi(m|\PEhyperparam)$, we use either of the following models
\begin{enumerate}
	\item \textsc{Power} model:
\begin{equation}
	\pi(m|\PEhyperparam) \propto
	\begin{cases}
		m^{\alpha} & \text{if } m_{\min} \leq m \leq m_{\max}, \\
		0 & \text{otherwise.}
	\end{cases}
\end{equation}

\item \textsc{Peak} model:
\begin{equation}
	\pi(m|\PEhyperparam) \propto
	\begin{cases}
		\exp\left[ -\dfrac{(m - \mu)^2}{2\sigma^2} \right] & \text{if } m_{\min} \leq m \leq m_{\max}, \\
		0 & \text{otherwise.}
	\end{cases}
\end{equation}

See Table~\ref{tab:parameters_ns} for a description of the parameters used in the model and the corresponding prior ranges.

\end{enumerate}

\subsection{Binary Black Hole Mass Models}\label{appendix:BrokenPLTwoPeaks}

\BrokenPLTwoPeaks : The fiducial \ac{BBH} mass model is a mixture between a broken power law and two left-truncated Gaussian peaks, with low mass tapering applied to the full distribution. The broken power law is given by

\begin{eqnarray}
	p_\text{BP}(m_1 | \alpha_1, \alpha_2, m_\text{break}, m_\text{1,low}, m_\text{high}) = \frac{1}{N}
	\begin{cases}
		  \left(\dfrac{m_1}{m_\text{break}}\right)^{-\alpha_1} & m_\text{1,low} \leq m_1  < m_\text{break} \\
		  \left(\dfrac{m_1}{m_\text{break}}\right)^{-\alpha_2} & m_\text{break} \leq m_1  < m_\text{high},
	\end{cases}
\end{eqnarray}

\noindent where $\alpha_1$ and $\alpha_2$ are the power law indices, the transition between the low-mass and high-mass power law occurs at $m_\text{break}$, and the normalization constant is

\begin{equation}
	N = m_{\rm break} \left[
    \frac{1 - \left(\tfrac{m_{1,\rm low}}{m_{\rm break}}\right)^{1-\alpha_1}}{1-\alpha_1}
    + \frac{\left(\tfrac{m_{1,\rm high}}{m_{\rm break}}\right)^{1-\alpha_2} - 1}{1-\alpha_2}\right].
\end{equation}

\noindent The full mixture distribution $\pi(m_1|{\PEhyperparam})$ is

\begin{align}
 \pi(m_1 | {\PEhyperparam})\propto & \Biggl[ \lambda_0 p_{\rm BP}(m_1|\alpha_1, \alpha_2, m_{\rm break}, m_{1,{\rm low}}, m_{\rm high}) + \lambda_1 \mathcal{N}_{lt}(m_1 | \mu_1, \sigma_1, \text{low} = m_{1,{\rm low}}) \\ \nonumber
& + (1 - \lambda_0 - \lambda_1) \mathcal{N}_{lt}(m_1|\mu_2, \sigma_2, \text{low} = m_{1,{\rm low}}) \Biggr] S(m_1 | m_{1,{\rm low}}, \delta_{m,1}),
\end{align}

\noindent where $\mathcal{N}_{lt}$ is a left-truncated normal distribution. The Planck tapering function $S$ ensures a smooth turn-on of the distribution in the range $(m_{\rm 1,low}, m_{\rm 1,low} + \delta_{m,1}]$ and is given by

\begin{eqnarray}
S(m|m_{\rm low}, \delta_m) =  \begin{cases}
0 & m < m_{\rm low}, \\\
[1 + f(m - m_{\rm low}, \delta_m)]^{-1}  & m_{\rm low} \leq m < m_{\rm low} + \delta_m, \\
1 & m_{\rm low} + \delta_m \leq m,
\end{cases}
\end{eqnarray}

\noindent with
$$
f(m', \delta_m) = \text{exp} \Biggl( \frac{\delta_m}{m'} + \frac{\delta_m}{m' - \delta_m} \Biggr).
$$

We model the mass ratio as a power law with index $\beta_q$ and low-mass tapering applied to secondary mass $m_2$, conditioned on primary mass $m_1$,

\begin{equation}
p_{\rm PL}(q | m_1, \beta_q, m_{2,{\rm low}}, \delta_{m,2}) \propto q^{\beta_q} S(m_1 q | m_{2,{\rm low}}, \delta_{m,2}),
\end{equation}

\noindent with the same tapering function defined above. We numerically normalize the mass ratio distribution as a function of primary mass, and interpolate the normalization to arbitrary primary masses with a log-uniform grid across primary mass. We define $m_2 \leq m_1$, and therefore must enforce $m_{2, {\rm low}} \le m_{1, {\rm low}}$. We use a prior which is uniform in the two dimensional triangular space satisfying the inequality and between 3 and 10$\,\Msun$. Specifically, this defines the priors

\begin{align}\label{eq:m_low_prior}
& \pi(m_{1, {\rm low}}) = \frac{2}{({\rm max} - {\rm min})^2}(m_{1, {\rm low}} - {\rm min}), \\ \nonumber
& \pi(m_{2, {\rm low}}|m_{1, {\rm low}}) = \frac{1}{m_{1, {\rm low}} - {\rm min}},
\end{align}

\noindent where ${\rm min} = 3\,\Msun,{\rm max} = 10\,\Msun$. The priors used in this model can be found in Table \ref{tab:parameters_default}.

\ExtendedBrokenPLTwoPeaks : This model incorporates correlations between the primary mass and mass ratio, by allowing each primary mass mixture component in the \BrokenPLTwoPeaks to be associated with a separate power law mass ratio model. In terms of the functions defined above, the model is expressed as

\begin{align} \label{eq:exbp2p}
\pi(m_1, q | {\PEhyperparam})\propto \Biggl[ &\lambda_0 p_{\rm BP}(m_1|\alpha_1, \alpha_2, m_{\rm break}, m_{1,{\rm low}}, m_{\rm high}) p_{\rm PL}(q | m_1, \beta_q^{\rm BP}, m_{2,{\rm low}}^{\rm BP}, \delta_{m,2}^{\rm BP})  \\ \nonumber
& + \lambda_1 \mathcal{N}_{lt}(m_1 | \mu_1, \sigma_1, \text{low} = m_{1,{\rm low}}) p_{\rm PL}(q | m_1, \beta_q^{\rm peak1}, m_{2,{\rm low}}^{\rm peak1}, \delta_{m,2}^{\rm peak1}) \\ \nonumber
& + (1 - \lambda_0 - \lambda_1) \mathcal{N}_{lt}(m_1|\mu_2, \sigma_2, \text{low} = m_{1,{\rm low}}) p_{\rm PL}(q | m_1, \beta_q^{\rm peak2}, m_{2,{\rm low}}^{\rm peak2}, \delta_{m,2}^{\rm peak2}) \Biggr] S(m_1 | m_{1,{\rm low}}, \delta_{m,1}),
\end{align}

\noindent where the superscripts (BP, peak1, and peak2) denote the different mass ratio hyperparameters ($\beta_q$, $m_{2,{\rm low}}$, $\delta_{m,2}$) for the broken power law, first Gaussian, and second Gaussian components. The priors on these hyperparameters are the same as those used in the \BrokenPLTwoPeaks model, listed in Table \ref{tab:parameters_default}, except for the power law index parameters $\beta_q$, which assume U$(-10,13)$.

\begin{deluxetable}{ccc}
\tablecaption{Summary of \BrokenPLTwoPeaks model parameters and priors.}
\tablehead{
\colhead{Parameter} & \colhead{Description} & \colhead{Prior}
}
\label{tab:parameters_default}
\startdata
$\alpha_1$ & Spectral index of 1st primary mass power law & U$(-4, 12)$ \\
$\alpha_2$ & Spectral index of 2nd primary mass power law & U$(-4, 12)$ \\
$m_{\rm break}$ & Power law break location & U$(20, 50)$ \\
$\mu_1$ & Location of the first peak & U$(5, 20)$ \\
$\sigma_1$ & Width of the first peak & U$(0,10)$ \\
$\mu_2$ & Location of the second peak & U$(25, 60)$ \\
$\sigma_2$ & Width of the second peak & U$(0,10)$ \\
$m_{\rm 1, low}$ & Lower edge of taper function & see \eqref{eq:m_low_prior} \\
$\delta_{\rm m,1}$ & Mass range of low mass tapering & U$(0,10)$ \\
$\lambda_0, \lambda_1$ & Mixing fractions between power law and peaks & ${\rm Dir}(\mathbf{\alpha}=(1,1,1))$ \\
$m_{\rm high}$ & Maximum mass for distribution, which is pinned to $m_{\rm high} = 300\,\Msun$ by default & $\delta(m_{\rm high} - 300)$ \\
\tableline
$\beta_q$ & Spectral index of mass ratio power law & U$(-2,7)$ \\
$m_{\rm 2, low}$ & Lower edge of taper function in $m_2$ & see \eqref{eq:m_low_prior} \\
$\delta_{\rm m,2}$ & Mass range of low mass tapering in $m_2$ & U$(0,10)$ \\
\enddata
\tablecomments{The priors for the \ExtendedBrokenPLTwoPeaks model are the same except for the $\beta_q$ parameters, which assume a prior of U$(-10,13)$.}
\end{deluxetable}

\subsection{\powerlawRedshift Model}\label{appendix:redshiftModels}

We model redshift evolution by the comoving merger rate density (Equation~\ref{eq:comoving_merger_rate}). In particular, we use the model
\begin{equation}
\pi(z|\kappa) \propto \frac{1}{1+z}\frac{\mathrm{d} V_c}{\mathrm{d} z}(1+z)^\kappa
\label{eq:powerlawredshift}
\end{equation}
where the prefactor converts from a rate density in comoving volume and source frame time to detector frame time and redshift. 
In other words, the comoving rate density scales as $\mathcal{R} \propto (1+z)^\kappa$.
We use a prior on $\kappa$ as specified in Table~\ref{tab:powerlawredshift}.

\begin{deluxetable}{ccc}
\tablecaption{Summary of \powerlawRedshift model parameter and prior.}
\tablehead{
\colhead{Parameter} & \colhead{Description} & \colhead{Prior}
}
\label{tab:powerlawredshift}
\startdata
$\kappa$ & Power-law index on comoving merger rate evolution & U($-10, 10$) \\
\enddata
\end{deluxetable}

\subsection{Spin Models}\label{appendix:spinModels}

BBH spins can be parameterized in several ways which are useful for GW data analysis. 
Here, we use dimensionless spin magnitudes for the \acp{BH} $\chi_1$ and $\chi_2$, and cosine tilt angles $\cos\theta_1$ and $\cos\theta_2$, as well as the effective spin parameters $\chieff$~\citep{Racine:2008qv,Ajith:2009bn,Damour:2001tu} and $\chip$~\citep{Schmidt:2010it,Schmidt:2012rh,Schmidt:2014iyl}. 
The effective inspiral spin $\chieff$ used because it is typically the most precisely measured \ac{BBH} spin parameter, due to its lower-order appearance post-Newtonian expansions~\citep{Arun:2008kb}. 
Other parametrizations of spin precession exist~\citep[e.g.,][]{Fairhurst:2019vut,Gerosa:2020aiw,Thomas:2020uqj} but for reasons of convention, we only work with the typical parameterization given in Equation 16 of \cite{GWTC:Introduction}.  
When modeling the effective spins, we use the analytic per-event parameter-estimation prior on $\chieff$, $\chip$, and $q$~\citep{Iwaya:2024zzq} in the calculation of the hierarchical likelihood (Equation~\ref{eq:hierarchical_likelihood}).
For all models presented in this work, we assume that azimuthal angles $\phi_i$ are distributed uniformly between $0$ and $2\pi$---the same as their parameter estimation prior---due to their typically uninformative individual-event posteriors.
See Table 3 of \cite{GWTC:Introduction} for more information about spin parameters.

We report results with spin vectors defined at the reference frequencies used for inference on each signal~\citep{GWTC:Methods}.
The effective spin $\chi_\mathrm{eff}$ is approximately conserved throughout the inspiral~\citep{Racine:2008qv,Gerosa:2015tea}, so its population distribution should be unaffected by the choice of reference frequency.
While the effective precessing spin $\chi_\mathrm{p}$ and the spin angles are dependent on reference frequency, they are comparatively weakly constrained by the data.
Additionally, their parameter estimation priors are invariant under time evolution, meaning measurements are robust against different reference frequencies.
At the current number of events and because of model dependence while measuring tilt distributions, we do not expect significant differences between the inferred tilt or $\chip$ distributions at different reference frequencies~\citep{Mould:2021xst}.
Our approach is consistent with past \ac{LVK} analyses, where evolved spins have not been used~\citep{LIGOScientific:2018jsj,LIGOScientific:2020kqk,KAGRA:2021duu}.

\subsubsection{Component Spin Models}\label{appendix:componentSpinModels}

\begin{deluxetable}{ccc}
\tablecaption{Summary of \default model parameters for spin magnitudes (Equation~\ref{eqn:spin_mag_model}) and tilt angles (Equation~\ref{eqn:spin_tilt_model},~\ref{eqn:spin_tilt_model_with_min}).}
\tablehead{
\colhead{Parameter} & \colhead{Description} & \colhead{Prior}
}
\label{tab:componentSpinPriors}
\startdata
$\mu_\chi$ & Location of the $\chi$ distribution & U($0$, $1$) \\
$\sigma_\chi$ & Width of the $\chi$ distribution & U($0.005$, $1$) \\
$\mu_t$ & Location of the Gaussian component of the $\cos\theta$ distribution & U($-1$, $1$) \\
$\sigma_t$ & Width of the Gaussian component of the $\cos\theta$ distribution & U($0.01$, $4$) \\
$\zeta$ & Fraction in the Gaussian component of the $\cos\theta$ distribution & U($0$, $1$) \\
\tableline
$t_{\rm min}$ & Minimum of the $\cos\theta$ distribution & U($-1$, $1$) \\
\enddata
\tablecomments{U stands for a uniform prior.}
\end{deluxetable}

\default:
We model the spin magnitudes ($\chi_i$) as a truncated Gaussian distribution between $0$ and $1$, assuming they are identically and independently distributed: 
\begin{equation}
    \pi(\chi_i|\mu_\chi,\sigma_\chi) = \mathcal{N}_{[0,1]}(\chi_1 | \mu_\chi, \sigma_\chi)\mathcal{N}_{[0,1]}(\chi_2 | \mu_\chi, \sigma_\chi)\,.
\label{eqn:spin_mag_model}
\end{equation}
This choice attempts to rectify shortcomings of the non-singular Beta distribution used to model the spin magnitudes in previous work~\citep{LIGOScientific:2018jsj, LIGOScientific:2020kqk, KAGRA:2021duu}; see Equation~\eqref{eqn:beta_distribution} below.
The non-singular Beta distribution is forced to go to $\pi(\chi)=0$ at $\chi=0,1$, which crucially does not allow for measurements of contributions to the population at $\chi=0$ or $\chi=1$, even though non-trivial contributions may exist~\citep{Callister:2022qwb,Galaudage:2021rkt,Hussain:2024qzl}.
Allowing the Beta distribution to be singular would add limited additional model flexibility, only adding the option for $\pi(\chi) = \infty$ at the $\chi=0,1$, but not anywhere in between $0$ and $\infty$. 
Thus, we opt for the truncated Gaussian model, which can take on a continuous range of values at $\chi$'s boundaries.

We model the distribution of the cosine spin tilt angle ($\cos\theta_i$) as a mixture between a Gaussian distribution truncated on $-1$ to $1$ and an isotropic distribution, assuming they are identically but \textit{not} independently distributed:
\begin{equation}
    \pi(\cos\theta_i| \mu_t, \sigma_t, \zeta ) = \zeta \,\mathcal{N}_{[-1,1]}(\cos\theta_1 | \mu_t, \sigma_t)\mathcal{N}_{[-1,1]}(\cos\theta_2 | \mu_t, \sigma_t) + \frac{1-\zeta}{4}\,.
\label{eqn:spin_tilt_model} 
\end{equation}
We here allow for the location of the Gaussian sub-population to vary, following~\cite{Vitale:2022dpa}, rather than fixing it at $\mu_t=1$ as was done in previous work~\citep{LIGOScientific:2018jsj, LIGOScientific:2020kqk, KAGRA:2021duu}.
Priors on the \default~hyperparameters are given in Table~\ref{tab:componentSpinPriors}.

{\sc Minimum Tilt Model}:
For the spin tilts, we also use a model in which $t_{\rm min}$ (where $t\equiv\cos\theta$), a hard lower truncation of $\pi(\cos\theta)$, is informed by the data. 
In this case the population distribution becomes 
\begin{equation}
    \pi(\cos\theta_i| \mu_t, \sigma_t, \zeta, t_{\rm min} ) =
	\begin{cases} 
        \zeta \,\mathcal{N}_{[t_{\rm min},1]}(\cos\theta_1 | \mu_t, \sigma_t)\mathcal{N}_{[t_{\rm min},1]}(\cos\theta_2 | \mu_t, \sigma_t) + \displaystyle \frac{1-\zeta}{(1-t_{\rm min})^2} \qquad\,\cos\theta_i > t_{\rm min}\,\,, \\
        0 \qquad\qquad\qquad\qquad\qquad\qquad\qquad\qquad\qquad\qquad\qquad\qquad\qquad\quad 
		\cos\theta_i \leq t_{\rm min}\,\,.
	\end{cases}
	\label{eqn:spin_tilt_model_with_min}
\end{equation}
Equation \eqref{eqn:spin_tilt_model} is the same as Equation~\eqref{eqn:spin_tilt_model_with_min} with $t_{\rm min} = - 1$.
The prior on $t_{\rm min}$ and other hyperparameters are given in Table~\ref{tab:componentSpinPriors}.

{\sc Beta Distribution Spin Magnitude}:
In Appendix~\ref{appendix:model_comparison_spins}, we compare the \default~model to alternatives.
For the spin magnitudes, these are the Constrained and Unconstrained Beta Distributions, which both follow the form:
\begin{equation}
	\pi(\chi_i | \alpha, \beta) \propto {\chi_i}^{\alpha-1}{(1-\chi_i)}^{\beta-1} \,\,.
	\label{eqn:beta_distribution}
\end{equation}
The Constrained Beta was the {\sc Default} model for \gwtcthree~\citep{KAGRA:2021duu} and requires the shape parameters $\alpha, \beta > 1$, which forces $\pi(\chi_i=0,1) = 0$~\citep{Wysocki:2018mpo, LIGOScientific:2020kqk}.
The Unconstrained Beta relaxes this constraint on the the shape parameters. 
If $0< \alpha,\beta < 1$, the Beta distribution becomes singular, meaning $\pi(\chi_i=0,1) = \infty$.
Following \cite{KAGRA:2021duu}, we sample the mean and standard deviation of the Beta distribution, rather than $\alpha$ and $\beta$, and use the same priors for $\mu_\chi$ and $\sigma_\chi$ as given in Table~\ref{tab:componentSpinPriors} for the \default model.
For the Constrained Beta case, we additionally impose the cut $\alpha, \beta > 1$.
The relationship between $\{\mu_\chi,~\sigma_\chi\}$ and $\{\alpha,~\beta\}$ is given in Equation 5 of \cite{LIGOScientific:2018jsj}.

\subsubsection{Identical versus Non-identically Distributed Spin Magnitudes and Tilts}
\label{appendix:iid}

In Appendix~\ref{appendix:model_comparison_spins}, we investigate whether or not the primary and secondary spins are identically distributed, assuming the \default model.
\textit{Identical} distribution means that the two distributions share the same set of hyperparameters, while \textit{non-identical}  means that each is described by different hyperparameters.
Spin magnitudes are also \textit{independently} distributed, meaning that $\pi(\chi_1, \chi_2)$ is separable in terms of $\chi_1$ and $\chi_2$.
The acronym IID means they are independently and identically distributed, while IND means independently and non-identically distributed:
\begin{align}
	\chi_i~\mathrm{IID} & \implies \pi(\chi_{1,2} | \mu_\chi, \sigma_\chi) = \mathcal{N}_{[0,1]}(\chi_1 | \mu_\chi, \sigma_\chi) \, \mathcal{N}_{[0,1]}( \chi_2 | \mu_\chi, \sigma_\chi) \,\, , \label{eqn:iid_mags} \\
	\chi_i~\mathrm{IND} & \implies \pi(\chi_{1,2} | \mu_{\chi,1}, \sigma_{\chi,1}, \mu_{\chi,2}, \sigma_{\chi,2}) = \mathcal{N}_{[0,1]}(\chi_1 | \mu_{\chi,1}, \sigma_{\chi,1}) \, \mathcal{N}_{[0,1]}( \chi_2 | \mu_{\chi,2}, \sigma_{\chi,2}) \,\,.
	\label{eqn:ind_mags}
\end{align}
The tilt angles, on the other hand, are \textit{non-independently} distributed, meaning that $\pi(\cos\theta_1, \cos\theta_2)$ is not separable.
The acronynm NID means they are non-independently but identically distributed, while NND means non-independently and non-identically distributed:
\begin{align}
	\cos\theta_i~\mathrm{NID} & \implies \pi(\cos\theta_{1,2}|\zeta, \mu_t, \sigma_t) = \zeta \,\mathcal{N}_{[-1,1]}(\cos\theta_1 | \mu_t, \sigma_t)\,\mathcal{N}_{[-1,1]}(\cos\theta_2 | \mu_t, \sigma_t) + \frac{1-\zeta}{4} \label{eqn:nid_tilts} \,\, ,\\
	\cos\theta_i~\mathrm{NND} & \implies \pi(\cos\theta_{1,2}|\zeta, \mu_{t,1}, \sigma_{t,1}, \mu_{t,2}, \sigma_{t,2}) = \zeta \,\mathcal{N}_{[-1,1]}(\cos\theta_1 | \mu_{t,1}, \sigma_{t,1})\,\mathcal{N}_{[-1,1]}(\cos\theta_2 | \mu_{t,2}, \sigma_{t,2})  + \frac{1-\zeta}{4} \,\,.
	\label{eqn:nnd_tilts}
\end{align}
For the non-identically distributed cases, each component's hyperparameters have the same priors as those listed in Table~\ref{tab:componentSpinPriors} for the identically distributed case. 
Table~\ref{tab:modelComparisonSpins} gives Bayes factors between different combinations of IID/IND spin magnitudes and NID/NND spin tilts; see associated dicsussion in Appendix~\ref{appendix:model_comparison_spins}.

\subsubsection{Effective Spin Models}\label{appendix:effectiveSpinModels}

\begin{deluxetable}{cccc}
\tablecaption{Summary of \effectiveSpinModel~(Equation~\ref{eqn:chi_eff_chi_p_gaussian_model}) and \skewnormalChiEff~(Equation~\ref{eqn:chi_eff_skewnormal_model}) spin parameters.}
\tablehead{
\colhead{Parameter} & \colhead{Description} & \colhead{Prior G} & \colhead{Prior SN}
}
\label{tab:effectiveSpinPriors}
\startdata
$\mu_\mathrm{eff}$ & Location of the $\chieff$ distribution & U($-1$, $1$) & U($-1$, $1$) \\
$\sigma_\mathrm{eff}$ & Width of the $\chieff$ distribution & U($0.05$, $1$) & LU($0.01$, $4$) \\
$\mu_p$ & Location of the $\chi_\mathrm{p}$ distribution & U($0.05$, $1$) & U($0.01$, $1$) \\
$\sigma_p$ & Width of the $\chi_\mathrm{p}$ distribution & U($0.07$, $1$)& LU($0.01$, $1$)\\
\tableline
$\rho$ & Degree of correlation between $\chieff$ and $\chi_\mathrm{p}$ & U($-0.75$, $0.75$) & N/A \\
\tableline
$\epsilon $ & Skew of the $\chieff$ distribution & N/A & U($-1$, $1$) \\
\enddata
\tablecomments{The first column of priors ({\bf G}) gives those used for the \effectiveSpinModel results, the second column ({\bf SN}) is \skewnormalChiEff. U stands for a uniform prior; LU for log-uniform.}
\end{deluxetable}

\effectiveSpinModel: We here assume that the distribution of $\chieff$ and $\chip$ across the \ac{BBH} population is a bivariate truncated Gaussian~\citep{Miller:2020zox, Roulet:2018jbe} characterized by the location and width of the $\chieff$ and $\chip$ distributions, and the covariance between them:
\begin{equation}
    \pi(\chieff, \chip | \mathbf{\mu}, \mathbf{\Sigma}) \propto \mathcal{N}(\chieff, \chip | \mathbf{\mu}, \mathbf{\Sigma}) \,\, ,
\end{equation}
where $\mathbf \mu =(\mu_{\rm eff}, \mu_{\rm p})$ and
\begin{equation}
    \mathbf{\Sigma} =
    \begin{pmatrix}
        \sigma_{\rm eff}^2 & \rho\,\sigma_{\rm eff}\,\sigma_{\rm p} \\
        \rho\,\sigma_{\text{eff}}\,\sigma_{\rm p} & \sigma_{\rm p}^2
    \end{pmatrix}   \,\,.
\end{equation}
This expands to: 
\begin{equation}
    \pi(\chieff, \chip | \mathbf{\mu}, \mathbf{\Sigma}) \propto 
    \exp\left[ -\frac{1}{2(1 - \rho^2)}
    \left( \frac{(\chi_{\rm eff} - \mu_{\rm eff})^2}{\sigma_{\rm eff} ^2} 
    - \frac{2\rho (\chi_{\rm eff} - \mu_{\rm eff})(\chi_{\rm p} - \mu_{\rm p})}{\sigma_{\rm eff}\sigma_{\rm p}}
    + \frac{(\chi_{\rm p} - \mu_{\rm p})^2}{\sigma_{\rm p} ^2} \right) \right]\,\,.
    \label{eqn:chi_eff_chi_p_gaussian_model} 
\end{equation}
The bivariate Gaussian is truncated over the range $\chieff \in [-1,1]$, $\chip\in[0,1]$ and is normalized numerically.
Priors on the hyperparameters are given in Table \ref{tab:effectiveSpinPriors}.

\skewnormalChiEff: To account for observational evidence that the $\chieff$ distribution is not symmetric~\citep{Callister:2021fpo,Adamcewicz:2022hce,Banagiri:2025dxo}, we additionally model $\chieff$ as a skewed, truncated Gaussian distribution:

\begin{equation}
    \pi(\chieff|\mu_{\rm eff}, \sigma_{\rm eff}, \epsilon) \propto 
	\begin{cases} 
        (1 + \epsilon) \, \mathcal{N}_{[-1,1]}(\chieff | \mu_{\rm eff}, \sigma_{\rm eff} (1 + \epsilon)) \qquad \chieff \leq 0 \,\, ,\\ 
        (1 - \epsilon) \, \mathcal{N}_{[-1,1]}(\chieff | \mu_{\rm eff}, \sigma_{\rm eff} (1 - \epsilon)) \qquad \chieff \geq 0 \,\,.
	\end{cases}
    \label{eqn:chi_eff_skewnormal_model} 
\end{equation}
This model includes a parameter $\epsilon$ which describes the skew of the distribution, such that  $\epsilon > 0$ characterizes a distribution with more support for $\chieff<\mu_{\rm eff}$, while $\epsilon < 0$ has more support when $\chieff>\mu_{\rm eff}$. 
The \skewnormalChiEff~distribution reduces to a standard, symmetric Gaussian in the case that $\epsilon=0$.
We here model $\chip$ as a truncated normal distribution and infer its location and width. 
Priors on the hyperparameters are given in Table \ref{tab:effectiveSpinPriors}.

\subsection{Copula Correlation Models}\label{appendix:copulaCorrelationModels}

\begin{deluxetable}{ccc}
\tablecaption{
    Summary of parameters exclusive to the \copulacorrelation, \linearcorrelation, and \splinecorrelation correlated models, along with their priors.
}
\tablehead{
    \colhead{Parameter} & \colhead{Description} & \colhead{Prior}
}
\label{tab:parameters_correlations}
\startdata
$\kappa_{x,y}$ & Level of correlation between parameters $x$ and $y$ inferred with a copula & U$(-20, 20)$ \\
\tableline
$\mu_{\mathrm{eff}|q}$ & $q=1$ intercept in linear $\mu_\mathrm{eff}(q)$ & U$(-1, 1)$ \\
$\delta \mu_{\mathrm{eff}|q}$ & Gradient in linear $\mu_\mathrm{eff}(q)$ & U$(-2, 2)$ \\
$\ln \sigma_{\mathrm{eff}|q}$ & $q=1$ intercept in linear $\ln \sigma_\mathrm{eff}(z)$ & U$(-5, 0)$ \\
$\delta \ln \sigma_{\mathrm{eff}|q}$ & Gradient in linear $\ln \sigma_\mathrm{eff}(q)$ & U$(-12, 4)$ \\
\tableline
$\mu_{\mathrm{eff}|z}$ & $z=0$ intercept in linear $\mu_\mathrm{eff}(z)$ & U$(-1, 1)$ \\
$\delta \mu_{\mathrm{eff}|z}$ & Gradient in linear $\mu_\mathrm{eff}(z)$ & U$(-1, 1)$ \\
$\ln \sigma_{\mathrm{eff}|z}$ & $z=0$ intercept in linear $\ln \sigma_\mathrm{eff}(z)$ & U$(-5, 0)$ \\
$\delta \ln \sigma_{\mathrm{eff}|z}$ & Gradient in linear $\ln \sigma_\mathrm{eff}(z)$ & U$(-3, 5)$ \\
\tableline
$\mu_{\mathrm{eff}|q}^i$ & $i^\mathrm{th}$ node in spline $\mu_\mathrm{eff}(q)$ & U$(-1, 1)$ \\
$\ln \sigma_{\mathrm{eff}|q}^i$ & $i^\mathrm{th}$ node in spline $\ln \sigma_\mathrm{eff}(q)$ & U$(-5, 0)$ \\
\tableline
$\mu_{\mathrm{eff}|z}^i$ & $i^\mathrm{th}$ node in spline $\mu_\mathrm{eff}(z)$ & U$(-1, 1)$ \\
$\ln \sigma_{\mathrm{eff}|z}^i$ & $i^\mathrm{th}$ node in spline $\ln \sigma_\mathrm{eff}(z)$ & U$(-5, 0)$ \\
\enddata
\tablecomments{    There are four variations of the \copulacorrelation~model in which $(x,y) = (q,\chieff)$, $(m_1,\chieff)$, $(z,\chieff)$ and finally, $(m_1,z)$.
    Within the spline models, we use four nodes $i$.
}
\end{deluxetable}

Copulas allow for a variable correlation between two parameters $x$ and $y$ while keeping the marginal distributions for the respective parameters fixed.
This is done using the fact that the cumulative distribution function (CDF) of any randomly distributed variable is itself a uniformly distributed variable between $0$ and $1$, in order to model the CDFs of $x$ and $y$ with a correlated two-dimensional uniform distribution.
This correlated two-dimensional uniform distribution, known as a copula density function, is dependent on a (hyper)parameter $\kappa_{x,y}$, which determines the level (or strength) of the correlation.
A coordinate transformation (which is determined by the chosen marginal distributions for $x$ and $y$) can then be applied to provide a correlated two-dimensional model in the desired coordinates.
Conveniently, the Jacobian for this transformation turns out to be the product of the chosen marginal distributions, meaning the two-dimensional population model can be written as
\begin{equation}
    \pi_{xy}(x,y|\PEhyperparam,\kappa_{x,y}) = \pi_c \left(u(x|\PEhyperparam),v(y|\PEhyperparam)|\kappa_{x,y}\right) \pi_x(x|\PEhyperparam) \pi_y(y|\PEhyperparam).
\end{equation}
Here, $\PEhyperparam$ are the set of hyperparameters governing the marginal distributions $\pi_x$ and $\pi_y$, while $\pi_c$ denotes the chosen copula density function.
Finally,
\begin{equation}
    u(x|\PEhyperparam) = \int_{x_{\min}}^x \mathrm{d} x' \ \pi_x(x'|\PEhyperparam),
\end{equation}
and
\begin{equation}
    v(y|\PEhyperparam) = \int_{y_{\min}}^y \mathrm{d} y' \ \pi_y(y'|\PEhyperparam),
\end{equation}
which we refer to as $u$ and $v$ for the sake of brevity, are the CDFs of $x$ and $y$ respectively.

The \copulacorrelation~models all assume the same marginal distributions and copula density functions, but differ in which pairs of parameters they correlate (i.e., which parameters $u$ and $v$ are functions of in the above notation).
Namely, the mass distribution is assumed to follow the default \BrokenPLTwoPeaks~model, redshift follows the default power-law redshift model, and $\chieff$ and $\chip$ are Gaussian distributed, but uncorrelated (i.e., they follow the \effectiveSpinModel~model with $\rho=0$).
We assume a Frank copula density function
\begin{equation}
    \pi_c\left(u, v | \kappa_{x,y}\right) = 
    \frac{-\kappa_{x,y} e^{-\kappa_{x,y} (u+v)} (e^{-\kappa_{x,y}} - 1)}{\left( e^{-\kappa_{x,y}} - e^{-\kappa_{x,y} u} - e^{-\kappa_{x,y} v} + e^{-\kappa_{x,y}(u+v)} \right)^2},
\end{equation}
where $\kappa_{x,y} \in (-\infty, \infty)$.
Positive values of $\kappa_{x,y}$ imply a correlation, while negative values of $\kappa_{x,y}$ imply an anti-correlation.
The Frank copula density function is not defined at $\kappa_{x,y} = 0$, but becomes uncorrelated as $\kappa_{x,y} \rightarrow 0$ from above and below, so we assume $\pi_c\left(u, v | \kappa_{x,y}=0\right) = 1$ (technically, making this a piece-wise function).
The Frank copula density function is chosen as it allows for positive and negative correlations that are symmetric about $\kappa_{x,y}=0$, and gives rise to correlated distributions that we believe appear physically reasonable \citep[][]{Adamcewicz:2022hce, Adamcewicz:2023mov}.
The prior on $\kappa_{x,y}$ (which is the same for all copula model variations), is given in Table~\ref{tab:parameters_correlations}.

Copulas are advantageous as they allow for a potential correlation to be quantified by a single variable $\kappa_{x,y}$, that otherwise has no influence on the distribution of the population.
However, relative to the linear and spline models explored below, they suffer from a lack of flexibility when it comes to covariance.
There are a limited number of copula density functions available, all of which introduce a correlation in a unique, but rigid way \citep[e.g,][]{Adamcewicz:2022hce, Adamcewicz:2023mov}.
Furthermore, known two-dimensional copulas depend on a single parameter $\kappa_{x,y}$, and cannot infer, for example, a separable broadening and correlation with the mean of a distribution simultaneously.
As a result, using copulas to probe for more complex structure in two-dimensions (assuming fixed marginal distributions), requires model comparison between a number of different copula density functions \citep[e.g.,][]{Adamcewicz:2023mov}.

\subsection{Linear Correlation Models}\label{appendix:LinearCorrelationModels}
The $(q,\chieff)$ and $(z,\chieff)$ \linearcorrelation~models begin by assuming that $\chieff$ is Gaussian distributed for any given value of mass ratio $q$ and redshift $z$ respectively.
As the two models are otherwise identical, for the remainder of this Section, we substitute $x$ for $q$ and $z$.
This parameter-dependent Gaussian distribution for $\chieff$, $\pi(\chieff|x)$, is truncated at unphysical values of $|\chieff| \geq 1$.
From here, the mean and (natural log) width of the $\chieff$ distribution are allowed to evolve with the chosen variable $x$ linearly:
\begin{equation}
    \mu_\mathrm{eff}(x) = \mu_{\mathrm{eff}|x} + \delta \mu_{\mathrm{eff}|x} x,
\end{equation}
and
\begin{equation}
    \ln \sigma_\mathrm{eff}(x) = \ln \sigma_{\mathrm{eff}|x} + \delta \ln \sigma_{\mathrm{eff}|x_0} x.
\end{equation}
Here, $\mu_{\mathrm{eff}|x}$, $\delta \mu_{\mathrm{eff}|x}$, $\ln \sigma_{\mathrm{eff}|x}$, and $\delta \ln \sigma_{\mathrm{eff}|x_0}$ are all hyperparameters fit to the data, where $\delta \mu_{\mathrm{eff}|x}$ and $\delta \ln \sigma_{\mathrm{eff}|x_0}$ quantify the strength of the correlation between $x$ and the mean and width of the $\chieff$ distribution respectively.
The priors for these parameters are given in Table~\ref{tab:parameters_correlations}.
Meanwhile, masses follow the default \BrokenPLTwoPeaks~model and redshift is distributed according to the default power-law redshift model.

These models \citep[in the style of those presented in][]{Safarzadeh:2020mlb, Callister:2021fpo, Biscoveanu:2022qac}, therefore allow us to infer separable trends in the mean and width of the $\chieff$ distribution with another parameter.
The inferred correlations should be interpreted with care as, unlike the copula models defined in Section~\ref{appendix:copulaCorrelationModels}, the correlation hyperparameters here will also affect the shape of the marginal $\chieff$ distribution.
As such, it can be difficult to be certain whether an inferred correlation is entirely due to a trend between $\chieff$ and another parameter, or is in part due to a better fit to the marginal $\chieff$ distribution \citep{Adamcewicz:2022hce}.

\subsection{Spline Correlation Models}\label{appendix:SplineCorrelationModels}
The \splinecorrelation~models~\citep{Heinzel:2023hlb} are similar to the \linearcorrelation~models defined in Section~\ref{appendix:LinearCorrelationModels}.
The key difference is that the \splinecorrelation~models more flexibly model the parameter-dependent mean $\mu_\mathrm{eff}(x)$ and (natural log) width $\ln \sigma_\mathrm{eff}(x)$ of the $\chieff$ distribution as cubic splines dependent on $x$.
Each spline has four nodes $\mu_\mathrm{eff}^i$ and $\ln \sigma_{\mathrm{eff}|x}^i$ that are placed uniformly in $\log_{10} x$, and are inferred from the data.
The priors for these nodes are given in Table~\ref{tab:parameters_correlations}.

%% file: appendix_non_parametric_models.tex
We describe \nonparametrics---\BSpline, \ac{BGP}, \ac{AR}, \vamana---used in this work. Results in the main text make use of \BSpline and \ac{BGP} models. \ac{BGP} was chosen since it was the only model capable of modeling the full mass spectrum, while \BSpline was chosen for most \ac{BBH} analyses for its ease of description and flexibility, as it simultaneously models all parameters with B-Splines. See Appendix~\ref{appendix:non_parameteric_comparison} for comparison of these \nonparametrics.
\subsection{\BSpline models} \label{appendix:BSpline}

\BSpline: The \BSpline model~\citep{Edelman:2022ydv} simultaneously fits all population parameters with Basis-splines (B-splines). 
A $k^\mathrm{th}$ order B-spline of variable $x$ consists of a linear combination of $n$ basis functions $\{B_{k,n}(x)\}$, each of which are degree $k-1$ piecewise polynomials joined at a set of $m$ locations called knots $\{x_m\}$. 
The number of basis functions $n$ defined across any range of $x$ values is determined solely by the order of the spline $k$ and the total number of knots $m$ as $n = k + m$.
This forms a basis that spans the space of possible interpolants between the knots $\{x_m\}$. Given a vector of coefficients $\pmb{\alpha}$, any function $f$ can then be approximated as 
\begin{equation}
\tilde{f}(x) = \sum^n_{i=1} B_{k,i}(x) \alpha_i.
\label{eqn:basis_splines}
\end{equation}
In the case of this work, $f$ is a probability distribution $p(\theta|\pmb{\alpha}_\theta)$ for a population-level parameter $\theta$, where the B-spline coefficients $\pmb{\alpha}_\theta$ are the hyperparameters of the distribution that are inferred during parameter estimation. 

Given an adequate number of knots, a B-spline is highly flexible and therefore capable of identifying sharp features that could be present in the data. 
This does, however, naturally make the \BSpline model prone to overfitting. 
To combat this, we include a smoothing prior $\pi_\text{BS}$ that penalizes large differences between neighboring coefficients. 
The prior is defined as 
\begin{equation}
p(\pmb{\alpha}|\tau) = \text{exp}\left( -\frac{1}{2} \tau \pmb{\alpha}^{\rm T} \pmb{D}_r^{\rm T} \pmb{D}_r\pmb{\alpha}\right),
\end{equation}
where $D_r$ is the $r$-order difference matrix with shape $(n - r) \times n$ and $\tau$ is a scalar that controls the level of smoothing. 
Ideally, $\tau$ is also inferred during parameter estimation. We found that $\tau$ consistently railed against prior boundaries, which meant that the limits imposed on the likelihood uncertainty was the main driver of smoothness. We therefore fix $\tau$ to a reasonable value (roughly between $5{-}10$) for each population distribution. 
With a sufficient number of bases, typically $n \sim 30{-}40$, this penalty prior will prevent the B-spline from overfitting the data while the large number of bases will provide enough flexibility to fit sharp features. 

The mass and spin distributions are modeled as B-splines, and the redshift distribution is modeled as a power law modulated by a B-spline \citep{Edelman:2022ydv}, with all components inferred simultaneously.
One of the \nonparametrics used in previous analyses~\citep{KAGRA:2021duu} was the \textsc{Powerlaw + Spline} model. This model assumed that the primary BBH mass followed a power-law distribution with moderate deviations controlled by a cubic spline.
The \BSpline model does not assume an underlying shape for the primary mass distribution, instead allowing for full model flexibility.

The \BSpline model infers the separable components of the mass distributions, $p(m_1)$ and $p(q)$, wherein the primary mass is defined over the range $3{-}300 \, \Msun$ and the mass ratio is defined over the range $0.03{-}1$. Unlike the \BrokenPLTwoPeaks model, a minimum secondary mass is not enforced during parameter estimation. To provide a more direct comparison to the \parametric, the \BSpline mass distributions shown in Section \ref{subsec:bbh_masses} are not the separable components $p(m_1)$ or $p(q)$ but instead the marginal distributions conditioned on $m_2 > 3 \, \Msun$, that is, $p(q|m_2 > 3 \, \Msun) = \int p(q)p(m_1)\Theta(m_1q-3)\mathrm{d}m_1$. We include the separable distributions along with the conditional marginal distributions in the mass ratio plot in Figure \ref{fig:bbh_mass_non_parametrics} to illustrate how this assumption affects the shape of the mass ratio distribution. 

\textsc{Isolated Peak}: This model is defined by two subpopulations~\citep{Godfrey:2023oxb}. One assumes a primary mass distribution described by a log-Gaussian peak while the other infers the mass distribution with a B-spline. Mass ratio, spin magnitude, and spin tilt distributions are inferred separately for each subpopulation, also using B-splines. The redshift distribution is the same for each subpopulation and is inferred with a power law modulated by a B-spline. 

\subsection{\acf{BGP} Model}\label{appendix:BGPModel}

In previous population analyses~\citep{KAGRA:2021duu}, we used the \ac{BGP} model to study the joint distribution of primary and secondary masses. Here, we extend it to model the joint distributions of mass, spin, and redshift.
The \ac{BGP} approach models the rate of \ac{BBH}, \ac{BNS} and \ac{NSBH} mergers as a piecewise constant function over a set of fixed bins across the one-, two-, or three-dimensional joint space. 
The comoving merger rate density in each bin is a hyperparameter of the model.
In addition, the \ac{BGP} model couples the logarithm of the rate density in each bin with a Gaussian process covariance, assuming an exponential quadratic kernel (also known as a radial basis function (RBF) kernel). 
The exponential quadratic kernel has a hyper-hyperparameter length scale $\lambda$ for each parameter in the joint space.

The covariance between two bins is proportional to the exponential of the negative squared distance between the bin centers, in units of the length scales along each parameter.
In addition, there is one more hyper-hyperparameter $\sigma$, which acts as an overall multiplicative scaling of the covariance matrix~\citep{Ray:2023upk}.
The \ac{BGP} approach directly infers the rate density in each bin, as well as the hyper-hyperparameters of the covariance kernel.
The prior on the parameter $\sigma$ is a half-normal with width $1$, and the length scale priors were tuned to the mean and variation in the set of distances between the bin centers.

We used 22 bins spaced uniformly in log-mass for the mass \ac{BGP} models, and 15 bins spaced uniformly between $\chi_{\rm eff} \in [-0.7,0.7]$ for the effective spin \ac{BGP} models.

The \ac{BGP} approach has the advantage of being able to model nontrivial correlations in the population of compact binaries. 
However, it assumes an arbitrary binning scheme, which reduces the resolution of the constraints and leads to unphysical discontinuities at the boundary between bins.
Furthermore, measurements of the rate density in some bins may be driven by the \textit{a priori} assumption of a Gaussian covariance, and not the data. This can also cause smoothing near sharp features e.g., near the $m_1=m_2$ boundary.

\subsection{\acf{AR} Model}

The \ac{AR} model \citep{Callister:2023tgi} is a highly flexible model for one dimensional marginal merger rate densities. 
The Monte Carlo integrals for estimating the likelihood in Equation~\eqref{eq:hierarchical_likelihood} involve a large set of parameter estimation samples [Equation~\eqref{eq:single_event_estimators}] and found injections for the selection efficiency [Equation~\eqref{eq:selection_estimator}].
The merger rate density at each sample is its own hyperparameter, which is directly inferred from the data. 
Without any further assumptions, such a model is severely underconstrained and will converge on the maximum likelihood functional distribution~\citep{Payne:2022xan}.
In order to \textit{a priori} favor smoother distributions, this finite list of samples is ordered along the dimension of interest like, for example, their primary mass. 
The marginal merger rate density is then assumed to be an autoregressive Gaussian random walk in log-space.
There are two hyper-hyperparameters, $\sigma_{\rm AR}$ and $\tau_{\rm AR}$, which control the scale of variability and the autocorrelation length along the Gaussian random walk respectively.
These hyper-hyperparameters have half-normal and log-normal priors respectively, tuned to the scale of the data: see Appendix B in \cite{Callister:2023tgi} for details. These hyper-hyperparameters are jointly inferred along with the logarithmic merger rate density at each sample.

The \ac{AR} model is particularly well-suited to distributions with sharp features and nontrivial evolution, complementing the \ac{BGP} approach. 
However, the \ac{AR} model used here cannot model correlations in the population. Furthermore, as in the \ac{BGP} approach, the inferred merger rate may extrapolate based off nearby constraints in regions of little information.
When evaluating uncertainties in the \ac{AR} model, one should consider the impact of the \ac{AR} smoothing prior. 
In particular, lower bounds on the merger rate in some regions may not be an accurate lower bound on the true distribution.

Like the \BSpline model, the \ac{AR} model does not enforce a minimum secondary mass through a mass-dependent mass ratio.

\subsection{\acf{FM} Model}

The \vamana model \citep{Tiwari:2020vym} is a flexible mixture model framework for modeling the \ac{BBH} population. 
\vamana models the population as a Gaussian mixture model in chirp mass and aligned spins, and a power law in mass ratio and redshift \citep{Tiwari:2021yvr}. 
Furthermore, \vamana is able to model correlations between parameters in the population, and additionally has a variable model dimension, sampling the trans-dimensional posterior using a reversible-jump MCMC method. For the analyses presented here, \vamana uses $11$ mixture components.

%% file: appendix_validation_studies.tex
In the following sub-appendices, we describe various methods used to select and validate the data, models, and approaches presented in the main text. 
In Appendix~\ref{appendix:model_comparison_mass} and Appendix~\ref{appendix:model_comparison_spins} we discuss the process for selecting the fiducial mass and spin models, respectively, under the \parametric and present results from other models which were under consideration.
Appendix~\ref{appendix:model_comparison_effective_spins} describes how the specific convergence cut biases the inferred effective spin distribution. 
In Appendix~\ref{appendix:non_parameteric_comparison}, we compare various types of methods for the \nonparametric, which were presented in Appendix~\ref{appendix:non_parametric_models_summary}.
Appendix~\ref{appendix:FGMC_rates} gives compact object merger rates when sub-threshold triggers are considered, i.e., using a lower~\ac{FAR} threshold. 
Finally, Appendix~\ref{appendix:supplementary_correlations} provides supplementary results for population-level correlations between masses, spins, and redshifts.

\begin{deluxetable}{lcc}
\tablecaption{Bayes factor comparison of selected models from the mass model comparison study.}
\tablehead{
    \colhead{Model} & \colhead{Abbreviation} & \colhead{$\log_{10}\mathcal{B}$}
}
\label{tab:model_comparison_mass}
\startdata
\BrokenPLTwoPeaks $\bigstar$ & \BPTP & $0$ \\
\tableline
\textsc{Broken Power Law + 1 Peak} & \BPOP & $\BBHMassModelBayesFactors[pl2pk1]$ \\
\textsc{Broken Power Law + 3 Peaks} & \BPHP & $\BBHMassModelBayesFactors[pl2pk3]$ \\
\textsc{Power Law + Peak} & \PP & $\BBHMassModelBayesFactors[plp]$ \\
\enddata
\end{deluxetable}

\begin{figure*}[t]
	\centering
	  \includegraphics[width=\textwidth]{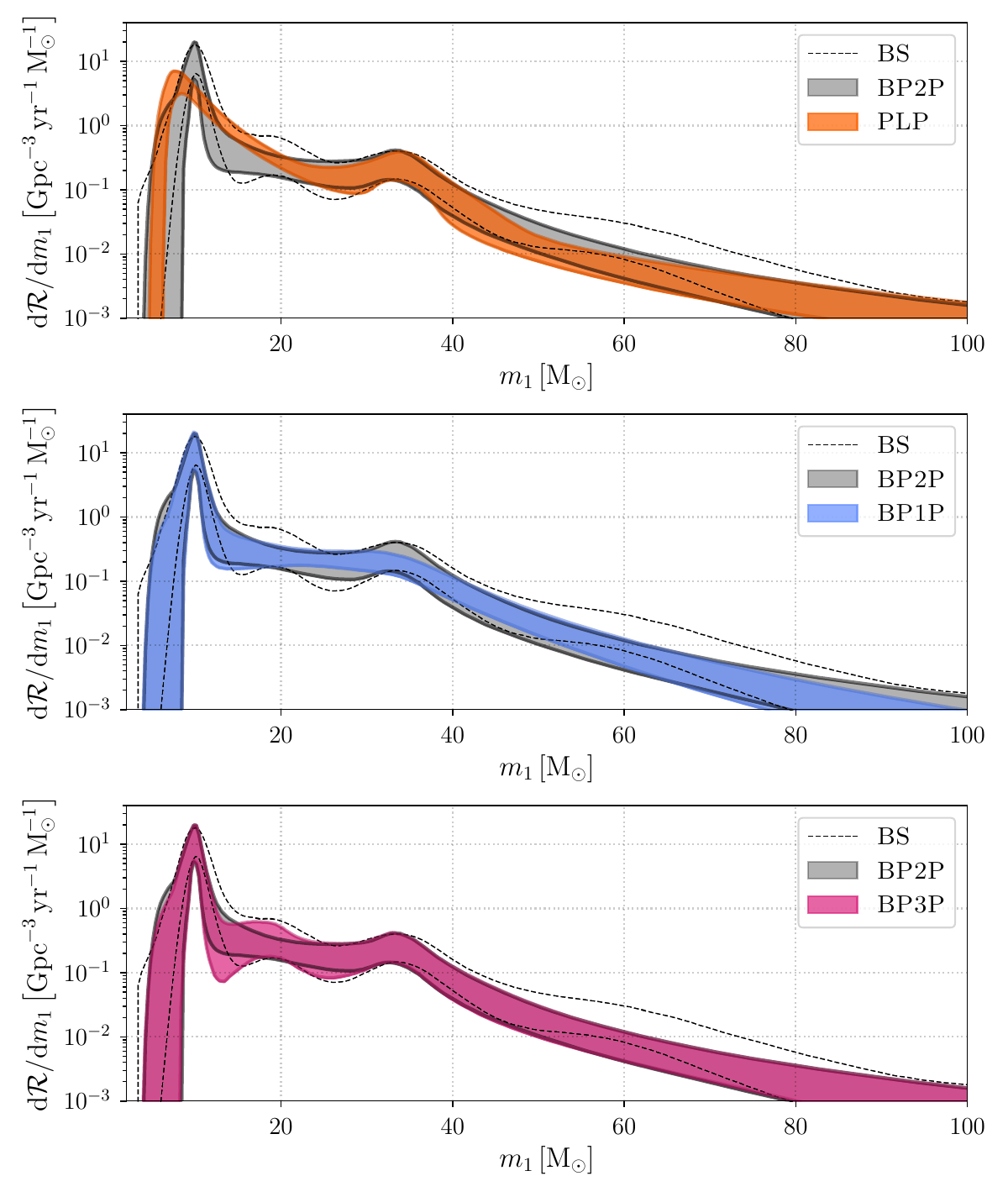}
	  \caption{Inferred mass distributions of various \parametrics compared to the fiducial mass model \BrokenPLTwoPeaks (gray shaded) and 
	  the \BSpline model (black dashed). The fiducial model from \gwtcthree, \PLPeak, is shown in orange, a broken power law with a single peak in blue, and a broken power law with three peaks in pink.}
	  \label{fig:mass_model_comp_1}
  \end{figure*}

\subsection{Model Comparison Study: Mass}\label{appendix:model_comparison_mass}

\begin{figure}
	\centering
	  \includegraphics[width=120mm]{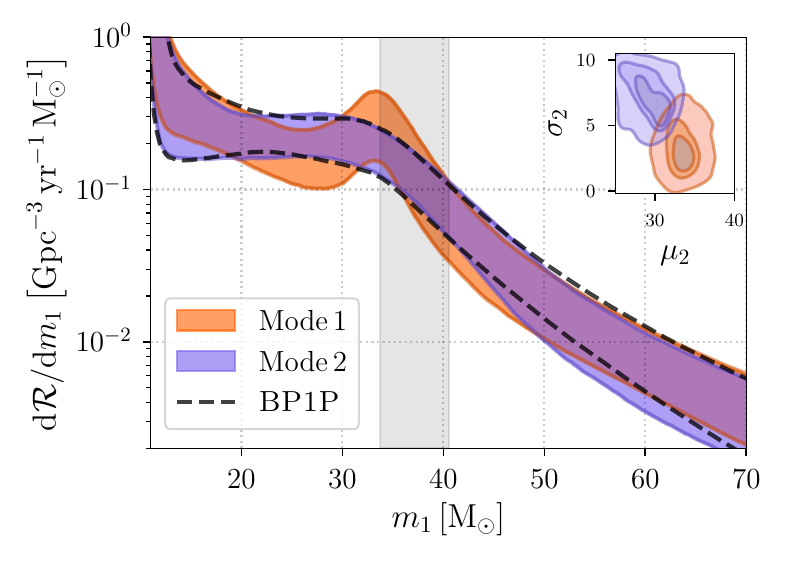}
	  \caption{Differential merger rate as a function of primary mass (evaluated at $z=0.2$) of the two different modes recovered by the \BrokenPLTwoPeaks model. The orange shaded region shows the 90\% credible interval for the dominant mode ($\BBHMassModelBayesFactors[default_dominant_mode]\%$ of posterior), reflecting a distinct peak at $35 \,\Msun$, and the purple shaded region shows the $90\%$ credible interval for the subdominant mode ($\BBHMassModelBayesFactors[default_subdominant_mode] \%$ of posterior), reflecting a broken power law morphology without a distinct $35 \,\Msun$ peak. The black dashed lines show the $90\%$ credible bounds of the \textsc{Broken Power Law + 1 Peak} model for comparison. The inset figure shows the joint posterior of the peak mean $\mu_2$ and width $\sigma_2$ for each mode. The vertical grey shaded region indicates the $90\%$ credible interval of the sum $\mu_2 + \sigma_2$, which is consistent between both modes.}
	  \label{fig:primary_mass_modes}
  \end{figure}

The fiducial mass model is a \parametric that is intended to provide a minimal but accurate description of the data with a parametrization that is more readily interpretable than the \nonparametrics.  Table \ref{tab:model_comparison_mass} shows the Bayes factors between the models that performed the best in our study and our fiducial model \BrokenPLTwoPeaks. We also compare to our previous fiducial model from \citet{KAGRA:2021duu}, the \PLPeak model. In Figure \ref{fig:mass_model_comp_1}, we see that \PLPeak infers the peak component at $35\Msun$, as in \citet{KAGRA:2021duu}, but a broken power law plus 1 peak (\BPOP) model infers the peak component at $10\Msun$ and the power law break at $35\Msun$. In Table \ref{tab:model_comparison_mass}, we see that all models that include a broken power law are strongly favored over the \PLPeak model. Among models with a broken power law, the Bayes factors do not show strong support for one model over another. While Bayes factors are an important model comparison statistic to consider, they can be influenced by prior assumptions. In the case of the models in Table \ref{tab:model_comparison_mass} where Bayes factors are not decisive, we ultimately chose as our fiducial model the model that was a minimal extension of the \PLPeak model and probed features shared among all of the \nonparametrics shown in Figure \ref{fig:bbh_mass_non_parametrics}. The \BrokenPLTwoPeaks model best fit these criteria, though with some nuances that we discuss below.

In \gwtcthree, the \PLPeak model identified an overdensity in the merger rate at $m_1 =$ \muTOT relative to a global power law. Evidence for this feature first appeared in \citet{LIGOScientific:2018jsj} and was strengthened in \citet{LIGOScientific:2020kqk} and \citet{KAGRA:2021duu}. As mentioned in Section \ref{subsec:bbh_masses}, this feature at $35\Msun$ may be an overdensity relative to an underlying mass continuum or could mark the onset of a decline in the merger rate. In addition to the Bayes factor comparison shown in Table \ref{tab:model_comparison_mass}, this ambiguity is also present in the \BrokenPLTwoPeaks posterior, which includes two modes that correspond to two different morphologies, shown in Figure \ref{fig:primary_mass_modes}. The dominant mode is correlated with a narrower ($\sigma_2 = \CIPlusMinus{\DefaultBBHDominantMode[sigpp_2]} \,\Msun$) peak at $\mu_2 = \CIPlusMinus{\DefaultBBHDominantMode[mpp_2]} \,\Msun$ and accounts for $\BBHMassModelBayesFactors[default_dominant_mode] \%$ of the posterior volume, while the subdominant mode is correlated with a wider ($\sigma_2 = \CIPlusMinus{\DefaultBBHSubdominantMode[sigpp_2]} \,\Msun$) peak at $\mu_2 = \CIPlusMinus{\DefaultBBHSubdominantMode[mpp_2]} \,\Msun$ and accounts for $\BBHMassModelBayesFactors[default_subdominant_mode]\%$ of the posterior volume. This latter mode has a morphology nearly identical to the inferred \BPOP distribution (see Figure \ref{fig:primary_mass_modes}). The inflection point of the Gaussian, $\mu_2 + \sigma_2 = \CIPlusMinus{\defaultbbh[mu_plus_sig]} \,\Msun$, is consistent between both modes, which indicates that the right half of the Gaussian must fall below the power law to match the declining rate in this region (the gray vertical shaded region in Figure \ref{fig:primary_mass_modes}).

The power law indices, $\alpha_1$ and $\alpha_2$ are not correlated with the two modes present in the \BPTP posterior, though the measured slope of the mass distribution between $18{-}19 \,\Msun$ is not equivalent to $\alpha_1$ in the subdominant mode. In this mode, the slope (calculated by finite difference) is $\CIPlusMinus{\DefaultBBHSubdominantMode[mass_1][powerlaw_slope_18-19]}$, which includes more support for negative values compared to the inferred power law index, $\alpha_1 = \CIPlusMinus{\DefaultBBHSubdominantMode[alpha_1]}$. We infer the power law break location at $m_\text{break} = \CIPlusMinus{\defaultbbh[break_mass]} \,\Msun$. The \BPOP model, which does not include a second Gaussian component, measures the break location much more precisely at $m_\text{break} = \CIPlusMinus{\PowerLawTwoPeakOneBBH[break_mass]} \,\Msun$. The uncertainty in the \BPTP power law break is due to the presence of the second Gaussian component, which likely obscures the break most of the time. Because the peak is narrower in the dominant \BPTP mode, one might expect the break location to be better constrained in this mode; however, this is not the case. The break location does not appear to be correlated with either mode. 

\subsection{Model Comparison Study: Spin Magnitudes and Tilt Angles}\label{appendix:model_comparison_spins}

\begin{deluxetable}{llllc}
\tablecaption{
    Comparison of different parametric models for spin magnitudes $\chi_i$ and tilt angles $\cos\theta_i$.
}
\tablehead{
    \colhead{$\chi_i$ model} & \colhead{} & \colhead{$\cos\theta_i$ model} & \colhead{} & \colhead{$\log_{10}\mathcal{B}$}
}
\label{tab:modelComparisonSpins}
\startdata
Truncated Gaussian $\bigstar$ & IID & Isotropic + Truncated Gaussian & NID & $\spinModelComparisonLogTenBayesFactors[MagTruncnormIidTiltIsotropicTruncnormNid][log10_bayes_factor_over_new_default]$ \\
Constrained Beta & IID & Isotropic + Aligned Gaussian & NID & $\spinModelComparisonLogTenBayesFactors[MagBetaConstrainedIidTiltIsotropicAlignedNid][log10_bayes_factor_over_new_default]$ \\
\tableline
Constrained Beta & IID & Isotropic + Truncated Gaussian & NID & $\spinModelComparisonLogTenBayesFactors[MagBetaConstrainedIidTiltIsotropicTruncnormNid][log10_bayes_factor_over_new_default]$ \\
Unconstrained Beta & IID & Isotropic + Truncated Gaussian & NID & $\spinModelComparisonLogTenBayesFactors[MagBetaUnconstrainedIidTiltIsotropicTruncnormNid][log10_bayes_factor_over_new_default]$ \\
Truncated Gaussian & IID & Isotropic + Aligned Gaussian & NID & $\spinModelComparisonLogTenBayesFactors[MagTruncnormIidTiltIsotropicAlignedNid][log10_bayes_factor_over_new_default]$ \\
\tableline
Truncated Gaussian & IND & Isotropic + Truncated Gaussian & NID & $\spinModelComparisonLogTenBayesFactors[MagTruncnormIndTiltIsotropicTruncnormNid][log10_bayes_factor_over_new_default]$ \\
Truncated Gaussian & IID & Isotropic + Truncated Gaussian & NND & $\spinModelComparisonLogTenBayesFactors[MagTruncnormIidTiltIsotropicTruncnormNnd][log10_bayes_factor_over_new_default]$\\
Truncated Gaussian & IND & Isotropic + Truncated Gaussian& NND & $\spinModelComparisonLogTenBayesFactors[MagTruncnormIndTiltIsotropicTruncnormNnd][log10_bayes_factor_over_new_default]$ \\
\enddata
\tablecomments{The first row is the Default model for \gwtcfour (\default), while the second is the Default that was used for \gwtcthree and earlier~\citep{LIGOScientific:2018jsj,LIGOScientific:2020kqk,KAGRA:2021duu,Talbot:2017yur}.
The third through final are other models explored. All log Bayes factors ($\log_{10}\mathcal{B}$) are with respect to the first row.}
\end{deluxetable}

We consider two additional spin magnitude models beyond the \default~model from Section~\ref{subsubsec:bbh_spin_mag_and_tilts}: the Constrained Beta and Unconstrained Beta models, both defined by Equation~\eqref{eqn:beta_distribution}. 
We also consider one additional spin tilt model, the Isotropic + Aligned Gaussian, which fixes $\mu_t = 1$ in the \default~model and served as the {\sc Default} in \gwtcthree~\citep{Talbot:2017yur,KAGRA:2021duu}.
The first five rows of Table~\ref{tab:modelComparisonSpins} give the log Bayes factors between these models.
Within the \default~model, we also test whether the primary and secondary spins are identically and independently distributed (see Appendix~\ref{appendix:iid}), with log Bayes factors given in the bottom three rows of Table~\ref{tab:modelComparisonSpins}.
The \default~model with IID spin magnitudes and NID spin tilts performs the best, and is adopted as the default component spin model in the main text.

\begin{figure*}
	\centering
	\includegraphics[width=0.8\linewidth]{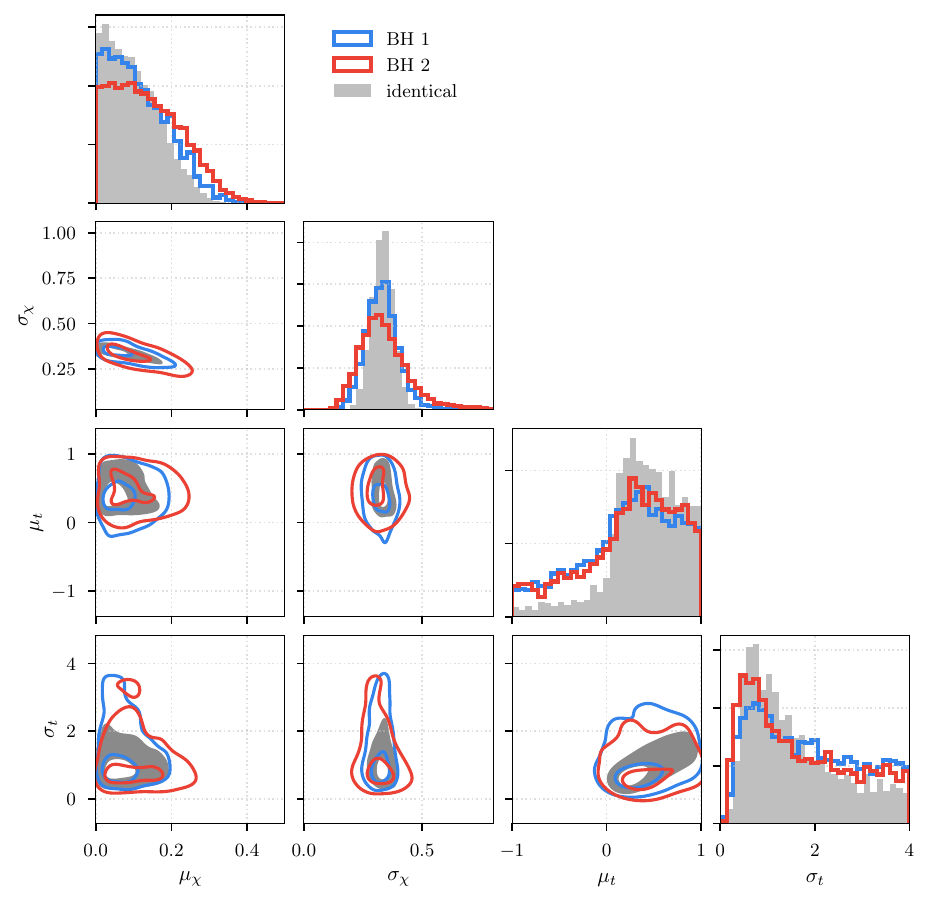}
	\caption{Distribution of \default~hyperparameters for the primary \ac{BH} (blue) and secondary \ac{BH} (red) when neither magnitudes nor tilts are assumed identical (last row of Table~\ref{tab:modelComparisonSpins}), compared to when they are both assumed identical (gray, first row of Table~\ref{tab:modelComparisonSpins}).
	The contours of the two dimensional distributions mark the 50th and 90th percentiles.}
	\label{fig:spin_mags_tilts_comparison_cornerplot}
  \end{figure*}

Figure \ref{fig:spin_mags_tilts_comparison_cornerplot} shows posteriors for the magnitude and tilt hyperparameters assuming that $\chi_i$ and $\cos\theta_i$ are both identically distributed (gray) versus both non-identically distributed (blue and red). 
The primary and secondary spins have consistent hyperparameter distributions.
Assuming identical distribution makes the hyperparameters more-precisely constrained---yielding tighter 90\% bands on the resultant population distributions---but does not affect the overall shape of the distributions. 
The only difference of note is that $\chi_2$ has less support for a small $\mu_\chi$ than $\chi_1$, potentially hinting at more highly spinning secondary BHs on a population level. 

\subsection{Dependence of the Effective Spin Distribution on the Likelihood Variance Cut}\label{appendix:model_comparison_effective_spins}

We next investigate the effect of the likelihood variance threshold (see Appendix~\ref{sec:likelihood-estimator-variance}) on our \parametricAdj effective spin results. 
To ensure that posteriors generated using Monte Carlo estimation are trustworthy, i.e., the likelihood estimate is converged, we enforce that all hyperparameter samples yield a log-likelihood variance ${\sigma^2}_{\ln\cal \hat L} < 1$ for all analyses presented in the main text~\citep{Talbot:2023pex}.
In \gwtcthree, however, a different quantity was used to assess the Monte Carlo uncertainty: the effective number of independent samples, $N_{\rm eff}$~\citep{Farr:2019rap}. 
It was there imposed that $N_{\rm eff}$ be greater than $4 N_{\rm events}$ for the sensitivity injections used in Equation~\eqref{eq:selection_efficiency} and greater than 10 for every event in the catalog of CBCs. 
This $N_{\rm eff}$ cut is generally less stringent than the approach taken for \gwtcfour. 
The $\chip$ distribution is sensitive to which method is used to cut out samples.

\begin{figure*}
	\centering
	\includegraphics[width=\linewidth]{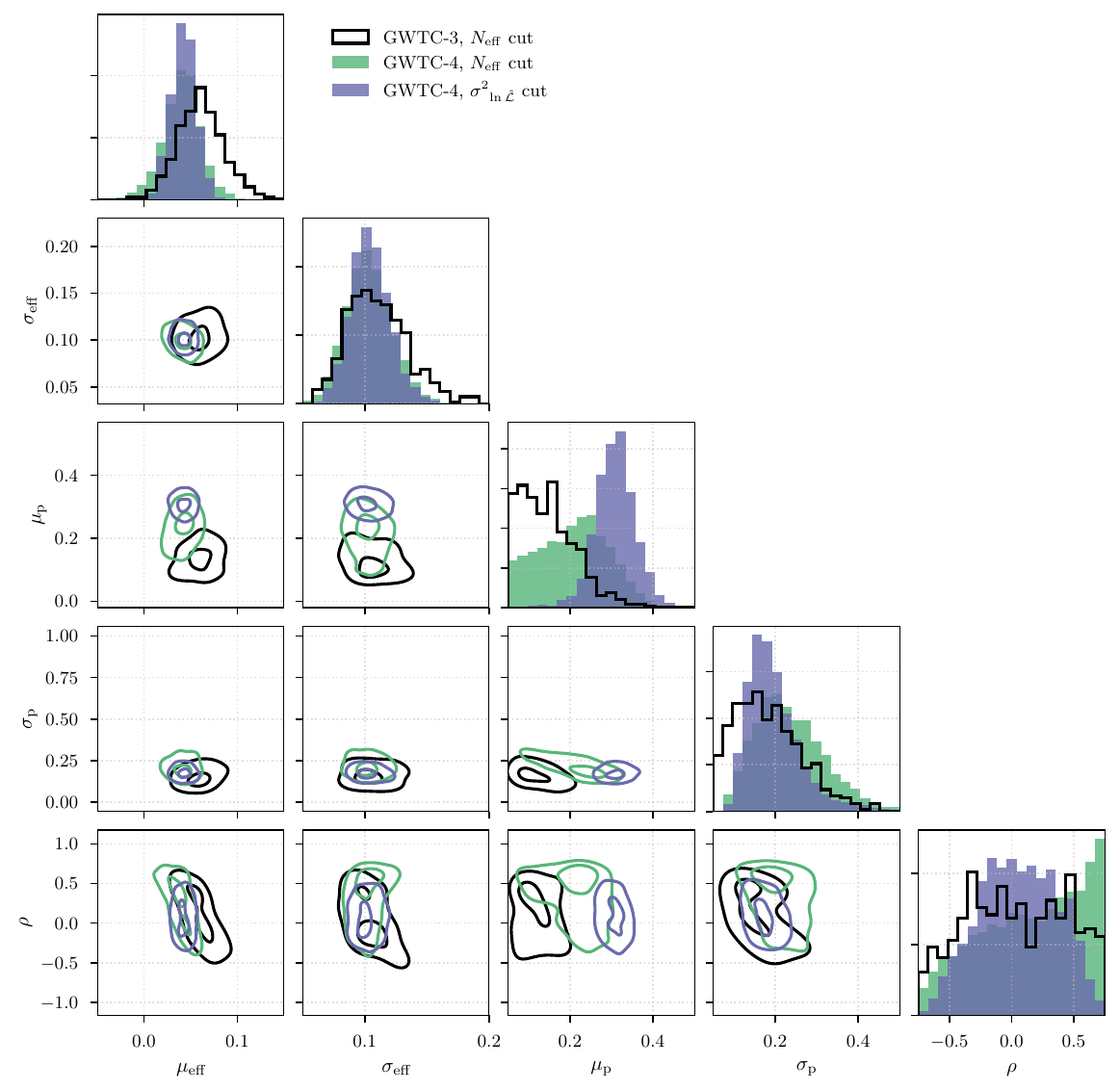}
	\caption{
	Posterior for the \effectiveSpinModel model hyperparameters, as defined in Table~\ref{tab:effectiveSpinPriors}, for \gwtcfour under two methods of cutting out samples that may lead to an unconverged likelihood.
	Posteriors excluding samples with a substantially large log-likelhood variance (${\sigma^2}_{\ln\cal \hat L}$ cut) are shown in purple; those excluding samples with a substantially small number of effective samples ($N_{\rm eff}$ cut) are in green.
	In black are the \gwtcthree results with the $N_{\rm eff}$ cut for comparison.
	The $\chieff$ hyperparameters are not affected by the cuts, while the $\chip$ and joint-distribution hyperparameters are.
	The contours of the two dimensional distributions mark the 50th and 90th percentiles.}
	\label{fig:effective_spins_comparison_cornerplot}
  \end{figure*}

\begin{figure*}
	\centering
	\includegraphics[width=\linewidth]{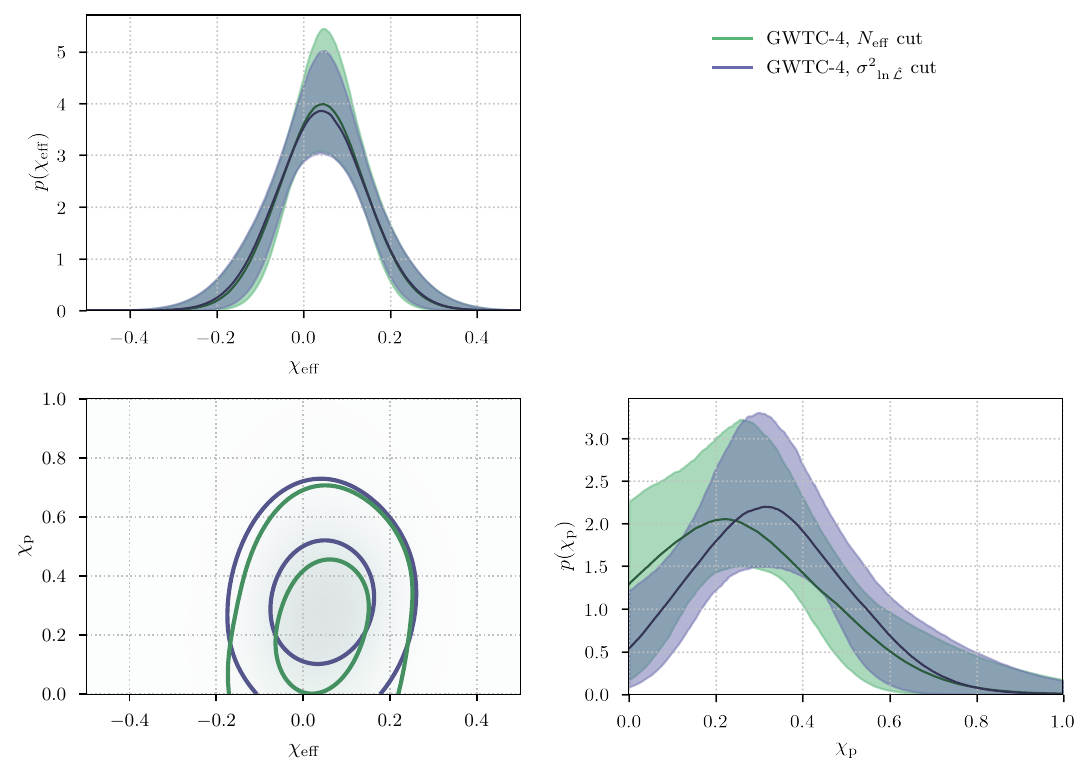}
	\caption{
	Marginal and joint $\chieff$ and $\chip$ distributions under the \effectiveSpinModel model using two methods of cutting out hyperparameter samples that may lead to an unconverged likelihood (see Figure~\ref{fig:effective_spins_comparison_cornerplot}).
	The marginal distributions show the median (solid line) and 90\% credible interval on the probability density for each population distribution (shaded).
	The joint distribution is the \ac{PPD} with contours marking the 50th and 90th quantiles.}
	\label{fig:effective_spins_comparison_dists}
  \end{figure*}

For the most direct comparison to \gwtcthree, we here use the \effectiveSpinModel model.
Figure~\ref{fig:effective_spins_comparison_cornerplot} shows the posteriors for the \effectiveSpinModel parameters where the ${\sigma^2}_{\ln\cal \hat L}$ (purple) versus  $N_{\rm eff}$ (green) cuts are done on the \gwtcfour results. 
The LVK \gwtcthree results (with the $N_{\rm eff}$ cut) are shown in comparison (black). 
Figure~\ref{fig:effective_spins_comparison_dists} shows the resultant marginal and joint $\chieff$--$\chip$ population distributions. 
There are three main differences between the results with the two cuts on \gwtcfour data: 
\begin{enumerate}
	\item The ${\sigma^2}_{\ln\cal \hat L}$ cut yields a $\mu_{\rm p}$ posterior which is constrained away from zero, while the the $N_{\rm eff}$ cut does not.
	\item The $N_{\rm eff}$ cut allows for wider $\chip$ distributions (larger $\sigma_{\rm p}$) than the ${\sigma^2}_{\ln\cal \hat L}$ cut. 
	\item Under the $N_{\rm eff}$ cut, $\chieff$ and $\chip$ are preferentially positively correlated, while under the ${\sigma^2}_{\ln\cal \hat L}$ cut we remain agnostic, with preference for small-to-zero correlation.
\end{enumerate}
Thus, the claims that the $\chip$ distribution does not peak at zero, and that the data prefer $\chieff$ and $\chip$ being uncorrelated at the population level, are not necessarily astrophysical in origin.
Rather, they are driven by regions of the parameter-space that our events and sensitivity injections allow us reliably probe with Monte Carlo likelihood estimators. 

The particular sensitivity of the $\chip$ distribution to these cuts can be at-least partly attributed to the use of a uniform and isotropic spin prior in \gwtcfour~parameter estimation, which induces a $\chip$ prior that goes to $0$ at $\chip = 0,1$.
Thus, Equation~\eqref{eq:hierarchical_likelihood} is prone to yield a small $N_{\rm eff}$ and/or large ${\sigma^2}_{\ln\cal \hat L}$ when $\chip$ is near-minimal or near-maximal.
Using alternative spin priors in parameter estimation could aid in getting more effective samples at small $\chip$, improving our ability to probe the presence or absence of negligibly precessing \acp{BH} at the population level.
The Monte Carlo uncertainty from working with a finite number of samples to estimate the likelihood will only grow more problematic as the number of observed \acp{CBC} increases~\citep{Talbot:2023pex}; development of mitigation techniques is an area of active research~\citep[e.g.,][]{Gerosa:2020pgy,Rinaldi:2021bhm,Talbot:2020oeu,Callister:2022qwb, Talbot:2023pex, Leyde:2023iof,Hussain:2024qzl,Callister:2024qyq,Lorenzo-Medina:2024opt,Mancarella:2025uat}.

\subsection{Comparison of Weakly Modeled Approach Results}\label{appendix:non_parameteric_comparison}

\begin{figure*}[t]
	\centering
	  \includegraphics[width=\textwidth]{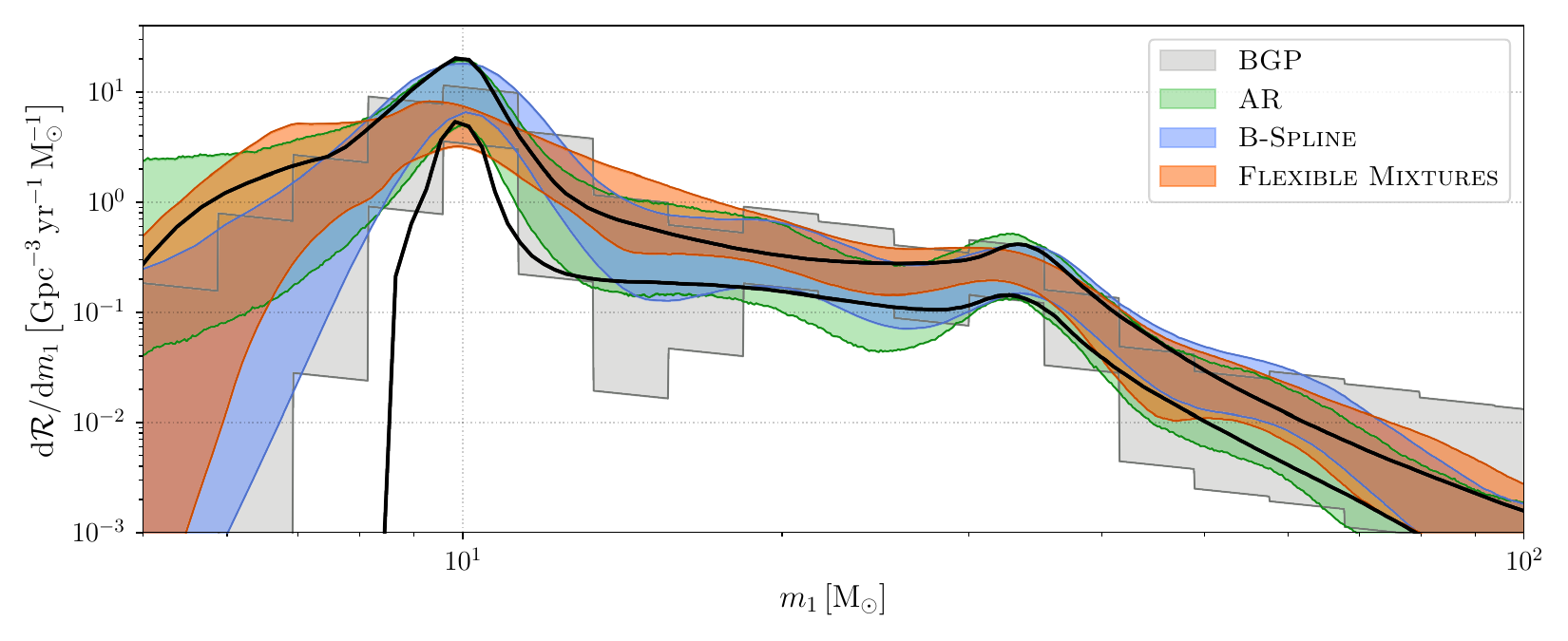}
	  \includegraphics[scale=0.7]{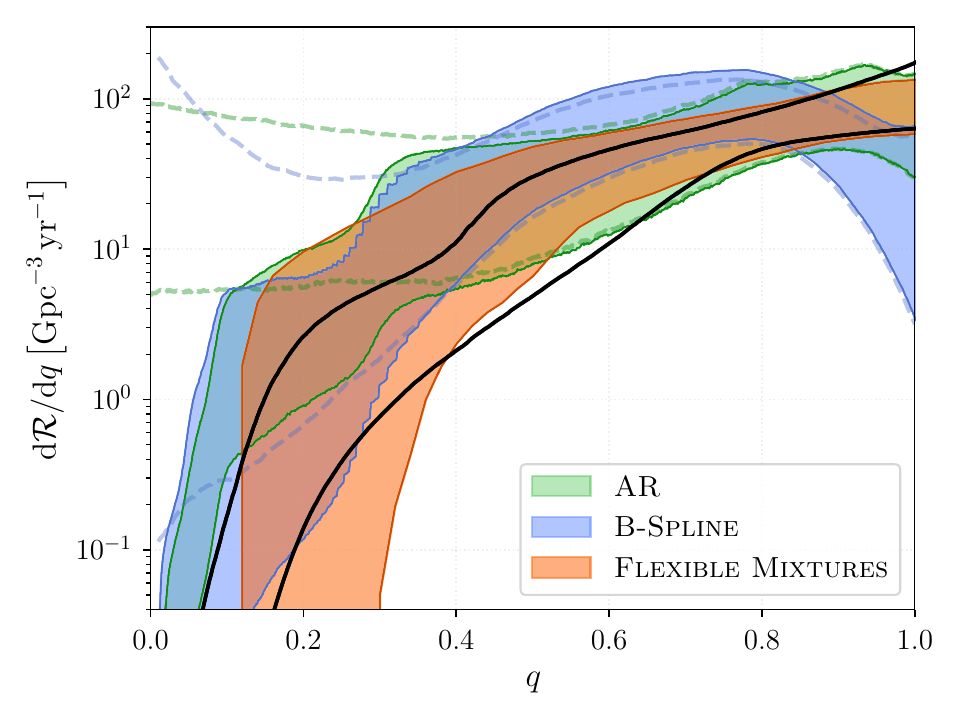}
	  \caption{(\textit{Top}) Primary mass distributions using the \nonparametrics outlined in Appendix \ref{appendix:non_parametric_models_summary}. The \BrokenPLTwoPeaks result is shown in black for comparison. All distributions show rates evaluated at $z=0.2$, except for BGP, which shows the rate evaluated on the $z=0.1-0.25$ bin. Note that Flexible Mixtures does not infer primary mass directly, but instead derives it from the chirp mass Gaussian mixture model and the mass ratio as a power law distribution. The lack of substructure in the Flexible Mixtures mass distribution relative to the other models is likely due to model misspecification from assuming a power law mass ratio model. (\textit{Bottom}) Mass ratio distributions using the \nonparametrics outlined in Appendix \ref{appendix:non_parametric_models_summary}. The \BrokenPLTwoPeaks result is shown in black for comparison. Flexible Mixtures models the chirp mass as a Gaussian mixture model and the mass ratio as a power law. For the \BSpline and \ac{AR} models, the separable distribution $p(q)$ is shown by dashed lines and the conditional marginal distribution $p(q|m_2 > 3\Msun) = \int p(q)p(m_1)\Theta(m_1q-3){\rm d}m_1$ is shown by the shaded regions, highlighting that the low mass ratio truncation seen in the \parametric is largely a prior effect.}
	  \label{fig:bbh_mass_non_parametrics}
  \end{figure*}

\textit{Mass distributions}: We supplement the \parametric with various \nonparametrics. Figure \ref{fig:bbh_mass_non_parametrics} (top panel) shows the inferred primary mass distribution using the \BSpline, AR, Flexible Mixtures, and BGP models compared to the fiducial \BrokenPLTwoPeaks model. All models agree within their $90\%$ credible regions, though the Flexible Mixtures model is the only model that does not exhibit a prominent peak at ${\sim}10 \Msun$. The Flexible Mixtures model does not directly infer the primary mass, but instead it models the chirp mass as a Gaussian mixture model and the mass ratio with a power law. The primary mass distribution shown in Figure \ref{fig:bbh_mass_non_parametrics} is then derived from these two distributions. The discrepancy between this model and the other \nonparametrics could be due to model misspecification, i.e., assuming the mass ratio distribution follows a power law for all primary masses. The fact that the discrepancy exists primarily in the ${\sim}10 \Msun$ region supports the \textsc{Isolated Peak} result in Section \ref{subsec:bbh_ratio}, which suggests the ${\sim}10 \Msun$ peak disfavors equal mass mergers (i.e., is inconsistent with a power law in mass ratio) compared to higher mass \acp{BBH}. The mass ratio distributions inferred by the \BSpline, AR, and Flexible Mixtures is shown Figure \ref{fig:bbh_mass_non_parametrics} (bottom panel). The $90\%$ credible regions are consistent between each model. The \BSpline model infers a peak away from $q=1$, which is not present in the AR or Flexible Mixtures results. This may be due to the \BSpline model's greater flexibility compared to the AR and Flexible Mixtures models, as it infers all parameters simultaneously with B-Splines. The AR model assumes an auto-regressive process only in primary mass and mass ratio, while other parameters are inferred with the fiducial models listed in Table \ref{tab:summary_of_models}. As noted in Appendix \ref{appendix:BSpline}, the figure includes the marginal distributions conditioned on $m_2 > 3\Msun$ from the \BSpline and AR models in addition to the fully separable components (dashed lines). The two distributions are essentially identical above $q \sim0.4$, below which the marginal distribution falls off sharply. This implies that the behavior of the \parametric below $q \sim 0.4$ is primarily due to the prior assumption of a minimum mass cutoff, and not due to information inferred about the shape of the distribution in that regime. 

\begin{figure}
	\centering
	\includegraphics[width=0.9\linewidth]{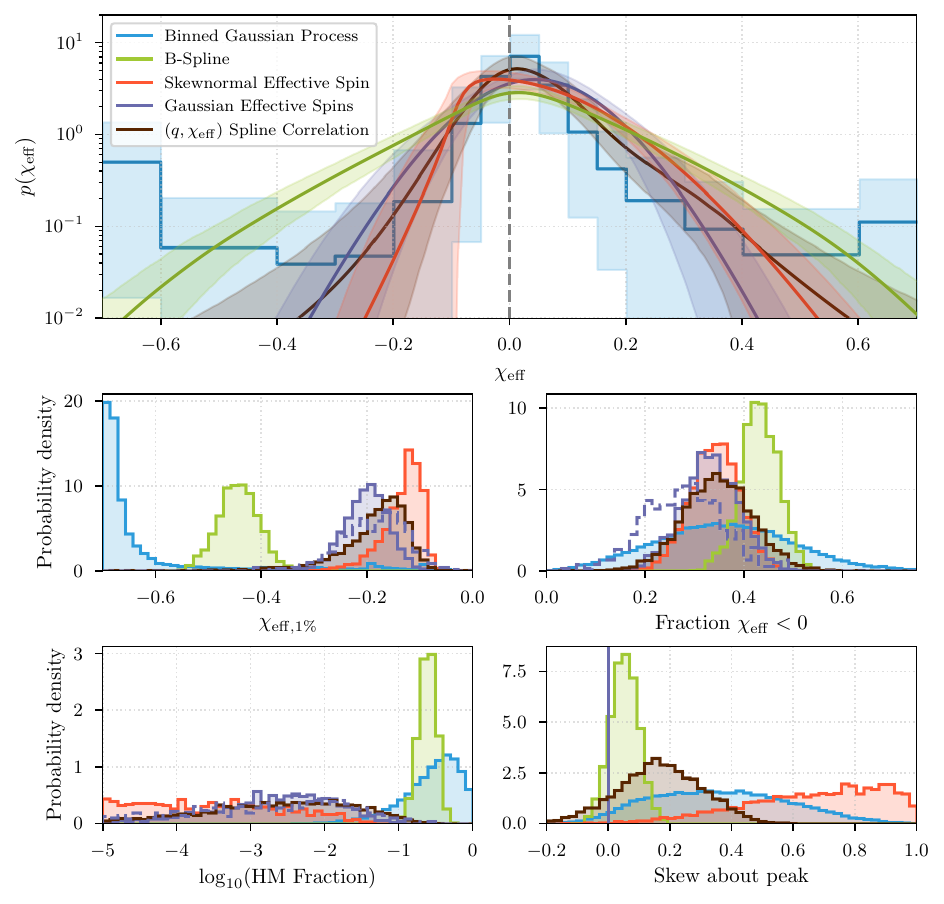}
	\caption{
		\textit{First row:} Marginal $\chieff$ distributions with BGP (blue) and \BSpline (green) models, as compared to the \skewnormalChiEff (red-orange), \effectiveSpinModel (purple), and $(q,\chieff)$ \splinecorrelation (maroon) models.
		The \ac{PPD} (average) is shown with a dark line, and the 90\% confidence intervals are shaded.
		\textit{Second and third rows:} Posteriors on the first percentile of the $\chieff$ distribution (upper left), the fraction of $\chieff<0$ (upper right),  HM fraction (lower left), and the skew (lower right).
		See Section~\ref{subsubsec:bbh_effective_spins} for a discussion of these quantities.
		Colors correspond to the top panel; purple dashed is \gwtcthree with the \effectiveSpinModel model for comparison.
	} 
	\label{fig:effective_spins_nonparametric}
  \end{figure}

\textit{Effective spin distributions}: In addition to the \skewnormalChiEff model and various correlation models presented for $\chieff$ in the main text (Figures~\ref{fig:chi_eff_chi_p_pdf} and \ref{fig:q_chieff_ppd}), we use the \nonparametric for a consistency check and find the distributions shown in the top panel of Figure~\ref{fig:effective_spins_nonparametric}. 
We plot the $\chieff$ distribution directly inferred with a binned Gaussian process (BGP; blue) as well as that reconstructed from the spin magnitude and tilt \BSpline model results (green), and compare them to the \skewnormalChiEff (red-orange), \effectiveSpinModel (purple), and $(q,\chieff)$ \splinecorrelation (maroon) results from the main text. 
The two approaches paint the same qualitative picture: the $\chieff$ distribution peaks at small values.
However, there is variation between the \nonparametric and \parametric; the \nonparametricAdj distributions are wider than the \parametricAdj distributions.
In Figure~\ref{fig:effective_spins_nonparametric}, we additionally plot posteriors distributions on statistics derived from each $\chieff$ distribution (c.f., Table~\ref{tab:min_chieff}), with results summarized as follows:
\begin{itemize}
	\item \textit{First percentile of the $\chieff$ distribution} ($\chi_{\mathrm{eff}, 1\%}$): 
	The three \parametricAdj distributions find consistent first-percentiles, around $\sim 0.2$, while the \nonparametricAdj find that the $\chieff$ distribution extends to lower values. 
	In the case of the BGP, the first-percentile distribution rails against the minimum allowed value of $-0.7$. 
	\item\textit{Fraction of \acp{BBH} with negative} $\chieff$: 
	All models yield consistent posteriors with one another and find that the fraction of \acp{BBH} with negative $\chieff$ is greater than zero and less than $\sim 0.8$. 
	The BGP posterior on this fraction is the widest, and the \BSpline posterior peaks at a higher values than the near-identical \parametricAdj posteriors.
	\item \textit{HM fraction}:
	The HM fraction is a heuristic for the upper limit of the fraction of \acp{BBH} coming from hierarchical mergers; it is calculated as $0.16$ times the fraction of $\chieff < -0.3$~\citep{Fishbach:2022lzq,Baibhav:2020xdf}. 
	We consistently constrain the HM fraction to be small, with the three parametric models finding it $\lesssim \mathcal{O}(10^{-1})$.
	The two non-parametric models, on the other hand, find it to be $\gtrsim \mathcal{O}(10^{-2})$.
	\item \textit{Skew of the $\chieff$ distribution about its peak}: 
	Each model finds fairly different values of the skew, which is here defined as the difference between the fraction of $\chieff > \chi_0$ and $\chieff < \chi_0$, where $\chi_0$ is the value of $\chieff$ at which the distribution peaks~\citep[][Equation 5]{Banagiri:2025dxo}.
	By design, the \effectiveSpinModel~model will always have a skew of $\sim 0$: truncated normal distributions are approximately symmetric about their peak if the truncation occurs substantially far from the distribution's bulk, which is here the case.
	All other models find a preference for positive skew, meaning that the distribution has more support for $\chieff$ above its peak than below. 
	This is the most pronounced for the \skewnormalChiEff, which is entirely inconsistent with a skew of 0. 
\end{itemize}

\subsection{Merger Rates Including Subthreshold Triggers}\label{appendix:FGMC_rates}

Search pipelines compute \acp{SNR} of detector data to identify portions of 
data with above-threshold \ac{SNR}, i.e., ``triggers", which may contain 
\ac{GW} events~\citep{GWTC:Methods}.
Based on pipeline-specific ranking statistics, triggers are assigned 
significances quantified by \acp{FAR}. 
In the main text, merger rates were calculated from population analyses that 
only used triggers surpassing a fixed-significance \ac{FAR} threshold, which was 
motivated to introduce minimal contamination from noise events.
By design, the list of triggers included in those analyses is threshold-dependent. 
Here, we explore the merger rates by including subthreshold triggers 
~\citep{Farr:2013yna, Kapadia:2019uut}, which ensures that the inferred rates 
are free of biases due to arbitrary significance thresholds and mitigates loss 
of information from excluding subthreshold GW candidates.
Specifically, we consider the full set of available triggers from a 
matched-filtering search, 
\GSTLAL~\citep{Messick:2016aqy, Sachdev:2019vvd, Hanna:2019ezx, Cannon:2020qnf, Ewing:2023qqe, Tsukada:2023edh, Sakon:2022ibh, Ray:2023nhx, Joshi:2025nty, Joshi:2025zdu}. 

The method used by \GSTLAL to self-consistently classify triggers and compute 
$p_{\rm astro}$~\citep{LIGOScientific:2018mvr, LIGOScientific:2021usb, KAGRA:2021vkt, Kapadia:2019uut, GWTC:Methods, GWTC:Results, Ray:2023nhx} 
values provides the rate densities described here for \acp{BNS}, \acp{NSBH}, 
and \acp{BBH}~\citep{GWTC:Methods, Kapadia:2019uut}.
The approach is a simplified version of the methods described in Appendix C 3 of 
\citet{KAGRA:2021duu}, as well as the discovery of GW200105 and 
GW200115~\citep{LIGOScientific:2021qlt} and GW230529~\citep{LIGOScientific:2024elc}. 
The simplification involves fixing the mass distribution to the Salpeter 
model~\citep{Salpeter:1955it} as opposed to marginalizing over population 
uncertainties inferred by the above-threshold event analyses explored in the main 
text.
We do not expect the marginalization over population uncertainties to have a 
significant impact on the inferred merger rates, given the uncertainty ranges.
When we apply the simpler model to \gwtcthree~\citep{KAGRA:2021duu}, the estimated
rates are comparable to those derived when marginalizing over the inferred population.
For effective inspiral spins of each trigger, we use a uniform distribution. 
Consistent with how searches categorize triggers and compute 
\VT~\citep{GWTC:Methods}, we set the boundary between \acp{NS} and 
\acp{BH} to be $3~\Msun$. 
For triggers categorized as \ac{BNS} or \ac{NSBH}, we set the bounds of the 
spins corresponding to a \ac{NS} to $\pm 0.4$.
These bounds are consistent with what the analyses in the main text adopt 
(see Section~\ref{subsec:event_selection} and Section~\ref{sec:ns}).

Using the fixed population distribution and the Poisson mixture model of 
~\cite{Kapadia:2019uut}, we  infer the merger rate densities of the different 
source categories (\ac{BNS}, \ac{NSBH}, and \ac{BBH}) from the full list of 
available \GSTLAL triggers with \ac{FAR} $<$ $1$ $\text{hour}^{-1}$.
This is done while self-consistently accounting for the 
possibility that some of these triggers are likely noise artifacts. 
We construct the posterior of astrophysical counts of \ac{BNS}, \ac{NSBH}, 
and \ac{BBH} events by utilizing mass-based binning template weights~\citep{Ray:2023nhx}.
We estimate the time--volume sensitivity~\citep{Tiwari:2017ndi, GWTC:Methods} 
$\VT$ for each event category $\alpha$ 
($\alpha = $ \ac{BNS}, \ac{NSBH}, \ac{BBH}), using injections~\citep{Essick:2025zed}, 
and their contributions to the counts posterior~\citep{Kapadia:2019uut}. 
Finally, we compute the rates posterior from marginalized counts posterior and 
the estimated $\VT$~\citep{LIGOScientific:2018mvr, LIGOScientific:2020kqk, KAGRA:2021duu, GWTC:Results}:
\begin{equation}
	p\left(R_\alpha\right) = p\left(\Lambda_{1 \alpha} | \mathbf{x} \right) \VT_{\text{O1-O4a}, \alpha} .
	\label{eq:fgmc}
\end{equation}
Here, $p\left(\Lambda_{1 \alpha} | \mathbf{x} \right)$ is the marginalized counts 
posterior where $\Lambda_{1 \alpha}$ is the astrophysical count for event 
category $\alpha$ and 
$\Lambda_{1 \alpha} = R_\alpha \VT_{\text{O1-O4a}, \alpha}$.
Here, $\text{O1-O4a}$ indicates that data from \ac{O1} throughout \ac{O4a} are used. 
The vector $\mathbf{x}$ represents the set of triggers using in this analysis, 
where each $x_i$ consists of the ranking statistics information, \ac{SNR}, and 
an identifier for the template associated with the trigger.

A Jeffreys prior, $\propto N^{-1/2}$, is imposed on the astrophysical counts 
for \acp{BBH} to construct their posterior from the mixed Poisson likelihood 
and compute the merger rate. 
The merger rate of \acp{BBH} is computed to be 
$\CIBoundsDash{\FgmcRates[RatesBBH]}$ $\perGpcyr$. 
This is consistent with the \ac{BBH} merger 
rate provided in the analyses presented in the main text, and the rate has been 
further narrowed compared to the results of \citet{KAGRA:2021duu} while still 
being consistent within uncertainties.  
A uniform prior is used for \acp{NSBH} and \acp{BNS}, as the number of detected 
events containing a \ac{NS} is small. 
We compute the \ac{NSBH} merger rates to be 
$\CIBoundsDash{\FgmcRates[RatesNSBH]}$ $\perGpcyr$, 
which is consistent with the \ac{NSBH} merger rate presented in main text. 
Similarly to the rate for \acp{BBH}, the rate of \acp{NSBH} is consistent with 
the rate obtained from the analyses of \gwtcthree but has been further narrowed.
The \ac{BNS} merger rate is calculated to be 
$\CIBoundsDash{\FgmcRates[RatesBNS]}$ $\perGpcyr$, 
which is consistent with the \ac{BNS} merger rate presented in the main text and 
has been narrowed down compared to the rate obtained from the analyses of 
\gwtcthree.

\subsection{Supplementary Results: BBH Correlations}\label{appendix:supplementary_correlations}

\begin{figure*}
	\centering
	\includegraphics[width=0.9\columnwidth]{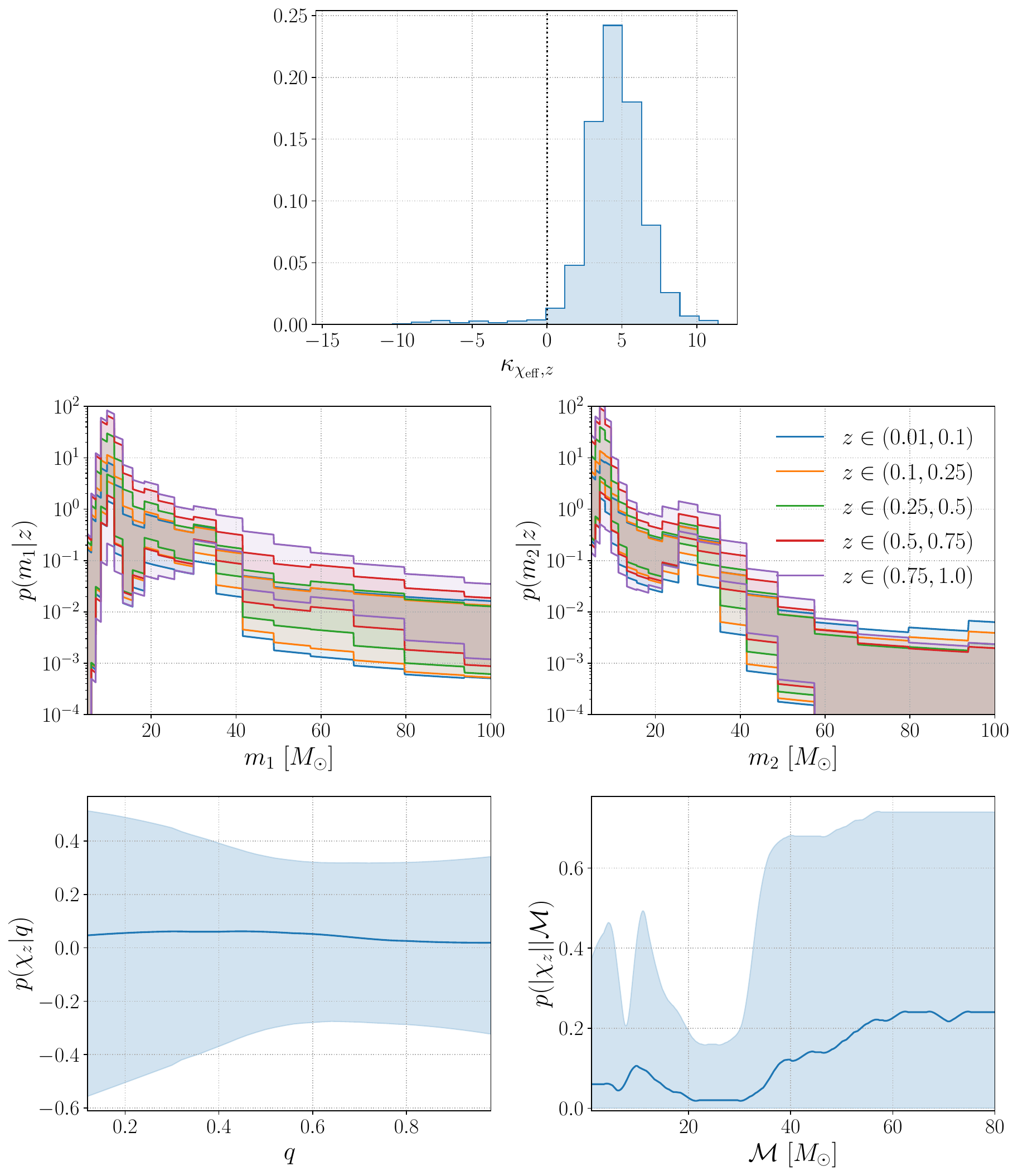}
	\caption{
	\textit{Top:} Posterior distributions for the level of correlation between redshift and effective inspiral spin $\kappa_{\chieff, z}$ inferred using the \copulacorrelation~model.
	The vertical black dashed line in each plot indicates a value of $\kappa_{\chieff, z}$, at which no correlation is implied.
	\textit{Middle:} Inferred distributions of primary mass (left), and secondary mass (right) in the redshift and spin correlated \ac{BGP}~analysis.
	The solid lines bound the \confidenceLevel~credible intervals for each redshift bin.
	We can see that all redshift bins from $z=0.01$ to $z=1$ are consistent within \confidenceLevel~credibility.
	\textit{Bottom:} Correlated mass and spin \acp{PPD} from the \vamana~model.
	Solid lines give the medians, while the shaded regions encompass 90\% of the \ac{PPD} volume.
	The left panel gives the inferred distribution of aligned spin $\chi_z$ given mass ratio.
	We do not see any evidence for or against a correlation.
	The right panel gives the inferred distribution of the aligned spin magnitude $|\chi_z|$ as a function of chirp mass.
	We see that the uncertainty is very large, with a notable drop in the ${\sim} 20{-}30 \,\Msun$ region.
	For reference, this region very roughly corresponds to the $30 \,\Msun$ peak observed in the component mass distributions (a $30 \,\Msun + 30 \,\Msun$ \ac{BBH} has a chirp mass of $\mathcal{M} \approx 26 \,\Msun$).
	Following the mass- and spin-correlated \ac{BGP} results then, it is unsurprising that the \vamana results prefer a local minimum in aligned spin magnitude in this region.
	}
	\label{fig:correlation_appendix}
\end{figure*}

Here, we supplement the correlated \ac{BBH} population results presented in Section~\ref{sec:Population-level correlations between parameters} with additional results and figures.
In the top panel of Figure~\ref{fig:correlation_appendix}, we plot the posterior distribution for the level of correlation $\kappa_{\chieff, z}$ inferred in the \copulacorrelation~model for $(z,\chieff)$.
Note the peak at positive values, and long tail into negative values.
The middle panels show the inferred mass distributions, using a mass-redshift correlated \ac{BGP}~analysis, binned by redshift.
We see that both the primary and secondary mass distribution appear to be broadly consistent across redshifts up to $z = 1$.
Finally, we investigate mass and spin correlations with the data-driven \vamana~analyses.
In the bottom panels of Figure~\ref{fig:correlation_appendix}, we plot the inferred distribution of orbital aligned spin $\chi_z$ as a function of both chirp mass and mass ratio, obtained with the \vamana~analysis.
We see that the results have a large degree of uncertainty, and are broadly consistent with findings explored here and in Section~\ref{sec:Population-level correlations between parameters}.

